\documentclass[a4paper,12pt]{article}
\usepackage{epsfig}
\usepackage{amssymb}

\newcommand{\nappend}{\setcounter{equation}{0}
\def\theequation{\rm{A}.\arabic{equation}}\section*}

\newcommand{\be}{\begin{equation}}
\newcommand{\ee}{\end{equation}}
\newcommand{\br}{\begin{eqnarray}}
\newcommand{\er}{\end{eqnarray}}
\newcommand{\ba}{\begin{array}}
\newcommand{\ea}{\end{array}}
\newcommand{\bi}{\begin{itemize}}
\newcommand{\ei}{\end{itemize}}
\newcommand{\bn}{\begin{enumerate}}
\newcommand{\en}{\end{enumerate}}
\newcommand{\bc}{\begin{center}}
\newcommand{\ec}{\end{center}}

\newcommand{\eps}{\epsilon}

\def\Ord{\buildrel{\scriptscriptstyle <}\over{\scriptscriptstyle\sim}}
\def\OOrd{\buildrel{\scriptscriptstyle >}\over{\scriptscriptstyle\sim}}

\newcommand{\gsim}{\lower.7ex\hbox{$\;\stackrel{\textstyle>}{\sim}\;$}}
\newcommand{\lsim}{\lower.7ex\hbox{$\;\stackrel{\textstyle<}{\sim}\;$}}

\newcommand{\plb}[3]{Phys.\ Lett.\ {\bf B#1} (19#2) #3}

\newcommand{\npb}[3]{Nucl.\ Phys.\ {\bf B#1} (19#2) #3}
\newcommand{\prd}[3]{Phys.\ Rev.\ {\bf D#1} (19#2) #3}

\newcommand{\prl}[3]{Phys.\ Rev.\ Lett.\ {\bf #1} (19#2) #3}

\newcommand{\cpc}[3]{Comp.\ Phys.\ Commun.\ {\bf #1} (19#2) #3}

\def\sign{\mathrm{sign}}

\def\eps{\varepsilon}

\def\M{ \overline{|\mathcal{M}|^2} }

\def\ar{\to}

\begin{document}
\tolerance=100000
\thispagestyle{empty}
\setcounter{page}{0}

\begin{flushright}
{\rm RAL-TR-1999-033}\\
{\rm April 1999} \\
\end{flushright}

\vspace*{\fill}

\begin{center}
{\Large \bf Higgs production in association with squark pairs\\[0.25cm]
in the Minimal Supersymmetric Standard Model\\[0.35cm]
at future hadron colliders}\\[2.cm]

{{\Large A. Dedes} {\large and} {\Large S. Moretti}} \\[3mm]
{\it Rutherford Appleton Laboratory,
Chilton, Didcot, Oxon OX11 0QX, UK} \\[10mm]
\end{center}

\vspace*{\fill}

\begin{abstract}{\small\noindent
We study neutral and charged Higgs boson production in 
association with stop and sbottom squarks at the Large Hadron
Collider, within  the Supergravity  inspired
Minimal Supersymmetric Standard Model.
The phenomenological 
relevance of such reactions is twofold. 
Firstly, they constitute a novel production mechanism of  
 Higgs particles, either through a decay of a heavier (anti)squark
into a lighter one or via a Higgs bremsstrahlung process. 
Secondly, their production rates are extremely sensitive to the values
assumed by the five input parameters of the model, this possibly allowing
one to put stringent constraints on the latter.
After an exhaustive scan of  the  parameter space, 
we find that the majority of such processes could be detectable at 
high luminosity, provided $\tan\beta$ is large, $\tan\beta\OOrd  30$ 
(except in
the case of $\tilde{t}_1\tilde{t}_1^*h$ and $\tilde{t}_1\tilde{t}_2^*h$ final 
states, whose
detection is also possible for smaller values), that the universal
soft Supersymmetric breaking masses are in the ranges $M_0 \Ord  500$ GeV and
$M_{1/2} \Ord  220$ GeV,   and that the trilinear couplings are negative,
$A_0 <  0$.
We also point out some sizable decay signatures and discuss their
Standard Model (SM) backgrounds. 
Finally,  
we derive compact analytical formulae
of the corresponding scattering matrix elements.
}\end{abstract}

\vspace*{\fill}
\newpage
\setcounter{page}{1}

\section{Introduction and motivation}
\label{sec:intro}

`If Supersymmetry (SUSY) exists, it will be discovered at the next generation
of hadronic machines', has been  a recurring motto so far. Indeed
sooner (at the Tevatron, $\sqrt s=2$ TeV) or later (at the Large
Hadron Collider (LHC),  $\sqrt s=14$ TeV), depending on the  mass scale
of the Higgs bosons and of the 
Superpartners of ordinary matter, several Supersymmetric  `signatures' 
should clearly be viable\footnote{For some reviews, see, e.g., 
\cite{Tevareview} and \cite{LHCreview}.}. 
Typical SUSY events at hadron colliders will involve either the production
and decay of heavy spartons, squarks and gluinos, whose
foreseen mass range is expected to be around the TeV scale  \cite{signals}, 
or of Higgs bosons \cite{Higgs},
primarily of the lightest one, for which the SUSY theory imposes a stringent
mass bound of the order of the electroweak (EW) scale.

However, even assuming that such a discovery will take place, there might 
well be little to learn about the
fundamental dynamics of SUSY from such new events.
In fact, although, e.g., the LHC
is able to produce gluinos and squarks with masses up to 2 TeV or so
and their detection has been shown to be feasible with rather little 
effort \cite{HPSSY}, 
it is much more difficult to determine exactly the SUSY masses involved, 
because in most models (i.e., those 
assuming $R$-parity conservation) there are at least two 
missing SUSY particles in each event. Clearly, failing the knowledge 
of the SUSY mass spectrum, other typical SUSY quantities, such as couplings, 
decay rates, etc., cannot be assessed either. Needless to say,
their measurement would be of paramount importance 
in order to constrain the free parameters entering the SUSY Lagrangian. 
However, by resorting to specific kinematic distributions \cite{HPSSY},
it is at least possible to make precision measurements of some `combinations'
of SUSY masses, but only in a few fortunate cases these can lead to strong
constraints on the theory and its parameters. 
Besides, in minimal SUSY theories, the Higgs sector typically  
(i.e., at tree level) depends
on only two such parameters, the ratio of the vacuum expectation values (VEVs)
of the Higgs fields and one of the masses of the five physical states 
corresponding to the latter, 
as all others SUSY inputs enter through higher perturbative orders. Therefore,
even the detection of a SUSY Higgs
signal would carry very poor information in terms of the underlying SUSY
model.
As a matter of fact,  a second question about SUSY has to  
legitimately be risen. 
Namely, `Which Supersymmetric model will one discover ?'

Thus, the key task for the Tevatron and the LHC is 
{not} only to find SUSY,  but also to assess which model is
behind it and the value of its parameters. 
For example, in the context of Supergravity (SUGRA) inspired  models
\cite{SUGRA}, with the minimal particle content  of the  MSSM 
(henceforth denoted as M-SUGRA, that we take to be the reference framework
of our analysis) \cite{MSSM,MSUGRA},  
the dynamics of the theory can be specified by only five 
entries: (i) a universal scalar mass $M_0$; (ii) similarly,
that for the gauginos $M_{1/2}$; (iii) the 
universal trilinear breaking terms $A_0$
(all defined at the Grand Unification scale $M_{\mathrm{GUT}}$ \cite{graham}). 
After the radiative EW symmetry breaking has taken place, 
two further parameters are needed
to describe the low energy dynamics: (iv) the mentioned
ratio of the VEVs of the two Higgs fields, 
denoted by $\tan\beta\equiv v/v'$ and defined at the EW scale; and (v) 
a discrete parameter, ${\mbox{sign}}(\mu)=
\pm1$, being $\mu$ the Higgsino mass term. 

Assuming universal soft breaking terms at the GUT scale, one is then 
able to calculate the masses of SUSY (s)particles, their couplings, decay 
rates, etc., at the EW scale, through the evolution of the renormalisation 
group equations (RGEs), the latter involving $M_0$, $M_{1/2}$, $A_0$, 
$\tan\beta$ and ${\mbox{sign}}(\mu)$ as inputs. 
Ultimately, a comparison of such predictions 
with the corresponding experimental measurements, as reconstructed 
from the actual data via dedicated Monte Carlo (MC) 
simulations \cite{isajet}--\cite{herwig},
should allow one to impose indirect constraints on the above parameters.
Indeed, an additional procedure to follow in order to determine the
latter could well be to search for the evidence of some more exotic signals 
of SUSY, in which, however, the dependence on such parameters is 
somewhat more manifest.

Elementary processes of the type
\begin{equation}\label{proc}
g + g \longrightarrow
{\tilde{q}}_{\chi} + {\tilde{q}}^{'*}_{\chi'} + \Phi,
\end{equation}
where $q^{(')}=t,b$, $\chi^{(')}=1,2$ and $\Phi=H,h,A,H^\pm$, in all
possible combinations, as appropriate in the MSSM,
serve the double purposes of:
\begin{enumerate}
\item furnishing production mechanisms of Higgs bosons of the MSSM,
both neutral and charged, in addition to the Standard Model (SM)-like 
channels \cite{Higgs};
\item yielding production rates, for particular combinations of 
$q^{(')}$, $\chi^{(')}$ and $\Phi$, strongly dependent
on some of the fundamental SUSY parameters of the M-SUGRA model.
\end{enumerate}

The importance of the first point should be understood in the
following terms. On the one hand, the detection of all neutral 
Higgs particles
$H,h$ and $A$ of the theory is not certain, neither at the
Tevatron \cite{Tevareview} nor even at the LHC \cite{LHCreview}. In 
addition,  the discovery potential of heavy charged Higgs bosons
$H^\pm$ at both the above colliders  has been proved to
be extremely limited \cite{charged}. 
Under these circumstances, the possible existence of 
novel and detectable Higgs production channels represents a phenomenologically 
important result per se. (Notice that, for certain choices of
$\sign(\mu)$, $A_0$ and $\tan\beta$, the squark-squark-Higgs coupling
can become the largest EW coupling of the SUSY theory, even exceeding the
standard Yukawa ones.) On the other hand, 
the fact that in processes (\ref{proc}) the Higgs bosons are produced
in association with squarks via a Yukawa bremsstrahlung or in 
`non-dominant' squarks decays\footnote{So that the corresponding 
partial widths are
significantly different from the total widths, thus retaining the
dynamics of the squark-squark-Higgs production vertices also at decay 
level.}, 
implies that the Higgs mechanism
can be probed in the  sparticle sector too. In fact, other known
production and decay mechanisms used to detect MSSM Higgs bosons 
mainly involve Higgs
couplings to ordinary matter. The only exceptions are the squark 
loop-contributions to neutral Higgs boson production via
 gluon-gluon fusion and to Higgs boson decays through pairs of 
photons/gluons \cite{xggh}, which are however 
swamped in both modes by the dominant terms involving
ordinary heavy particles.

As for the second point that we put forward, we should remind the reader of the
actual form of the mentioned squark-squark-Higgs vertices in the 
MSSM (which can be 
found, e.g., in \cite{HHG}). In many of these,  namely when 
$\chi\ne\chi'$, {both} the low-energy SUSY parameters $\mu$ and $\tan\beta$
enter {explicitly} in the Feynman rules, other than {implicitly}
in the scalar masses. Furthermore, those vertices also contain 
$A_{{{q}}^{(')}}$, the trilinear couplings at the EW scale, which
depend critically on their common  value at the GUT scale, $A_0$.
(In the case $\Phi=A$ such a dependence is also not affected by the mixing
between the chiral, $\chi^{(')}=L,R$, and physical, $\chi^{(')}=1,2$,
squark states, so that no additional SUSY mixing parameters
enter the phenomenology of pseudoscalar Higgs production, this rendering
the latter  an ideal laboratory to study M-SUGRA effects \cite{short}).  
Therefore, one should expect a significant dependence of the production rates
of the scattering processes (\ref{proc}) 
on $\tan\beta$, 
$A_0$ and $\mu$ (particularly, its sign), this possibly yielding a new 
profitable mean to constrain
the underlying SUSY model. Even more so in the case $\tan\beta$ has 
previously been determined, for example, through a discovery in the 
MSSM Higgs sector. 

Concerning previous literature on the subject, we should mention that reactions
of the type (\ref{proc}) were were first
considered in Ref.~\cite{djouadi} for the case $g g \rightarrow \tilde{t}_1
\tilde{t}_1^* h$  in the so-called `decoupling' limit. 
Adopting the M-SUGRA scenario, associated production of both  
neutral and charged Higgs  bosons production with squark pairs --
with a special emphasis
on  CP-odd Higgs boson production -- was first consider by the
authors in \cite{short}. Furthermore, in Ref.~\cite{djouadi2} 
(and also \cite{fawzi})
light Higgs boson production in association with light top squarks was
reanalysed in the M-SUGRA scenario at both Tevatron and 
LHC\footnote{The same 
final state but produced
in $e^+e^-$ annihilations at the TeV scale, e.g., at the Next Linear Collider,
has been considered in Refs. \cite{djouadi2,ee}.}.
A general consensus on the possible detectability at the LHC of
$gg\ar {\tilde{t}}_1{\tilde{t}}_1^* h$ events emerged from
Refs.~\cite{djouadi,djouadi2,fawzi}, in the case of light top squarks and large
trilinear coupling. 

We generalise here those studies, 
as we consider the production of all possible Higgs
states, $\Phi=H,h,A$ and $H^\pm$, for a broader spectrum of their masses,
in conjunction with both squark flavours
that can
have sizable couplings to Higgs particles, ${\tilde{q}}={\tilde{t}}$ and
${\tilde{b}}$, the latter taken as not degenerate in mass, also
allowing for Higgs production in decay channels, when $\chi\ne\chi'$. 
On the other hand, to simplify the simulation, we restrict ourselves to the
case of gluon-gluon induced processes only, thus neglecting the case of 
quark-antiquark scatterings, which was instead considered in 
Refs.~\cite{djouadi,djouadi2}. This is however not restrictive. 
In fact, we have verified
that at the LHC the $gg$ contribution is around two order of magnitudes
larger than the $Q\bar Q$ one, in line with the findings of 
Refs.~\cite{djouadi,djouadi2}, well below the level of uncertainties 
arising in our computation from other sources (such as structure
functions, QCD $K$-factors,
etc.)\footnote{As for $\gamma,Z$ and $W^\pm$ $s$-channels, 
these are typically smaller by a factor of the order 
${\cal O}({\alpha_{\mathrm{em}}}/{\alpha_{\mathrm{s}}})^2$; whereas 
${\cal O}({\alpha_{\mathrm{s}}})^2$ gluino interactions in $t,u$-channels 
are suppressed by the small (EW induced) mixing between light quarks and
sbottom and stop squarks, as already remarked in \cite{djouadi}.}.
As for the Tevatron, we can anticipate  that the production cross sections
of processes (\ref{proc})
are generally very small (see also Ref.~\cite{djouadi2}), 
indeed below the level of detection over  most
of the M-SUGRA parameter space, so that we
neglect further consideration of this machine here. 

It is the purpose in this paper to assess the possible relevance of 
points 1. and 2. in phenomenological studies of SUSY to be carried
out at the LHC. In particular,  
the plan of the paper is as follows. In the next Section
we describe how we have proceeded in our calculations. In 
Sect.~\ref{sec:parameters} we illustrate the theoretical
model we have resorted to in our analysis.
Sect.~\ref{sec:results} presents some numerical results. 
A brief summary and our conclusions
are given in Sect.~\ref{sec:conclusions}. Finally, 
we collect  in 
Appendix the relevant analytical formulae that we have used.

\section{Calculation}
\label{sec:calc}
The leading-order
(LO) Feynman diagrams associated to processes of the type (\ref{proc})
in the unitary gauge are depicted in Fig.~\ref{fig:graphs}. 
The reader can find an analytical expression of the corresponding
matrix elements (MEs) in the Appendix. As a  test of 
the correctness of our
amplitudes, we have verified that they are gauge invariant by checking
various Ward identities of the theory, both analytically 
and numerically. 
The amplitudes squared have then been integrated over a three-body
phase space, using VEGAS \cite{VEGAS}, and convoluted with
gluon Parton Distribution Functions (PDFs), as provided by 
CTEQ(4L) \cite{CTEQ4}. The latter constitutes our default set,
taken at LO in order  to be consistent with our approximation in calculating
the scattering MEs. However,
in order to estimate the systematic error due the gluon behaviour
inside the proton, we also have resorted to other LO packages, such as 
MRS-LO(09A,10A,01A,07A) \cite{MRS98LO}. Typical differences among PDFs
were found to be less than 15--20\%. The centre-of-mass (CM) energy at 
the partonic level, $Q=\sqrt{\hat{s}}$, was the scale used to
evaluate both the structure functions
and the strong coupling constant (see next Section for the treatment of the
latter). 

Depending on the relative value of the final state masses in (\ref{proc}), 
$m_{{\tilde{q}}_{\chi}}$, $m_{{\tilde{q}}^{'*}_{\chi'}}$ and
$m_{\Phi}$, the production of Higgs particles can be regarded
as taking place either via a (anti)squark decay (if 
$m_{{\tilde{q}}_{\chi}} > m_{{\tilde{q}}^{'*}_{\chi'}} + m_{\Phi}$ or
$m_{{\tilde{q}}^{'*}_{\chi'}} > m_{{\tilde{q}}_{\chi}} + m_{\Phi}$)
or via a Higgs-strahlung (if 
$m_{{\tilde{q}}_{\chi}} < m_{{\tilde{q}}^{'*}_{\chi'}} + m_{\Phi}$ and
$m_{{\tilde{q}}^{'*}_{\chi'}} < m_{{\tilde{q}}_{\chi}} + m_{\Phi}$).
In the first case, to prevent our MEs from becoming singular, we need
to insert a finite width in the resonant
(anti)squark propagators, which we have 
done by adopting the Breit-Wigner expression given in the Appendix and the
appropriate numerical values for the widths, calculated as described in 
Sect.~\ref{sec:parameters}. Also in the second case, though no poles exist in 
the amplitudes, a finite width value has been retained in the propagators. 
Notice that we have treated the two processes on the same footing, without
making any attempt to separate them, as
for the time being we are only interested in the total production rates
of the 2 $\ar$ 3 processes (\ref{proc}), rather than in their 
subsequent decay distributions.

\section{The theoretical model and its parameters}
\label{sec:parameters}

In this paper we are going to display our results for 
squark-squark-Higgs production via processes
 (\ref{proc})  by assuming possibly
the simplest scenario in the choice of the soft SUSY breaking parameters
at the GUT scale. That is, the so-called minimal
Supergravity scenario or M-SUGRA inspired model, as already 
intimated in the Introduction. In this
scenario, the whole dynamics of the MSSM (which contains over
hundred  parameters in the case of conserved $R$-parity) at the GUT scale
is reduced to the three basic inputs already introduced: $M_0$, $M_{1/2}$
and $A_0$. The large top Yukawa coupling then triggers
the radiative EW breaking through the running 
of the soft Higgs breaking masses, from the GUT  scale down to EW regime.
{}From the minimisation conditions of the potential  
one can define the soft Higgs mixing parameter, $B$,
and the absolute value of the
Higgs mixing parameter of the SUSY potential, $\mu$. The model leaves
the sign($\mu$) and the value of $\tan\beta$ as further
undetermined parameters at the EW scale.

All  five M-SUGRA parameters enter into
the relevant Feynman rules for the squark-squark-Higgs vertices, 
either explicitly or implicitly (through the RGEs). 
These can be written in the physical squark basis $\tilde{q}_{1,2}$ as
\begin{eqnarray}\label{mixing}
\lambda_{\Phi\tilde{q}_1\tilde{q}'_1} &=& c_q c_{q'} \lambda_{\Phi
\tilde{q}_L\tilde{q}'_L} + s_q s_{q'} 
\lambda_{\Phi\tilde{q}_R\tilde{q}'_R}+
c_q s_{q'} \lambda_{\Phi\tilde{q}_L\tilde{q}'_R} + s_q c_{q'} 
\lambda_{\Phi\tilde{q}_R\tilde{q}'_L} ,\nonumber \\[2mm]
\lambda_{\Phi\tilde{q}_2\tilde{q}'_2} &=& s_q s_{q'} \lambda_{\Phi
\tilde{q}_L\tilde{q}'_L} + c_q c_{q'} 
\lambda_{\Phi\tilde{q}_R\tilde{q}'_R}
-s_q c_{q'} \lambda_{\Phi\tilde{q}_L\tilde{q}'_R} 
-c_q s_{q'}
\lambda_{\Phi\tilde{q}_R\tilde{q}'_L} ,\nonumber \\[2mm]
\lambda_{\Phi\tilde{q}_1\tilde{q}'_2} &=& -c_q s_{q'} \lambda_{\Phi
\tilde{q}_L\tilde{q}'_L} + s_q c_{q'} 
\lambda_{\Phi\tilde{q}_R\tilde{q}'_R}
+c_q c_{q'} \lambda_{\Phi\tilde{q}_L\tilde{q}'_R} 
-s_q s_{q'}
\lambda_{\Phi\tilde{q}_R\tilde{q}'_L} ,\nonumber \\[2mm]
\lambda_{\Phi\tilde{q}_2\tilde{q}'_1} &=& -s_q c_{q'} \lambda_{\Phi
\tilde{q}_L\tilde{q}'_L} + c_q s_{q'} 
\lambda_{\Phi\tilde{q}_R\tilde{q}'_R}
-s_q s_{q'} \lambda_{\Phi\tilde{q}_L\tilde{q}'_R} 
+c_q c_{q'}
\lambda_{\Phi\tilde{q}_R\tilde{q}'_L} , 
\label{fr}
\end{eqnarray}
where $\tilde{q}_{L,R}$ or $\tilde{q}'_{L,R}$ can in principle 
be any flavour of  chiral squarks. However, here we 
only focus our attention to the case of the 
the third generation of down and up squarks only, namely, sbottom and
stop scalars, whose physical mass eigenstates are denoted by
$\tilde{b}_{1,2}$ and $\tilde{t}_{1,2}$, respectively,
the subscript 1(2) referring to
the lightest(heaviest) of them. As usual, 
the Higgs fields are denoted by the generic symbol $\Phi$,
where $\Phi=H,h,A,H^\pm$.
All the $\lambda_{\Phi\tilde{q}_\chi\tilde{q}'_{\chi'}}$'s appearing in 
eq.~(\ref{mixing}) are function of $\mu, \tan\beta$ and $A_{t,b}$ and
can be read directly from the Appendix of Ref.~\cite{HHG}.
(We  
ignore the case of complex $\mu$ and $A_{q}$ ($q=t,b$) parameters by assuming
that their phases are very small, the preferred case following
 the measurements of the Electric Dipole Moments
\cite{nir}.)

Also the left-right squark mixing angles $s_q\equiv\sin\theta_q$ and
$c_q\equiv\cos\theta_q$ (here, $q=t,b$) depend on  the 
M-SUGRA parameters, since they read as
\begin{eqnarray}\label{angle}
\tan(2\theta_t) &=& \frac{2 m_t (A_t - \mu \cot\beta)}{M_{\tilde{Q}_3}^2-
M_{\tilde{U}_3^c}^2+(\frac{1}{2}-\frac{4 s_W^2}{3})M_Z^2 \cos2\beta}, 
\\[2mm] \nonumber
\tan(2\theta_b) &=& \frac{2 m_b (A_b - \mu \tan\beta)}{M_{\tilde{Q}_3}^2-
M_{\tilde{D}_3^c}^2+(-\frac{1}{2}+\frac{2 s_W^2}{3})M_Z^2 \cos2\beta},
\end{eqnarray}
with $M_Z$ the $Z$-boson mass and $s_W\equiv\sin^2\theta_W$ the sine
(squared) of the 
Weinberg angle, $m_t$ and $m_b$ the top  and bottom quark masses, 
where $A_t$ and $A_b$ are the trilinear couplings defined at the EW scale,
while 
$M_{\tilde{Q}_3}$, $M_{\tilde{U}_3^c}$ and $M_{\tilde{D}_3^c}$ are the running
soft SUSY breaking squark masses of the third generation,
for which we assume the values obtained
from their evolution starting from  a universal mass
at the GUT scale equal to $M_0$. 


Regarding  
the numerical values of the M-SUGRA parameters adopted in this paper,
we have proceeded as follows. For a start, we have set
$M_0=M_{1/2}=150$ GeV. Such rather
low values for the universal masses come as natural first choice,
if one is interested in detecting processes of the type
(\ref{proc}). For two simple reasons. On the one hand, 
these two quantities determine 
the actual $m_{\tilde q_\chi}$, $m_{\tilde q'_{\chi'}}$  
and $m_\Phi$ values entering processes
(\ref{proc}), through their intervention in the RGEs, in such a way that 
small values of $M_0$ and $M_{1/2}$ at the GUT scale convert into a rather
light squark and Higgs mass spectrum at the EW scale. On the other hand,
being $2\to3$ body processes, a strong suppression from the phase space
would arise in squark-squark-Higgs production if the masses in the final state
were too large\footnote{The additional depletion coming from the gluon PDFs,
which would be probed at much higher values of $Q^2$ (of the order
of the rest masses or more), where they are naturally smaller, would in part
be compensated by the rise of the quark-antiquark initiated subprocesses:
i.e., $Q+\bar Q\rightarrow
{\tilde{q}}_{\chi} + {\tilde{q}}^{'*}_{\chi'} + \Phi$,
where, again, one has that
 $q^{(')}=t,b$, $\chi^{(')}=1,2$ and $\Phi=H,h,A,H^\pm$.}.
For the above choice, the M-SUGRA model predicts squark and 
heavy Higgs masses in the region of 80--450 GeV (as we shall see in more
detail  below), so that the latter can
in principle materialise at LHC energies 
(further recall that the light Higgs mass is bound to be below 130 GeV). 

Then, we have varied the trilinear soft Supersymmetry
breaking parameter $A_0$ in a  region where it changes its
sign, e.g., $(-300,+300)$ GeV, while we spanned the $\tan\beta$ value
between 2 and 40.
As for $\mu$, whereas in our model its magnitude is constrained, its
sign is not. Thus, in all generality, we have explored both the possibilities
$\sign(\mu)=\pm1$.

As a further step of our analysis, we have then come back to
$M_0$ and $M_{1/2}$ and change them, while maintaining $\tan\beta$,
$A_0$ and sign$(\mu)$ fixed at some specific values.
We have done so only for those processes that we had already identified
to have not only a large cross section, but also a strong dependence on
one or more of these three M-SUGRA parameters.

Given the strong phase space suppression induced by the consequent increase
of $m_{\tilde{q}_\chi}$, $m_{\tilde{q}'_{\chi'}}$ 
and $m_\Phi$ in the final states,
we will cautiously maintain the universal scalar and gaugino masses
below 250--300 GeV (at least at first). 
However, the reader should not 
assume that this is a necessary condition to the experimental detection
of processes of the type (\ref{proc}). In fact, this need not be true, 
as we shall show that even for $M_0$ values as large as 500 GeV one
can find sizable squark-squark-Higgs production cross sections
for $M_{1/2}$ up to 200--250 GeV, as long
as  $A_0$ is strongly negative, $\tan\beta\OOrd  30$ and
 sign$(\mu)<0$.

Such unexpected behaviours  are strongly driven by the intervention
in the production rates of $\tan\beta$, $A_0$ and $\sign(\mu)$,
through the trilinear scalar couplings  
$\lambda_{\Phi\tilde{q}_\chi\tilde{q}'_{\chi'}}$, more than by the  
actual values of $m_{\tilde{q}_\chi}$, $m_{\tilde{q}'_{\chi'}}$ and 
$m_\Phi$. This is evident by a mere look at the standard Feynman rules,
as can be found, e.g., in Ref.~\cite{HHG}. 
We will make our concern in this paper that of guiding the reader 
through such delicate interplay between masses and couplings, by explicitely 
writing down the expression of the relevant vertices in those parameter
space domains where such complicated phenomenology manifests itself.

{Starting from the five M-SUGRA parameters  
$M_0$, $M_{1/2}$,  $A_0$,  $\tan\beta$ and
${\mbox{sign}}(\mu)$, we have
generated the spectrum of masses, widths, couplings
and  
mixings relative to squarks and Higgs particles entering reactions (\ref{proc})
by running the {\tt ISASUGRA/ISASUSY} programs contained in the latest
release of the package {\tt ISAJET} \cite{isajet}, version 7.40. 
The default value of the
top mass we used was 175 GeV.
Note that also typical EW parameters, such as $\alpha_{\mathrm{em}}$ and 
$\sin^2\theta_W$, were taken from this program, as they enter the 
RGEs of the SUSY theory. 
Concerning the value of the strong coupling constant,
$\alpha_{\mathrm{s}}$, entering the production processes
(\ref{proc}), we have proceeded as follows.
By using as inputs the extracted value of $\alpha_{\mathrm{s}}$
at the $M_Z$ scale,   we 
evolve it up to any scale $Q$ by making use of the 
two-loop renormalisation group equations and by taking into
account all the low-energy threshold effects from the various SUSY masses
by means of the theta function approximation, as discussed in
Ref. \cite{sakis}.

Finally, notice that in scanning over the M-SUGRA parameter space, one should 
make sure that the values generated for the Higgs boson and sparticle
 masses 
are in accordance with current experimental bounds.
Signs of the sort ``$\times$'' or shaded areas appearing in
our figures in forthcoming Sect.~\ref{sec:results} will correspond to
already prohibited areas for the parameter space of our SUSY model.
We nonetheless leave  them 
for illustrative purposes, in order to visualise the typical impact
of present and future experimental bounds on the phenomenology of our
reactions. For example, the M-SUGRA points individuated by the
combinations $M_0=M_{1/2}=130-150$ GeV, $\tan\beta=2$, 
$A_0=0$ GeV and $\sign(\mu) = -$, used in some of the 
tables and figures
in the next Section, contradict the limits on the
lightest Higgs boson mass from direct searches \cite{lep,ALEPH},
as they yield $m_h=72-80$ GeV. We will discuss the experimental bounds
in more detail in the next Section.

As for  theoretical constrains, these arise from two sources: namely,
the absence of charge and 
colour breaking minima and that of large contributions to the
EW observables
that are measured with high precision at LEP. The former is avoided when
the following inequalities hold tree level~\cite{lahanas}:
\begin{eqnarray}
A_t^2 \ &<& \ 3 \left (m_{\tilde{Q}_3}^2+m_{\tilde{U}^c_3}^2+\mu^2 +m_{H_2}^2
\right ) , \nonumber \\
A_b^2 \ &<& \ 3  \left (m_{\tilde{Q}_3}^2+m_{\tilde{D}^c_3}^2+\mu^2 +m_{H_1}^2
\right ) , \nonumber \\ 
A_{\tau}^2 \ &<& \ 3 \left (m_{\tilde{L}_3}^2+m_{\tilde{E}^c_3}^2
+\mu^2 +m_{H_1}^2
\right ) ,
\label{ccb}
\end{eqnarray}
where all masses appearing in eq.~(\ref{ccb}) are the soft Supersymmetry 
breaking masses except  the Higgsino mixing parameter $\mu$. When $A_0$
is below 1 TeV, as is the case in our analysis,  
the above constraints are always satisfied even for very light squark masses.
As for the contributions to the EW observables, we have found the region
$150~{\mathrm{GeV}}\Ord M_0,M_{1/2} \Ord 500$ GeV, 
$2\Ord \tan\beta \Ord 40$ and $|A_0| \Ord  900$ GeV covered by our analysis
in agreement with the most recent measurements 
 of the `effective' weak mixing 
angle, $\sin^2\theta_{\mathrm{eff}}=0.2321\pm 0.010$,
and of the $W^\pm$ mass,
 $M_W=80.388\pm 0.063$ GeV \cite{erler}. In particular, 
notice that the M-SUGRA prediction for $\sin^2\theta_{\mathrm{eff}}$
decreases for a lighter mass spectrum while it becomes constant
in the heavy mass region \cite{sakis1}. 

\section{Results}
\label{sec:results}

We begin this Section by analysing 
all reactions (\ref{proc}) in the low mass regime, i.e., that  induced
by values of $M_0$ and $M_{1/2}$ below 250--300 GeV.
This is done in Subsect.~\ref{lightspectrum}. In this scenario, we
will first present and discuss, 
for future reference,  the values of squark and Higgs
masses resulting from the RGE evolution:
 in \ref{squarks} and \ref{higgses}, respectively. Then,
we will move on to considering the production of, in
turn: neutral CP-even (in \ref{even}),
CP-odd  (in \ref{odd}) and charged (in \ref{charged})
Higgs bosons. Possible decay signatures of the latter will be
analysed in \ref{signatures}. Finally, Subsect.~\ref{heavyspectrum}
will pin-point those unusual cases discussed above, in which the suppression 
from very heavy scalar masses in the final 
states of reactions (\ref{proc}) can be overcome by strong vertex effects,
yielding in the end sizable cross sections.

\subsection{Light mass spectrum}
\label{lightspectrum}

Once  fixed $M_0=M_{1/2}=150$ GeV, one obtains the (sbottom and
stop) squark and Higgs masses reported in 
Figs.~\ref{fig:squarks} and \ref{fig:higgs}, respectively,
depending on 2 $\Ord\tan\beta\Ord$ 35 and for $A_0=-300,0,+300$~GeV. 
The two possible options for the sign of $\mu$ are also contemplated.

Far from willing to discuss exhaustively the dependence of
$m_{\tilde q_\chi}$, $m_{\tilde q'_{\chi'}}$ and $m_\Phi$
upon the five M-SUGRA parameters, we limit ourselves here to spotting 
in Figs.~\ref{fig:squarks}--\ref{fig:higgs} some interesting trends, that
will affect the overall behaviour of the squark-squark-Higgs cross sections
that we will be treating below. For a more complete overview, see, e.g.,
Refs. \cite{pokorski}.

\subsubsection{Squark masses}
\label{squarks}

The four squark flavours
of the first and second generation ($\tilde{q}=
\tilde{u} , \tilde{d} , \tilde{c} ,
\tilde{s}$) with left- and right-handed components are all 
nearly degenerate in mass and the latter is
given approximately by the following formula:
\begin{equation}
m_{\tilde{q}} \simeq \sqrt{M_0^2 + 6 M_{1/2}^2 },
\qquad\qquad\qquad
(\tilde{q}=
\tilde{u} , \tilde{d} , \tilde{c} ,
\tilde{s}).
\end{equation}
For our choice of $M_0$ and $M_{1/2}$, one gets 
$m_{\tilde q}\simeq 400$ GeV. The
light squark flavours are not our concern 
in processes (\ref{proc}), though they may well enter some of the
decay chains of the other (pseudo)scalar particles produced.

The two squark flavours of the third generation must be 
treated differently because the off-diagonal entries of their
mass matrices can be large, owing to the strength of the 
Yukawa couplings of the corresponding quarks 
 (in this respect, notice that
the bottom one becomes
comparable to that of the top in the large  $\tan\beta$ region \cite{HHG}).
It thus follows that
 the  mass eigenstates $\tilde{t}_1$, $\tilde{t}_2$,
$\tilde{b}_1$ and $\tilde{b}_2$ are all different and generally
smaller than $m_{\tilde q}$.  Among these,
$\tilde{t}_1$ is  most often significantly lighter than all the
other stop and sbottom states. 
At large $\tan\beta$, the same happens 
to the $\tilde{b}_1$ mass eigenstate, as
large values of $\tan\beta$ correspond to smaller sbottom masses, so that
one eventually gets that $m_{\tilde{t}_1} \simeq m_{\tilde{b}_1}$.

Variation of the trilinear couplings can also  cause significant
differences between the light stop and sbottom masses. Finally,
the $\sign$ of the Higgsino mass term
plays an important r{\^o}le when $\tan\beta$ is either small or large,
for the cases  $m_{\tilde{t}_{1,2}}$
and $m_{\tilde{b}_{1,2}}$, respectively.

Experimental limits on the squark masses come from searches at
Tevatron and LEP2.
The most stringent bound on the ${{\tilde t}_1}$ mass comes from the
hadron collider \cite{Tevatron}: in absence of mixing, values
of $m_{{\tilde t}_1}<120$ GeV are excluded 
for\footnote{Recall that in M-SUGRA the Lightest Supersymmetric Particle
(LSP) is the lightest neutralino, ${\tilde\chi}_1^0$.} 
$m_{\mathrm{LSP}}\equiv m_{\tilde{\chi}_1^0} <$ 38 GeV.
D\O ~exclude values of $m_{\tilde{b}_1}$ below 85 GeV 
for $m_{\tilde{\chi}_1^0} <$ 47 GeV \cite{d0} 
and ALEPH  do over
the region $m_{\tilde{b}_1}<83$ GeV for any value of LSP mass \cite{aleph1}. 
In addition, 
CDF exclude masses for the 
lightest top squark up to 120 GeV when the LSP is 
$m_{\tilde{\chi}_1^0} <$ 50 GeV \cite{Tevatron}. 
Finally, D\O, using 
data corresponding to an integrated luminosity of 79 pb$^{-1}$, contradicts
all models with $m_{\tilde{q}\ne\tilde{t}_1,\tilde{b}_1
}<$ 250 GeV for $\tan\beta\Ord2$, $A_0=0$ GeV and
 $\mu< 0$  \cite{d01} (in scenarios with equal squark and gluino masses 
the limit goes up to $m_{\tilde{q}\ne\tilde{t}_1,\tilde{b}_1}<260$ GeV). 

\subsubsection{Higgs boson masses}
\label{higgses}

The mass of the lightest 
 Higgs boson is here constrained to be less than 115 GeV
and it appears to exhibit constant values over the region
 $8\Ord\tan\beta \Ord 35$, for a given combination
of $A_0$ and $\sign(\mu)$.
To change either the trilinear coupling or the sign of the Higgsino mass
has the net effect of scaling $m_h$, lower with increasing $A_0$
and higher for positive $\mu$.

As for the masses of the other Higgs bosons,
they are nearly degenerate. They vary between  
180 GeV (when $\tan\beta$ is large) and 400 GeV 
(when $\tan\beta$ is small). The dependence on $\sign(\mu)$
is generally negligible, whereas the one on $A_0$ is very strong
in the intermediate $\tan\beta$ regime. One thing worth noticing here
is the existence of a point  
where all  three Higgs masses $m_H,m_A$ and $ m_{H^\pm}$ converge,
regardless of the values of $A_0$ and $\sign(\mu)$. This occurs
for $\tan\beta \approx 30$, where
\begin{equation}
m_H=m_A=m_{H^\pm} \approx 200~{\rm  GeV}.
\end{equation}

As for experimental bounds, 
LEP experiments have combined their results from data taken 
at CM energies from 91 to 183 GeV to place lower
 bounds on the masses of the light ($m_h$) and pseudoscalar ($m_A$)
Higgs bosons, of 78.8 and 79.1 GeV, respectively \cite{hocker}. In addition,
they exclude the range $0.8<\tan\beta <2.1$ for minimal stop mixing 
and $m_t=175$ GeV. 
Also, D\O\ \cite{d02} have recently removed at 95\%
confidence level the intervals $\tan\beta <0.97$ and $\tan\beta > 40.9$ for
$M_{H^\pm}=60$ GeV and $\sigma(t\bar{t})=5.5$ pb (again, with $m_t=175$ GeV).
However, the limits become less stringent with increasing $M_{H^\pm}$:
e.g., for $M_{H^\pm}>124$ GeV (as is the case here)
the available angular range is $0.3<\tan\beta<150$.

\subsubsection{CP-even Higgs boson production}
\label{even}

We present in 
Figs.~\ref{fig:lighthiggs}--\ref{fig:heavyhiggs} 
the $\tan\beta$ dependence of the
production cross sections of the two neutral scalar Higgs bosons,
$h$ and $H$, respectively, in association with any possible
combination of squarks of the third generation. Again, we parametrise
the dependence upon $A_0$ by adopting for the latter the discrete values
of $-300$, 0 and +300 GeV and we choose $\sign(\mu)=\pm1$.

Assuming an integrated luminosity of 100 inverse femtobarns 
over a twelve month period of running at the LHC
(i.e., $\int {\cal L} dt = 100 \; {\mathrm{fb}}^{-1}$), 
one can realistically hope for the detection of squark-squark-Higgs
processes only if the production cross section is above 1 fb or so.
In fact, we shall see in Sect.~\ref{signatures} how typical decay
fractions of clean signatures range at the level of 10\%
or below (see Subsect.~\ref{signatures} later on).

Under this assumption, one immediately sees that there are several
production channels of CP-even Higgs particles which could be observed,
over a large part of the M-SUGRA parameter space considered here.
Primarily, those involving the lightest  stop squark,  ${\tilde t}_1$,
particularly if also the lightest Higgs state is involved,
but not only.

For the case of $h$ production, 
there exists an approximate hierarchy of cross sections
which can  possibly be detected:
\begin{equation}\label{Xsections_h}
\sigma( \tilde{t}_1 \tilde{t}_1^* h )\OOrd 
\sigma( \tilde{t}_1 \tilde{t}_2^* h )\OOrd
\sigma( \tilde{b}_1 \tilde{b}_1^* h ).
\end{equation}
The cases  $\tilde{b}_2 \tilde{b}_2^* h$, 
$\tilde{t}_2 \tilde{t}_2^* h$
         and $\tilde{b}_1 \tilde{b}_2^* h$ 
never have significantly large rates.

In the case of $\tilde{t}_1 \tilde{t}_1^* h $ states, those with
largest production rates, one obtains 
\begin{equation}
\sigma (g g \rightarrow h \tilde{t}_1 \tilde{t}_1^* ) \OOrd 20 \; {\rm fb},
\end{equation}
for every combination of $\tan\beta$, $A_0$ and $\sign(\mu)$, in the case of 
$M_0=M_{1/2}=150$ GeV, thus a sort a lower limit over a
representative portion of the low mass regime of the M-SUGRA
scenario. Moreover, the largest production rate for
this  final state
(compatible with the current experimental constraints)
is obtained in the small $\tan\beta \sim 2 $ region and for 
the combinations  $A_0\Ord  -300$ and $\mu =+$, for which
\begin{equation}
\sigma ( \tilde{t}_1 \tilde{t}_1^* h) \OOrd  200 \; {\rm fb},
\end{equation}
corresponding to more than 20000 events per year 
running of the LHC.

The dominance of the production channel involving both the
lightest squarks and Higgs boson, above all other mechanisms
(\ref{proc}), was foreseeable. The reason is rather simple, 
in fact,   twofold. On the one hand, the sum of the rest masses in the
final states yields the smallest possible values, thus enhancing the volume
of the three-body  phase space, relatively to any other squark-squark-Higgs
combination. On the other hand, the cross section is also significantly 
enhanced when the trilinear coupling $A_0$ assumes negative values.
From the analysis of the RGEs we find that $A_t \sim -400 (-250)$ GeV
when $A_0=-300 (300)$ GeV. Now, the coupling of the light Higgs boson to
top squarks is driven by $A_t$ when the latter takes on large values.
More specifically, the corresponding vertex reads as
(recall that we are in the kinematic limit 
$m_{X}\approx m_H\approx m_A\approx m_{H^\pm}\gg M_Z$: see
Fig.~\ref{fig:higgs}) 
\begin{eqnarray}
\lambda_{h\tilde{t}_1\tilde{t}_1}^{m_{X}\gg M_Z}
 \ &\simeq& \
 \frac{i g_W M_Z}{c_W} \biggl \{ \biggl [
-\frac{1}{2} \cos 2 \beta \cos^2\theta_{\tilde{t}}
+ \frac{2}{3} s_W^2
\cos 2\beta \cos 2\theta_{\tilde{t}} \biggr ] \nonumber \\[2mm]
&-& \frac{m_t^2 }
{M_Z^2 } -\frac{m_t \sin 2 \theta_{\tilde{t}}}{2 M_Z^2}  
\left ( A_t + \mu \cot \beta \right ) \biggr \} \;,
\label{ht1t1aprox}
\end{eqnarray}
(here and in the following, $g_W^2=4\pi\alpha_{\mathrm{em}}/s^2_W$, 
$c_W=\sqrt{1-s^2_W}$ and $M_W^2=M_Z^2(1-s^2_W)$)
where  for large $A_t$ (note that
$\sin\theta_{\tilde{t}}$ is maximal in such a case)
the coupling goes like $\lambda_{h\tilde{t}_1\tilde{t}_1}
\propto \frac {m_t A_t}{M_Z^2}$. 
As a consequence, because of the presence of the
trilinear term $A_t$, the coupling of the light Higgs boson to 
light stop squarks
could be much larger compared to that to the top quark,
 which behaves like
 $\lambda_{h t t} \propto
\frac{m_t}{M_Z}$, as already
recognised in \cite{djouadi,djouadi2}. Over the parameter space
that we have chosen here, the cross section
of $g g \rightarrow \tilde{t}_1 \tilde{t}_1^* h$ can be either
larger than or of the same order as that of 
$g g \rightarrow t \bar{t} h$ \cite{djouadi,djouadi2}.

Thus, the subprocess $gg\rightarrow  \tilde{t}_1 \tilde{t}_1^* h$ can
well boast the status of an additional discovery mechanism of the
lightest scalar Higgs boson of the MSSM, as remarked in 
Refs.~\cite{djouadi,djouadi2}.
(This is true also in non M-SUGRA models, where however one still has that
$m_{H}\approx m_A\approx m_{H^\pm}\gg m_h$ \cite{djouadi,djouadi2}.)
In this respect, the reader should further notice the stability of its
production cross section against variations of $\tan\beta$ (see
also, e.g., Fig. 4 in \cite{djouadi2})\footnote{The $\tan\beta$ dependence
of $\sigma(\tilde{t}_1\tilde{t}_1^* h)$ at low values of such a parameter
is mainly a phase space effect, as it can be deduced by comparing
Fig.~\ref{fig:squarks} to Fig.~\ref{fig:lighthiggs}.
In addition, in the low $\tan\beta$ domain, there are residual effects
onto the production rates induced by the term $\mu \cot\beta$ arising 
from the off-diagonal elements of the squark mass matrices
and affecting the $\lambda_{h {\tilde t}_1{\tilde t}_1}$ vertex.}. 
This proves to be a
crucial point, as it is possible that one will have no narrow hints about
the actual value of this crucial parameter of the Higgs sector even after 
Run 2 at Tevatron (unless, of course,  
the lightest Higgs boson is discovered there !)
\cite{Tevatron}. In other terms,  
${\tilde t}_1{\tilde t}_1^* h$ would always be present at fixed
rate at the LHC, no matter whether $\tan\beta$ is large or small.
Similar arguments can be put forward concerning the $\sign(\mu)$ dependence.

Some care must instead be exercised with respect to the $A_0$ dependence.
In fact, to vary the universal trilinear coupling between, e.g., $-300$
and $+300$ depletes the cross section by a factor of about seven, as
shown in Fig.~\ref{fig:lighthiggs}. For even larger differences,
say, between $-500$ and $+500$ (not shown here for
reasons of space), the ratio between the cross sections become as big 
as 30 ! Not surprisingly then, Ref.~\cite{djouadi2} focused on the
choice $A_0=-2000$ GeV\footnote{Note that in this $A_0$ region the
M-SUGRA scenario clashes against the constraints from 
 the charge and colour breaking minima, {i.e.}, eq.~(\ref{ccb}).
This is the reason why we prefer to display our results in a rather 
more conservative range, i.e., $|A_0| \Ord 1$ TeV.}  (and $\sign(\mu)=+$).
Far from regarding this dependence as a shortcoming of
${\tilde t}_{1}{\tilde t}_{1}^*h$ events in helping in the quest
for the so far elusive lightest Higgs boson of the MSSM
(in fact, even for very large and 
positive $A_0$ values we found the cross section well above 1 fb),
this example allows us to enlighten that other aspect of
squark-squark-Higgs production that we have mentioned in the
Introduction: i.e., its potential in pinning down
some of the fundamental parameters of the M-SUGRA model. Needless
to say, variations of the cross section with $A_0$ as large as those 
mentioned above are well beyond the various sources of uncertainties
on the production rates (other than the theoretical ones
related to the PDFs and the effect of higher-order QCD corrections
also the experimental ones in their determination). 
To measure a production rate 
of ${\tilde t}_{1}{\tilde t}_{1}^*h$ events much larger
than about 50 fb (the value obtained 
in correspondence of $A_0=0$), in some specific decay channel, would 
unambiguously imply that the universal trilinear coupling at the
GUT scale is negative. 

As already mentioned, very little could be learn
about the actual values of $\tan\beta$  from this
specific process. However, if the latter is known beforehand
to be around 2 or so, one could use this information to constrain 
$\sign(\mu)$. In fact, for $A_0=-300$ GeV, one would get that 
\begin{eqnarray}
\sigma (\tilde{t}_1 \tilde{t}_1^* h) &\approx& 200  \; 
{\rm fb} \; \Rightarrow \mu = + \\[2mm]
\sigma (\tilde{t}_1 \tilde{t}_1^* h) &\approx& 50 \; 
{\rm fb} \; \Rightarrow \mu = - \;.
\end{eqnarray}

Let us proceed in this spirit to see whether other 
channels can be of some help in constraining the M-SUGRA model. 
Following the list 
of detectable $h$ production cross section given
in (\ref{Xsections_h}), we find the 
$\tilde{t}_1 \tilde{t}_2^* h$ final state \cite{short}.
This is not surprising either. 
In fact, the relevant coupling behaves like (again, $X=H,A,H^\pm$)
\begin{eqnarray}
 \lambda_{h\tilde{t}_1\tilde{t}_2}^{m_{X}\gg M_Z}
 \ &\simeq & \ 
 \frac{i g_W M_Z}{c_W} \biggl \{ 
\frac{1}{2} \cos 2 \beta\biggl (\frac{1}{2}-\frac{4}{3}s_W^2
\biggr ) \sin2\theta_{\tilde{t}}
 \nonumber \\[2mm]
&-&\frac{m_t \cos 2 \theta_{\tilde{t}}}{2 M_Z^2}  
\left ( A_t + \mu \cot \beta \right ) \biggr \} \;,
\label{ht1t2aprox}
\end{eqnarray}
becoming very large when $\cos 2\theta_{\tilde{t}}$ and
$ ( A_t + \mu \cot \beta ) $ reach their allowed maximum values. 
From  Fig.~\ref{fig:lighthiggs}, one can see that
this happens in the region of
small $\tan\beta$, negative sign of $\mu$ and $A_0\Ord -300$ GeV.

The intriguing aspect here, which was largely missing in the case
in which both squarks were the lightest, is that one could impose
severe constraints on the sign of the Higgsino mass term, other
than on $A_0$. In fact, in the detectable region, the curves corresponding
to $\sign(\mu)=-$ (higher) and $\sign(\mu)=+$ (lower) depart considerably.
For example, for relatively small $\tan\beta$ values, say 4,
the ratios as obtained by dividing the cross sections corresponding to  
negative $\mu$'s by those for positive Higgsino masses are quite large
indeed: about 7(5)[2] when $A_0=-300(0)[+300]$ GeV. At even lower
$\tan\beta$, say, equal to 2, one symbolically has:
\begin{eqnarray}
\sigma (\tilde{t}_1 \tilde{t}_2^* h) 
&\approx& 300  \; 
{\rm fb} \; \Rightarrow \mu = - \\[2mm]
\sigma (\tilde{t}_1 \tilde{t}_2^* h) 
&\approx& 2 \; 
{\rm fb} \; \Rightarrow \mu = + \;,
\end{eqnarray}
(e.g., for $A_0=-300$ GeV).
Luckily enough here, where the solid and dot-dashed curves start
getting closer (for $\tan\beta\OOrd15-20$) is precisely 
when the cross section is no longer observable. However, what just said
 makes the point that $\tan\beta$ ought to be known rather accurately
from some previous measurements, if one wants to constrain the other M-SUGRA
parameters by studying the production of the lightest Higgs scalar
of the theory produced in association with both stop mass eigenstates.

In our list of observable $h$ cross sections ${\tilde{b}}_1{\tilde{b}}_1^* h$
comes next. Here the potential is somewhat complementary to the
two above cases, in the sense that not to find any pairs of sbottom squarks
of the type ${\tilde b}_1$ produced in association with an $h$ scalar
once $\tan\beta$ is already known to be large could have powerful
consequences on the viability of M-SUGRA as the underlying model of SUSY.
To be specific, notice how the six curves corresponding to all the possible
combinations of the parameters $A_0=-300,0,+300$ GeV and $\sign(\mu)=\pm$
lie within a factor from 2 to 4 in cross section, in correspondence of
$\tan\beta=20$ and 35, respectively. Even the cases of $A_0=\pm500$ GeV do
not depart significantly from the central curve for $A_0=0$, in the above
$\tan\beta$ region. Unfortunately, contrary to the case of 
${\tilde{t}}_1{\tilde{t}}_2^* h$ production, here the most interesting region
is presumably below detection level. In fact, for $\tan\beta$ quite low,
a huge portion of M-SUGRA parameter plane collapses into a narrow stripe,
as the various curves tend to overlap,  all being contained within a
factor as small as 1.5 (e.g., at $\tan\beta=2$, also including the cases 
$|A_0|=500$ GeV, not shown in the figure).

As for $H$ production, one identifies as possible candidates for
detection the following cases:
\begin{equation}\label{Xsections_H}
\sigma( \tilde{b}_1 \tilde{b}_1^* H )\sim 
\sigma( \tilde{t}_1 \tilde{t}_1^* H )\OOrd
\sigma( \tilde{b}_2 \tilde{b}_2^* H )\sim 
\sigma( \tilde{t}_1 \tilde{t}_2^* H ).
\end{equation}
The remaining two combinations, i.e., $\tilde{b}_1 \tilde{b}_2^* H$
and $\tilde{t}_2 \tilde{t}_2^* H$, yield cross sections
that are hopelessly small.

Also some of the detectable $H$ production processes 
can have a significant impact in aiding
the determination of the M-SUGRA parameters, most notably those
yielding the final states ${\tilde{b}}_1{\tilde{b}}_1^* H$ and
${\tilde{b}}_2{\tilde{b}}_2^* H$. Here, if $\tan\beta$ is known to be, say,
35 or so, a detection of either the former or the latter
by the thousand or hundred, respectively, would
imply that $A_0$ is most certainly negative, since production rates 
corresponding
to $A_0$ values larger than zero are about a factor of 5 and 3 smaller
(rather irrespectively of $\sign(\mu)$). Somewhat less discrimination
power between positive and negative $A_0$ values have 
${\tilde{t}}_1{\tilde{t}}_1^* H$ and
${\tilde{t}}_1{\tilde{t}}_2^* H$ events, over the same (large) $\tan\beta$
region as above. The most interesting case would have been 
${\tilde{t}}_2{\tilde{t}}_2^* H$, as the collapse of the M-SUGRA
model that we already
noticed in the case of ${\tilde{b}}_1{\tilde{b}}_1^* h$ final states is
here even more striking, over a more
significant $\tan\beta$ region. Unluckily enough,
the corresponding production cross section never exceeds the femtobarn level.

A general remark in now in order, concerning the strength of the
coupling of sbottom squarks to neutral CP-even Higgs bosons.
The monotonic growth of the production rates of
sbottom squark processes with increasing $\tan\beta$,
as opposed to a much milder dependence of the stop ones, has a simple 
explanation.
For example, the coupling $\lambda_{h  \tilde{b}_1 \tilde{b}_1}$ is driven by
the term $\frac{m_b \mu \tan\beta}{M_Z^2}$ and its size
becomes large for large $\tan\beta$.
Another reason for an extra enhancement of the sbottom production rates comes
from the phase space available to the final states, as both
$m_{\tilde{b}_1}$ and $ m_{\tilde{b}_2}$ decrease very fast
when $\tan\beta$ gets large, whereas
this is much less the case for $m_{\tilde{t}_1}$ and $ m_{\tilde{t}_2}$
(see Fig.~\ref{fig:squarks}). There is however a point, for the
case of reactions involving the two sbottom mass eigenstates at once
(i.e., ${\tilde{b}_1}{\tilde{b}_2}^*h$ and
${\tilde{b}_1}{\tilde{b}_2}^*H$), in which the production cross sections
vanish altogether, somewhere in the vicinity of $\tan\beta=34-36$
(the zero for
${\tilde{b}_1}{\tilde{b}_2}^*H$ is actually beyond the $\tan\beta$ 
range plotted), 
the exact value depending upon  $A_0$ and $\sign(\mu)$. This is
clearly induced by the $\lambda_{h {\tilde{b}}_1{\tilde{b}}_2}$
and $\lambda_{H {\tilde{b}}_1{\tilde{b}}_2}$ vertices and their
typical  $\propto(\mu-A_b \tan\beta)$ behaviour, when
$|\mu|\ll | A_b \tan\beta|$ and $A_b$ changes its sign.

Before closing this Section, we investigate the residual dependence
of CP-even Higgs boson production in association with sbottoms and stops
on the input values of $M_{0}$ and $M_{1/2}$, when they are allowed
to deviate from their common default of 150 GeV assumed so
far. For illustrative purposes, we 
 do so by adopting two discrete values of $\tan\beta$, 2
and 40, and choosing the combination $A_0=0$ and negative $\mu$.
Anyhow, though not shown, we have verified that a similar pattern
to the one that we will outline below 
can be recognised also for the case of finite (positive and negative)
values of $A_0$ and positive Higgsino masses as well.
For this exercise, we focus our attention only to the dominant
production cross sections in either case, that is, 
$\tilde{t}_1 \tilde{t}_1^* h$,
$\tilde{t}_1 \tilde{t}_2^* h$ and 
$\tilde{b}_1 \tilde{b}_1^* h$ for light 
(see Tab.~\ref{table1}) and 
$\tilde{t}_1 \tilde{t}_1^* H$ and
$\tilde{b}_1 \tilde{b}_1^* H$ for heavy (see Tab.~\ref{table2}) Higgs bosons.

\begin{table}
\begin{eqnarray}
\begin{array}{|c|c|c|c|c|c|} \hline
M_0 ({\rm GeV}) & M_{1/2} ({\rm GeV}) & \tan\beta & 
\sigma (g g \rightarrow \tilde{t}_1 \tilde{t}_1^* h ) ({\rm fb}) &
\sigma (g g \rightarrow \tilde{t}_1 \tilde{t}_2^* h ) ({\rm fb})  &
\sigma (g g \rightarrow \tilde{b}_1 \tilde{b}_1^* h ) ({\rm fb}) \\ \hline
130 & 130 & 2 & 70.2 & 38 & 7.7 \times 10^{-2}
\\ \hline
200 & 150 & 2 & 32 & 150 & 2.9 \times 10^{-2}
\\ \hline
200 & 200 & 2 & 11 & 100 & 6.6 \times 10^{-3}
\\ \hline
300 & 250 & 2 & 2.9 &48 & 1.4\times 10^{-3}
\\ \hline
130 & 130 & 40 & 84 & 8.2  & 8.2 
\\ \hline
200 & 150 & 40 & 37 & 7.4 & 4.9
\\ \hline
200 & 200 & 40 & 13 & 3.4 & 1.3 
\\ \hline
300 & 250 & 40 & 3.5 & 1.8 & 0.51
\\ \hline
\end{array}
\nonumber 
\end{eqnarray}
\caption{The variation of the most significant cross sections 
of processes $g g \rightarrow \tilde{q}_\chi \tilde{q}_{\chi '}^* h $ with
$M_0$, $M_{1/2}$ and $\tan\beta$. The other M-SUGRA parameters are 
fixed as follows: $A_0=0$ GeV and $\sign(\mu) = -$.}
\label{table1}
\end{table}

We obtain that most of the cross sections
with the light Higgs particle involved decrease when either or both 
the parameters $M_0$ and $M_{1/2}$ increase. In this respect, however, 
it is well worth noticing that the total cross section for 
$\tilde{t}_1 \tilde{t}_2^* h $ production acquires a large statistic
significance  in the higher mass regime and maintains it even at the
very upper
end of it (some 5000 events/year can be produced at the LHC via this mode
if $\tan\beta=2$, 
$M_0=300$ GeV and $M_{1/2}=250$ GeV). In this area of the M-SUGRA
parameter space, $\tilde{t}_1 \tilde{t}_2^* h $ events are much more
numerous than $\tilde{t}_1 \tilde{t}_1^* h $ ones, the other way
round with respect to the low mass combination $M_0=M_{1/2}=130$ GeV,
despite of the more massive final state.
(However, this is only true al low $\tan\beta$.)
In fact, for $M_0=300$ GeV and $M_{1/2}=250$ GeV, the
squark masses are $m_{\tilde{t}_1}=472$ GeV, $m_{\tilde{t}_2}=591$ GeV,
$m_{\tilde{b}_1}=569$ GeV and $m_{\tilde{b}_2}=623$ GeV. The inversion
of  hierarchy between the two cross sections is induced by the
onset of the ${\tilde{t}_2}\to {\tilde{t}_1} h$ decay channel at large
$M_{0}$ and $M_{1/2}$ values, as $m_{\tilde{t}_2}\OOrd m_{\tilde{t}_1}
+m_h$, whose resonance enhancement in the $2\to2$ process 
$gg\to {\tilde{t}}_2{\tilde{t}}_2^*$ overcomes both the inner phase space
depletion and the strength of the Higgs-strahlung emission in 
$gg\to {\tilde{t}}_1{\tilde{t}}_1^*$ events.

As for the case $\tilde{b}_1 \tilde{b}_1^* h $ (and similarly
for $\tilde{b}_2 \tilde{b}_2^* h $, not shown in the table), 
there is no inversion of tendency here, in the interplay with the
light stop channel, as production rates are strongly dominated by the fact
that $m_{\tilde{b}_2}, m_{\tilde{b}_1} \gg m_{\tilde{t}_1}$.
They both are very much suppressed. On similar grounds, one can 
argue about the smallness of $\tilde{t}_2 \tilde{t}_2^* h $.
Finally, being $m_{\tilde{b}_2}\Ord m_{\tilde{b}_1} +m_h$
in most part of the ($M_0$,$M_{1/2}$) plane that we have spanned, 
the $\tilde{b}_2 \tilde{b}_1^* h $ final state never stands up
either as quantitatively interesting.

The results for the mass dependence of the
production rates for the heavy (CP-even) Higgs boson have a simpler pattern.
The all of their production phenomenology
is governed by the fact that over the ($M_0$,$M_{1/2}$) 
regions considered here one never finds the kinematic configuration 
$m_{\tilde{t(b)}_2}\OOrd  m_{\tilde{t(b)}_1}
+m_H$. That is, no production and decay channel can onset, and the
hierarchy of cross sections already seen for $M_0=M_{1/2}=150$ GeV 
replicates unaltered
in most cases, mainly governed by the size of the rest masses in the 
final state. Here cross sections remain sizable only if
neither $M_0$ nor $M_{1/2}$ exceed the value 150--200 GeV  
(for large $\tan\beta$, of course, see Fig.~\ref{fig:heavyhiggs}).
Even in such cases, though, presumably no more than a handful
of events can be selected in most decay channels.

\begin{table}
\begin{eqnarray}
\begin{array}{|c|c|c|c|c|} \hline
M_0 ({\rm GeV}) & M_{1/2} ({\rm GeV}) & \tan\beta & 
\sigma (g g \rightarrow \tilde{t}_1 \tilde{t}_1^* H ) ({\rm fb}) &
\sigma (g g \rightarrow \tilde{b}_1 \tilde{b}_1^* H ) ({\rm fb})  
\\ \hline
130 & 130 & 2 & 8.0 \times 10^{-3} & 2.1\times 10^{-3} 
\\ \hline
200 & 150 & 2 & 1.4 \times 10^{-3} & 4.2\times 10^{-4} 
\\ \hline
200 & 200 & 2 & 1.1 \times 10^{-4} & 7.5\times 10^{-5} 
\\ \hline
300 & 250 & 2 & 1.3 \times 10^{-5} & 8.2\times 10^{-6}
\\ \hline
130 & 130 & 40 & 3.3  & 14 
\\ \hline
200 & 150 & 40 & 1.3  & 3.5 
\\ \hline
200 & 200 & 40 & 0.38 & 0.84 
\\ \hline
300 & 250 & 40 & 0.068 & 0.014 
\\ \hline
\end{array}
\nonumber 
\end{eqnarray}
\caption{The variation of the most significant cross sections (in pb)
of processes $g g \rightarrow \tilde{q}_\chi \tilde{q}_{\chi '}^* H $ with
$M_0$, $M_{1/2}$ and $\tan\beta$. The other M-SUGRA parameters are 
fixed as follows: $A_0=0$ GeV and $\sign(\mu) = -$.}
\label{table2}
\end{table}

\subsubsection{CP-odd Higgs boson production}
\label{odd}

As discussed to some lenght in Ref.~\cite{short}, pseudoscalar
Higgs boson production in association with sbottom and stop squarks
of different mass (the only possible combination in absence of 
CP-violating phases in 
$\mu$ and $A_q$, with $q=t,b$), can boast a special attractiveness
because of the absence of mixing terms in the relevant squark-squark-Higgs
couplings. By making use of eq.~(\ref{fr}) and recalling that
if one reverts the chirality flow in the vertex $\lambda_{A\tilde{q}_L
\tilde{q}_R}$ the corresponding Feynman rule  changes its sign 
\cite{HHG}, one finds that those vertices 
reduce to
\begin{equation}
\lambda_{A\tilde{t}_1\tilde{t}_2} = -\frac{g_W m_t}{2 M_W} \left ( \mu -
A_t \cot\beta \right), \qquad\qquad
\lambda_{A\tilde{b}_1\tilde{b}_2} = -\frac{g_W m_b}{2 M_W} \left ( \mu -
A_b \tan\beta \right ). 
\label{cpoddfr}
\end{equation}
These are precisely the couplings entering 
the two processes of the type (\ref{proc}) inducing the final states 
${\tilde{q}}_1{\tilde{q}}_2^* A$, where $q=t,b$.

From this point of view, it is then clear the potential of squark
and pseudoscalar Higgs production in constraining the input values of all five 
M-SUGRA parameters. 
In other terms, to trace back (more technically, to fit) the shape
of the cross sections (if not of some differential distributions) 
in terms of the $\tan\beta$, $A_q$ (with $q=t,b$) and $\mu$ parameters 
entering eq.~(\ref{cpoddfr}) is presumably a much simpler job then doing
the same by using the more involved expressions in eq.~(\ref{fr}),
unless one exploits some asymptotic regime in either
$\tan\beta$, $A_q$ (with $q=t,b$) and/or $\mu$ in which the latter
reduce to the former.
It is under this perspective that we looked at the 
case of $A$ production in our Ref.~\cite{short}.

Rather than repeating the all discussion carried out there, we
summarise here the salient findings of \cite{short},
referring the reader to that paper for specific details.
The production cross sections can be found in Fig.~\ref{fig:cpoddhiggs}.
For $\tan\beta$ below 20 or so, the rates
 for pseudoscalar Higgs boson production are presumably 
too poor to be of great  experimental help. Furthermore, in the
 high $\tan\beta$ regime, pseudoscalar
Higgs boson production is in general less effective than other channels in 
constraining the sign of the Higgsino mass term: compare the overlapping
for the solid and dot-dashed curves (for each $A_0$) in the
detectable regions of $\tilde{t}_1 \tilde{t}_2^* A$ and
$\tilde{b}_1 \tilde{b}_2^* A$ production to the splitting occurring
in, e.g., $\tilde{t}_1 \tilde{t}_1^* H$. This, as far as it
concerns the flaws.  

As for the advantages, we would like to draw the attention of the reader to
the fact that reactions (\ref{proc}) with CP-odd Higgs bosons in the 
final state are quite sensitive to $\tan\beta$. The simple form of the
expressions for 
$\lambda_{A {\tilde q}_1{\tilde q}_2}$ ($q=t,b$) in eq.~(\ref{cpoddfr})
allows one to straightforwardly interpret 
the variation of the pseudoscalar rates with this parameter, namely,
the steep rise at high values of the latter. This can  in fact be 
understood as follows. For large 
$\tan\beta$,  the vertex couplings
 of eq.~(\ref{cpoddfr}) can be rewritten in the approximate form
\begin{equation}\label{limit}
\lambda_{A\tilde{t}_1\tilde{t}_2} \simeq  -\frac{g_W m_t}{2 M_W} 
\mu  ,  \qquad\qquad
\lambda_{A\tilde{b}_1\tilde{b}_2} \simeq  \frac{g_W m_b}{2 M_W} 
A_b \tan\beta ,
\label{largetanb}
\end{equation}
that is,
the coupling which is associated with the sbottom pair is 
proportional to $\tan\beta$, so that, eventually, the total
$\tilde{b}_1\tilde{b}_2^*A$ 
cross section will grow with $\tan^2\beta$
while the coupling related to 
the stop pair will assume constant values. In the 
latter, the enhancement of the $\tilde{t}_1\tilde{t}_2^*A$
cross section with $\tan\beta$ is
rather a phase space effect since, as $\tan\beta$ increases, the 
 CP-odd Higgs boson mass decreases considerably 
(the squark masses changing much less instead),
as we can see from Fig.~\ref{fig:higgs}. Of course, the same remains valid in
the former case as well, so that  our figure indicates a clear order
in the size of the cross sections,
$\sigma (\tilde{b}_1 \tilde{b}_2^* A )
 \OOrd  \sigma (\tilde{t}_1
 \tilde{t}_2^* A )$, at large $\tan\beta$.

But, let us now turn our attention to another peculiar dependence of
the production rates of $\tilde{t}_1 \tilde{t}_2^* A$ and 
$\tilde{b}_1 \tilde{b}_2^* A$:  the one on 
the common trilinear coupling $A_0$. Pretty much along the same lines
as for the combinations $\tilde{t}_1 \tilde{t}_1^* h$,
$\tilde{t}_1 \tilde{t}_2^* h$ and $\tilde{b}_1 \tilde{b}_1^* H$
one can make the case that the sensitivity
to $A_0$ of the $A$ production cross sections offers the chance of
constraining, possibly the sign, and hopefully the
magnitude, of this fundamental M-SUGRA parameter. 
This is presumably the best attribute of  
$\tilde{t}_1 \tilde{t}_2^* A$ and $\tilde{b}_1 \tilde{b}_2^* A$, under the
assumption  already made in few instances that the determination
of  $\tan\beta$ could come first from studies in the pure
Higgs sector.
Putting down some numbers in this respect, one may invoke 
 the following scenario, if $\tan\beta$ is, say, larger than 32:
\begin{eqnarray}
\sigma (\tilde{t}_1 \tilde{t}_2^* A) 
&\OOrd & 10  \; 
{\rm fb} \; \Rightarrow A_0 < -300 \\[2mm]
\sigma (\tilde{b}_1 \tilde{b}_2^* A) 
&\OOrd& 2 \; 
{\rm fb} \; \Rightarrow A_0 < -300,
\end{eqnarray}
quite independently of $\sign(\mu)$. Conversely, the non-observation
of pseudoscalar Higgs events in those regimes would imply most likely
a positive $A_0$ value. 

As for peculiar trends in Fig.~\ref{fig:cpoddhiggs}, it is worth mentioning 
(though we have not shown it here, as the reader can refer to \cite{short})
that the cross section for  sbottom production 
vanishes  too, at some (large) value
of $\tan\beta$, as it did for the case of CP-even Higgs production.
As the reader can appreciate in Fig.~2 of Ref.~\cite{short}, contrary
to the case of $\tilde{b}_1 \tilde{b}_2^* h$ and $\tilde{b}_1 \tilde{b}_2^* H$
final states, this however
happens in $\tilde{b}_1 \tilde{b}_2^* A$ events only
for positive and large values of $A_0$ (and both $\sign(\mu)=\pm$). This
is another
 consequence of the different nature of the couplings (\ref{fr}) to squarks
of CP-even versus CP-odd neutral Higgs bosons. Though we have failed
to find a point where this disappearance of $\tilde{b}_1 \tilde{b}_2^* A$
events for positive $A_0$ values corresponds to the survival of a detectable
cross section for negative $A_0$'s (at fixed $\tan\beta$), such matter
would presumably deserve further investigation in the future.

Before closing this Section, we study the dependence of pseudoscalar Higgs
boson production in association with stop and sbottom squarks on
the last two M-SUGRA independent parameters, $M_0$ and $M_{1/2}$: see
Tab.~\ref{table3}. The main effect of changing the latter is
onto the masses of the final state scalars, through the phase space volume
as well as via propagator effects in the scattering amplitudes.
(In fact, no decay channel of the heavier stop or sbottom into
the lighter one ever opens, at least for the
values of $M_0$ and $M_{1/2}$ that  we had looked at.)
In other terms, to increase one or the other depletes both
$\sigma (\tilde{t}_1 \tilde{t}_2^* A )$ and
$\sigma (\tilde{b}_1 \tilde{b}_2^* A )$ quite strongly, 
simply because
the values of all $m_{\tilde{q}_\chi}$'s, $m_{\tilde{q}_{\chi'}}$'s
 and $m_\Phi$'s get larger. 
For example, assuming $\tan\beta=40$: at $M_0=M_{1/2}=130$ GeV, one has
$m_{\tilde{t}_1}=248$ GeV,
$m_{\tilde{t}_2}=388$ GeV,
$m_{\tilde{b}_1}=256$ GeV,
$m_{\tilde{b}_2}=340$ GeV and
$m_{{A}  }=120$; whereas at $M_0=300$ GeV and $M_{1/2}=250$ GeV,
the figures are
$m_{\tilde{t}_1}=461$ GeV,
$m_{\tilde{t}_2}=611$ GeV,
$m_{\tilde{b}_1}=510$ GeV,
$m_{\tilde{b}_2}=591$ GeV and
$m_{{A}  }=292$ GeV.
In practice, the table shows that only rather light $M_0$ and $M_{1/2}$ masses
(say, below 200 and 150 GeV, respectively) 
would possibly allow for pseudoscalar production to be detectable at the LHC,
and only at large $\tan\beta$ \cite{short}.

\begin{table}
\begin{eqnarray}
\begin{array}{|c|c|c|c|c|} \hline
M_0 ({\rm GeV}) & M_{1/2} ({\rm GeV}) & \tan\beta & 
\sigma (g g \rightarrow \tilde{t}_1 \tilde{t}_2^* A ) ({\rm fb}) &
\sigma (g g \rightarrow \tilde{b}_1 \tilde{b}_2^* A ) ({\rm fb})  
\\ \hline
130 & 130 & 2 & 5.2 \times 10^{-2} & 4.1\times 10^{-4} 
\\ \hline
200 & 150 & 2 & 2.0 \times 10^{-2} & 8.8\times 10^{-5} 
\\ \hline
200 & 200 & 2 & 5.9 \times 10^{-3} & 2.8\times 10^{-5} 
\\ \hline
300 & 250 & 2 & 1.3 \times 10^{-3} & 3.9\times 10^{-6}
\\ \hline
130 & 130 & 40 & 79 & 13 
\\ \hline
200 & 150 & 40 & 1.4  & 2.4 
\\ \hline
200 & 200 & 40 & 0.31 & 0.6 
\\ \hline
300 & 250 & 40 & 0.048 & 0.098
\\ \hline
\end{array}
\nonumber 
\end{eqnarray}
\caption{The variation of the most significant cross sections 
of processes $g g \rightarrow \tilde{q}_\chi \tilde{q}_{\chi '}^* A $ with
$M_0$, $M_{1/2}$ and $\tan\beta$. The other M-SUGRA parameters are 
fixed as follows: $A_0=0$ GeV and $\sign(\mu) = -$.}
\label{table3}
\end{table}

\subsubsection{Charged Higgs bosons production}
\label{charged}

To have an additional source of charged Higgs bosons at the LHC, 
especially with masses
larger than the top mass $m_t$, would be very helpful from an experimental
point of view. In fact, it is well known the difficulty of detecting
charged Higgs scalars in that mass regime, not only because of a dominant
decay 
signature which suffers from very large QCD background (i.e.,
$H^+\to t\bar b \to b\bar b W^+\to b\bar b jj$), but also because
the production mechanisms are not many and with not very large 
rates \cite{charged}.
Unfortunately, as it turns out from Fig.~\ref{fig:chargedhiggs}, typical
production cross sections of $H^\pm$ scalars in association with
sbottom and stop pairs rarely exceed 10 fb. These rates compare rather
poorly with other mechanisms \cite{charged}, for the same
choice of $m_{H^\pm}$.  Thus, there is 
little to gain in exploiting processes (\ref{proc}) as discovery channels
of charged Higgs bosons.

Furthermore, their dependence on $\tan\beta$, $A_0$ and $\sign(\mu)$
replicates many of the tendencies already individuated in neutral
Higgs channels, for which the production cross sections are much larger.
Adding the fact that typical decay channels of the latter 
(e.g., in photon pairs) are much cleaner in the hadronic environment
of the LHC than those of the former, one would quite rightly 
conclude that the potential
of $\tilde{t}_1 \tilde{b}_1^* H^-$,
   $\tilde{t}_1 \tilde{b}_2^* H^-$,
   $\tilde{b}_1 \tilde{t}_2^* H^+$ and 
   $\tilde{b}_2 \tilde{t}_2^* H^+$ final states
in constraining the M-SUGRA parameter space is rather poor.

Nonetheless, it is worth recognising some of the typical trends
of the production cross sections, for the sake of future reference.
Let alone the last two combinations, for which the final state masses
are too heavy to be produced at detectable rate, we have a quick look
at the first two cases, which can indeed have cross sections significantly
above 1 fb, at least in the large $\tan\beta$ region. This enhancement
has a twofold explanation. Firstly, 
phase space effects, as for large $\tan\beta$ all scalar masses
(except $m_{{\tilde t}_1}$ and $m_h$) get smaller: 
see Figs.~\ref{fig:squarks}--\ref{fig:higgs}. Secondly, terms
in their couplings proportional to $m_b A_b \tan\beta$ are dominant
for most of the possible $A_0$ and $\sign(\mu)$ combinations. Finally,
notice also in the case of the $\tilde{t}_1 \tilde{b}_1^* H^\pm$
final state the vanishing of the cross section, this time at somewhat
lower values of $\tan\beta$ than in the case of the neutral Higgs bosons.

As for the $M_0$ and $M_{1/2}$ dependence, this is again realised
through the phase space and the propagators, 
as there is no significant enhancement from
resonant decays. In practice, only if $\tan\beta$ is extremely large
and both the universal scalar and gaugino masses are below 200 GeV, 
the two cross sections  for $\tilde{t}_1 \tilde{b}_1^* H^-$
and $\tilde{t}_1 \tilde{b}_2^* H^-$ remain detectable (indeed, those
containing the lightest stop squark): see  Tab.~\ref{table4}}.

\begin{table}
\begin{eqnarray}
\begin{array}{|c|c|c|c|c|} \hline
M_0 ({\rm GeV}) & M_{1/2} ({\rm GeV}) & \tan\beta & 
\sigma (g g \rightarrow \tilde{t}_1 \tilde{b}_1^* H^- ) ({\rm fb}) &
\sigma (g g \rightarrow \tilde{t}_1 \tilde{b}_2^* H^- ) ({\rm fb})  
\\ \hline
130 & 130 & 2 & 4.3 \times 10^{-4} & 1.4\times 10^{-3} 
\\ \hline
200 & 150 & 2 & 2.2 \times 10^{-5} & 3.0\times 10^{-4} 
\\ \hline
200 & 200 & 2 & 3.1 \times 10^{-5} & 9.8\times 10^{-5} 
\\ \hline
300 & 250 & 2 & 1.6 \times 10^{-6} & 1.4\times 10^{-5}
\\ \hline
130 & 130 & 40 & 2.2  & 4.8 
\\ \hline
200 & 150 & 40 & 1.5  & 4.5
\\ \hline
200 & 200 & 40 & 1.7 & 0.31
\\ \hline
300 & 250 & 40 & 0.064  & 0.023 
\\ \hline
\end{array}
\nonumber 
\end{eqnarray}
\caption{The variation of the most significant cross sections 
of processes $g g \rightarrow \tilde{q}_\chi \tilde{q}_{\chi '}^* H^\pm $ with
$M_0$, $M_{1/2}$ and $\tan\beta$. The other M-SUGRA parameters are 
fixed as follows: $A_0=0$ GeV and $\sign(\mu) = -$.}
\label{table4}
\end{table}



\subsubsection{Decay signatures}
\label{signatures}

So far we have only discussed production cross sections for processes
of the form (\ref{proc}) and made no considerations about possible decay
channels and relative branching fractions of either squarks
or Higgs bosons. 
Another related aspect is the typical kinematics of the signals, as
it appears in the detectors, and the size of the possible backgrounds.
Furthermore, the reader should appreciate how all channels
entering processes (\ref{proc}) are intertwined, in the sense
that any of these can act as a background to all  others. 

It is the aim of this Section that of indicating some possible 
decay signatures of the most relevant squark-squark-Higgs processes,
in which they show both large rates and their kinematics is  such
that they can hopefully be disentangled at the LHC.
In doing so, we distinguish between 
a small (see \ref{signatures}.1) and large (see \ref{signatures}.2)  
$\tan\beta$ regime, as we have shown that such a parameter is crucial
 in determining the actual size of
the production rates. As representative choice of the universal masses 
we adopt the combination with lowest values among those discussed in the
previous Subsections, i.e., $M_{0}=M_{1/2}=130$
GeV, further setting  $A_0$=0 and $\sign(\mu)$
negative.
\vskip0.5cm\noindent
{\bf \ref{signatures}.1 Small $\tan\beta$ regime}
\vskip0.5cm\noindent
For small $\tan\beta$'s the only relevant processes  are
$\tilde{t}_1 \tilde{t}_1^* h$ and $\tilde{t}_1 \tilde{t}_2^* h$
production. A possible decay signature for the
first case is the one contemplated
in Refs.~\cite{djouadi,djouadi2}. That is, 
${\tilde t}_1\to \chi_1^+b\to W^{+}b$ plus missing energy for the light
stop and $h\to \gamma\gamma$ for the light Higgs boson, with the $W^+$
decaying leptonically and/or hadronically.
 The final topology would be the same as 
in $t\bar t h$, with the only difference that 
for stop squark events there is a large amount of missing energy.

Since another option to tag the lightest MSSM Higgs boson at the
LHC is to use the more messy but dominant decay channel into
$b\bar b$ pairs (as opposed to exploiting the cleaner but suppressed
$\gamma\gamma$ mode) \cite{ATLAS,CMS},  
another possible decay sequence could be  the following:
$$
\arraycolsep=0pt 
\begin{array}{lllllll}\label{hchain}
{\tilde{t}_1} && &~~~~{\tilde{t}}_1^* &&h~~~~~ & \\
\downarrow &&& ~~~\downarrow &&\downarrow~~~~~ & \\
\chi_1^+ \,+\,b\; & &&~~~~\chi_1^- \,+\,\bar b &~&b \,+\,\bar b &  \\
\downarrow &&& ~~~\downarrow && \; & \\
q \,+\,\bar q'\,+\,\chi_1^0& && ~~~~\ell^-\,+\,\nu\,+\,\chi^0_1 &&&
\end{array}
$$
in which 
$q\bar q'=u\bar d,c\bar s$ and $\ell=e,\mu$. Considering also
the charge conjugated $\chi_1^+\chi_1^-$ decays, the final signature
would then be 
`$2~{\mathrm{jets}}~+~4b~+\ell^\pm+{E}_{\mathrm{miss}}$', 
where the missing energy/momentum is not only due to the 
 two $\chi^0_1$'s but also to the neutrinos.

The total branching ratio (BR)
 of such a decay sequence  is, for  $\tan\beta=3$ and 
according to Tab.~\ref{table5}, approximately 2.5\%. The production
cross section at the same $\tan\beta$ value is about 72 fb, 
so that about 176 events per year would survive. One may further assume 
a reduction factor of about 0.25 because of the overall 
efficiency $\varepsilon_b^4$
to tag four $b$-quarks (assuming $\varepsilon_b\approx0.7$). This
ultimately yields something more than 44 events per year.
 In addition, one should expect most of the signal events
to lie in the detector acceptance region, since leptons and jets
originate from decays of heavy objects. 

The same signature could well be exploited in the case of the 
${\tilde{t}_1} {\tilde{t}}_2^* h$ final state. Here, the 
production cross section is smaller than for 
${\tilde{t}_1} {\tilde{t}}_1^* h$ production, about 40 fb,
so is the ${\tilde t}_2\to \chi_1^+b$ decay rate as compared
to the ${\tilde t}_1\to \chi_1^+b$ one (see Tab.~\ref{table5}). However,
the final number of events per year is still quite large: 
about 14 after having already multiplied by  $\varepsilon_b^4$.

As for the kinematics of these two signatures, we may remark that they
have peculiar features that should help
in their selection: a not too large hadronic multiplicity, six jets in total,
each rather energetic (in fact, note that 
$m_{\chi^\pm_1}-m_{\chi^0_1}\approx58$ GeV and 
$m_{{\tilde{t}_2}}\approx378~{\mathrm{GeV}}\gg
 m_{{\tilde{t}_1}}\approx273~{\mathrm{GeV}}\gg 
 m_{\chi^\pm_1}\approx114$ GeV),  
so that their reconstruction from the detected 
tracks should be reasonably accurate;
high transverse momentum and isolated leptons to be used as
trigger; large $E_{\mathrm{miss}}$ to reduce non-SUSY processes;
four tagged $b$-jets that
can be exploited to suppress the `$W^\pm$ + light jet' background from QCD, 
and one $b\bar b$ pair resonating at the $h$ mass,
$m_h\approx90$ GeV. Moreover, the `irreducible' 
background from ${\tilde{t}_1}{\tilde{t}_1^*}Z$ events 
has been shown in Ref.~\cite{fawzi} to be under control,
even when $m_h\approx M_Z$, as it is the case here.

\begin{center}
\begin{table}
\begin{eqnarray}
\begin{array}{cc}   

\begin{array}{|ccc|}\hline
{\rm Particle} & {\mathrm{BR}} 
& {\rm Decay} \\ \hline   
\tilde{t}_1  & \stackrel{76\% 
}{\rightarrow}& \chi_1^+ b \\
&\stackrel{19\%
}{\rightarrow}& \chi_1^0 t
\\ \hline
\tilde{t}_2 & \stackrel{57\%
}{\rightarrow}& \chi_1^+ b \\
 & \stackrel{24\%
}{\rightarrow}  & \chi_2^+ b \\ 
 & \stackrel{11\%
}{\rightarrow}  & \chi_2^0 t \\ \hline


h & \stackrel{90\%
}{\rightarrow}& b \bar{b} \\
 & \stackrel{5\%
}{\rightarrow}  & \tau^+ \tau^- \\ 
 & \stackrel{0.0003\%
}{\rightarrow}  & \gamma \gamma \\ 
\hline




\end{array}
&
\begin{array}{|ccc|}\hline
{\rm Particle} & {\mathrm{BR}} 
& {\rm Decay} \\ \hline   

t & \stackrel{33\%
}{\rightarrow} & u \bar{d} b \\
& \stackrel{33\%
}{\rightarrow} & c \bar{s} b \\
& \stackrel{11\%
}{\rightarrow} & e^+ \nu b \\ 
& \stackrel{11\%
}{\rightarrow} & \mu^+ \nu b \\
& \stackrel{11\%
}{\rightarrow} & \tau^+ \nu b \\ \hline

\chi_1^+ & \stackrel{30\%
}{\rightarrow} & \chi_1^0 u \bar{d}^{\dagger} \\
& \stackrel{30\%
}{\rightarrow} & \chi_1^0 c \bar{s}^{\dagger} \\
& \stackrel{14\%
}{\rightarrow} & \chi_1^0  \tau^+ \nu^{\dagger} \\
& \stackrel{14\%
}{\rightarrow} & \chi_1^0 e^+ \nu^{\dagger} \\
& \stackrel{14\%
}{\rightarrow} & \chi_1^0 \mu^+ \nu^{\dagger} \\ \hline

\chi_2^0 &  \stackrel{28\%
}{\rightarrow} & \chi_1^0 \nu \bar{\nu} \\ 
& \stackrel{14\%
}{\rightarrow} & \chi_1^0 e^- e^+ \\ 
& \stackrel{14\%
}{\rightarrow} & \chi_1^0 \mu^- \mu^+ \\
& \stackrel{14\%
}{\rightarrow} & \chi_1^0 \tau^- \tau^+ \\
\hline 


\end{array}

\end{array}
\nonumber
\end{eqnarray}
\caption{Dominant decay channels and BRs
of the final state (s)particles in 
$g g \rightarrow \tilde{q}_\chi \tilde{q}_{\chi '}^* h $,
$q=t$ and $\chi,\chi'=1,2$,
for $M_0=M_{1/2}=130$ GeV, $A_0=0$, $\tan\beta=3$ and 
${\mathrm{sign}}(\mu)<0$ \cite{isajet}.
{~~$^\dagger$ \footnotesize{Via off-shell $W^{+}$.}}}
\label{table5}
\end{table}
\end{center}
%
%
%
%
%
\begin{center}
\begin{table}
\begin{eqnarray}
\begin{array}{cc}   

\begin{array}{|ccc|}\hline
{\rm Particle} & {\mathrm{BR}} 
& {\rm Decay} \\ \hline   
\tilde{t}_1 & \stackrel{94\% 
}{\rightarrow}& \chi_1^+ b \\ \hline
\tilde{t}_2 & \stackrel{40\%
}{\rightarrow}& \chi_2^+ b \\
 & \stackrel{26\%
}{\rightarrow}  & \chi_1^+ b \\
 & \stackrel{16\%
}{\rightarrow}  & {\tilde{b}}_1 W^+ \\ 
 & \stackrel{7\%
}{\rightarrow}  & {\tilde{t}}_1 Z \\ \hline

\tilde{b}_1 & \stackrel{61\%
}{\rightarrow}& \chi_2^0 b \\
 & \stackrel{32\%
}{\rightarrow}  & \chi_1^0 b \\ \hline
\tilde{b}_2 & \stackrel{42\%
}{\rightarrow}& \chi_3^0 b \\
 & \stackrel{31\%
}{\rightarrow}  & \chi_4^0 b \\ 
 & \stackrel{18\%
}{\rightarrow}  & \chi_2^0 b \\ \hline

h & \stackrel{94\%
}{\rightarrow}& b \bar{b} \\
 & \stackrel{6\%
}{\rightarrow}  & \tau^+ \tau^- \\ 
\hline

H & \stackrel{94\%
}{\rightarrow}& b \bar{b} \\
 & \stackrel{6\%
}{\rightarrow}  & \tau^+ \tau^-  \\ \hline

A & \stackrel{94\%
}{\rightarrow}& b \bar{b} \\
 & \stackrel{6\%
}{\rightarrow}  & \tau^+ \tau^- \\ \hline

H^\pm & \stackrel{91\%
}{\rightarrow} & \tau^\pm \nu \\
 & \stackrel{5\%
}{\rightarrow}  & \chi_1^0 \chi_1^\pm \\ \hline

\end{array}
&
\begin{array}{|ccc|}\hline
{\rm Particle} & {\mathrm{BR}} 
& {\rm Decay} \\ \hline   
\chi_1^+ & \stackrel{100\%
}{\rightarrow} & \tilde{\tau}_1^+ \nu \\ \hline 

\chi_2^+ & \stackrel{24\%
}{\rightarrow} & \chi_2^0 W^+ \\
& \stackrel{15\%
}{\rightarrow} & \chi_1^+ Z \\
& \stackrel{11\%
}{\rightarrow} & \chi_1^+ A \\ 
& \stackrel{10\%
}{\rightarrow} & \tau^+ \tilde{\nu} \\
 \hline

\chi_2^0 & \stackrel{100\%
}{\rightarrow} & \tilde{\tau}_1^\pm \tau^\mp   \\ 
\hline 

\chi_3^0 & \stackrel{24\%
}{\rightarrow} & \chi_1^+  W^- \\ 
& \stackrel{24\%
}{\rightarrow} & \chi_1^-  W^+ \\ 
& \stackrel{9\%
}{\rightarrow} & \chi_1^0  Z \\ \hline

\chi_4^0 & \stackrel{23\%
}{\rightarrow} & \chi_1^+  W^- \\ 
& \stackrel{23\%
}{\rightarrow} & \chi_1^-  W^+ \\
& \stackrel{6\%
}{\rightarrow} & \chi_1^0  h \\ 
& \stackrel{6\%
}{\rightarrow} & \chi_1^0  Z \\ \hline

\tilde{\tau}_1^+ & \stackrel{100\%
}{\rightarrow} & \chi_1^0 \tau^+ \\ \hline

\end{array}

\end{array}
\nonumber
\end{eqnarray}
\caption{Dominant decay channels and BRs
of the final state (s)particles in
$g g \rightarrow \tilde{q}_\chi \tilde{q}_{\chi '}^* X $, $X=h,H,A,H^\pm$,
$q=t,b$ and $\chi,\chi'=1,2$,
for $M_0=M_{1/2}=130$ GeV, $A_0=0$, $\tan\beta=40$ and 
${\mathrm{sign}}(\mu)<0$ \cite{isajet}.}
\label{table6}
\end{table}
\end{center}
%
%
{\bf \ref{signatures}.2 Large $\tan\beta$ regime}
\vskip0.5cm\noindent
In the large $\tan\beta$ regime there is a variety of
cross sections which can be significant:
$\tilde{t}_1 \tilde{t}_1^* X$, 
$\tilde{t}_1 \tilde{t}_2^* X$,
$\tilde{b}_1 \tilde{b}_1^* X$,
$\tilde{b}_2 \tilde{b}_2^* X$,
$\tilde{t}_1 \tilde{t}_2^* A$,
$\tilde{b}_1 \tilde{b}_2^* A$,
$\tilde{t}_1 \tilde{b}_1^* H^+$ and
$\tilde{t}_1 \tilde{b}_2^* H^+$, where $X=h,H$.
For reasons of space, however, we only focus our attention to one signature
for the  Higgs states not yet considered,  the one arising from 
the dominant production channel in all cases, with the only
exception of the charged Higgs bosons. In fact, we
will neglect analysing here their decay patterns, as we have already mentioned
the poor effectiveness of the charged Higgs production modes
both as discovery channels and probes of the
underlying M-SUGRA model. 

In the case of heavy scalar Higgs bosons, we consider the final state
$\tilde{b}_1 \tilde{b}_1^* H$. This yields a cross section of 14 fb
for $\tan\beta=40$. A possible decay chain is the one below.
$$
\arraycolsep=0pt 
\begin{array}{lllllll}\label{Hchain}
{\tilde{b}_1} && &~~~~{\tilde{b}}_1^* &&~~~~H~~~~~ & \\
\downarrow &&& ~~~\downarrow &&~~~~\downarrow~~~~~ & \\
\chi_1^0 \,+\,b\; & &&~~~~\chi_1^0 \,+\,\bar b &~&~~~~~b \,+\,\bar b &
\end{array}
$$
That is, a rather simple final state made up by four $b$-quarks and
missing energy. The BR of this sequence is about
9\% (see Tab.~\ref{table6}). 
Therefore, at high luminosity, one obtains 129 events per year,
before heavy flavour identification.

The main background is certainly ordinary QCD production of four jets.
However, the requirement of tagging four $b$-jets would reduce the
latter considerably. Furthermore, if the mass of the heavy scalar
Higgs boson is known, then one could impose
that two $b$-quarks reproduce  $m_H\approx 121$ GeV within a few GeV
(say, 5) in invariant mass. Finally, given the enormous mass difference
$m_{\tilde b_1}-m_b\approx 250$ GeV, compared to the rest mass of the LSP, 
$m_{\tilde \chi_1^0}\approx51$ GeV, one
should expect, on the one hand, a large amount of missing energy,
and, on the other hand, all four $b$-jets to be quite hard, both
aspects further helping to reduce the QCD noise. In the end, some
32 events could well be detected annually, having  already accounted for the
overall $b$-tagging efficiency $\varepsilon_b^4=0.25$.

For the case of the pseudoscalar Higgs particle, we choose the final state
${\tilde t}_1 {\tilde t}_2^* A$, which has a cross section of about
79 fb at $\tan\beta=40$. A possible signature 
could be\footnote{Additional examples can be found
in Ref.~\cite{short}.}:
$$
\arraycolsep=0pt 
\begin{array}{lllllll}\label{Achain}
{\tilde{t}_1} && &~~~~{\tilde{t}}_2^* &&~~~A~~~~~ & \\
\downarrow &&& ~~~\downarrow &&~~~\downarrow~~~~~ & \\
\chi_1^+ \,+\,b\; & &&~~~~\chi_1^- \,+\,\bar b &~&~~~~b \,+\,\bar b &  \\
\downarrow &&& ~~~\downarrow && \; & \\
{\tilde \tau}_1 \,+\,\nu& && ~~~~{\tilde \tau}_1^* \,+\,\bar\nu &&&\\
\downarrow &&& ~~~\downarrow && \; & \\
\tau^+\,+\,\chi_1^0& && ~~~~\tau^-\,+\,\chi^0_1 &&&
\end{array}
$$
Here, the final state is made up by four $b$-quarks and two $\tau$'s,
plus missing energy as usual. The decay fraction is 23\%
(again, see Tab.~\ref{table6}). That is, 1814 events per year before
tagging $b$'s and $\tau$'s.

The most dangerous backgrounds are probably $Z+4$~jet production and 
$t\bar t b\bar b$ events. The first can be rejected by asking, e.g.,
$M_{\tau^+\tau^-}\ne M_Z$, if $\tau$'s are reconstructed. In addition, both 
background processes  have (at least)
two $b$-quarks quite soft. As for the signal, given that the lightest
chargino mass is much smaller than the stop ones (in fact, 
$m_{{\tilde\chi}^\pm_1}\approx93$ GeV whereas
$m_{{\tilde t}_1}\approx248$ GeV and
$m_{{\tilde t}_2}\approx388$ GeV), all $b$'s are naturally energetic and
two of them also peak at $m_A\approx120$ GeV. Thus, to
require all $M_{bb}$ invariant masses sufficiently large with one
close to the $A$ mass
should help in enhancing considerably the
signal-to-background rates. Requiring large missing energy would
help further, especially against $t\bar t b\bar b$ events.
More difficult is to discern
differences in the $\tau$ behaviours (though, notice that
$m_{{\tilde \tau}_1}\approx76$ GeV $\gg m_\tau$). For $\varepsilon_b^4=0.25$,
and assuming leptonic decays of both $\tau^+$ and $\tau^-$ into
electrons and/or muons, one finally gets something of the order of
110 signal events per year.

\subsection{Heavy mass spectrum}
\label{heavyspectrum}

An attempt to summarise our findings  
is made in Fig.~\ref{fig:light}, where the most relevant cross sections
(see upper frame)
for the light mass regime (see lower frame)
are plotted for a choice of $M_{0}$, $M_{1/2}$, 
$A_0$, $\tan\beta$ and $\sign(\mu)$
which reflects their hierarchal order seen over most of the M-SUGRA 
parameter space discussed so far.

However, to assume that only light squark and Higgs masses can induce sizable
cross sections in events of the type (\ref{proc}) would be wrong.
This is a sufficient condition for many channels, but not a necessary one.
For example, even for very large universal masses, one could 
find a value of the soft trilinear coupling small enough to
overcome the loss of signal due to propagator and phase space effects.
In fact, as repeatedly shown in the previous Section, most of the
cross sections considered here grow quickly as  $A_0$ becomes negative.
Fig.~\ref{fig:heavy} makes eloquently this point (top insert), for the choice
$A_0=-900$ GeV, well consistent with the bounds imposed by 
the charge and colour breaking minima. There, 
 we have adopted a very large value for $M_0$, i.e., 500 GeV,
and varied $M_{1/2}$ between 100 and 500 GeV. The choice of a large 
$\tan\beta$ value, i.e., 35,  is necessary to obtain detectable rates,
except in those cases involving ${\tilde t}_1$ and $h$ in the same event.
In contrast, that of a negative  $\sign(\mu)$ never is.
The squark and Higgs masses produced by the above combinations of M-SUGRA
parameters can be found in the bottom frame of Fig.~\ref{fig:heavy}.

There are only a few production channels which survive 
the strong phase space suppression arising in
the heavy mass regime and yield cross sections 
 above 1 fb. Among these, other than those already encountered
${\tilde t}_1{\tilde t}_1^* h$, 
${\tilde t}_1{\tilde t}_2^* h$, 
${\tilde b}_1{\tilde b}_1^* h$, 
one notices the appearance of channels which had negligible rates
in the low mass regime, notably ${\tilde b}_1{\tilde b}_2^* h$
(compare to Fig.~\ref{fig:lighthiggs}). In this specific
instance, the effect is due to the enhancement induced by the onset of the 
$\tilde{b}_2 \rightarrow \tilde{b}_1 h$ decay mode in $gg\to
\tilde{b}_2 \tilde{b}_2^*$ events. Even the other two combinations
${\tilde t}_1{\tilde t}_1^* H$ and
${\tilde t}_1{\tilde b}_2^* H^-$, which had a rather low profile
in the light mass regime, now compete more closely with the dominant
modes. Finally, notice that pseudoscalar Higgs boson production is no longer
significant in this mass regime, in line with the results
presented in Ref.~\cite{short}.

In practise,
if $M_0=500$ GeV, events involving light CP-even Higgs bosons could be 
detected up to $M_{1/2}=400$ GeV or so, in either mode 
${\tilde t}_1{\tilde t}_1^* h$ or 
${\tilde t}_1{\tilde t}_2^* h$.  Final states of the type
${\tilde b}_1{\tilde b}_2^* h$ have surprisingly large rates
if $M_{1/2}$ is below 220 GeV. The maximum reach in $M_{1/2}$
via ${\tilde b}_1{\tilde b}_1^* h$,
${\tilde t}_1{\tilde b}_2^* H^-$ and
${\tilde t}_1{\tilde t}_1^* H$ is instead 220, 180 and 140 GeV,
respectively.

This is just one example where a  new phenomenology of
squark-squark-Higgs events arises for rather 
heavy $M_0$ and $M_{1/2}$ masses.
We have found several such combinations, and linked them to the fact that
$A_0$ ought to be significantly large and negative, $\tan\beta$ close
to $m_t/m_b$, but with small dependence on $\sign(\mu)$.

\section{Conclusions}
\label{sec:conclusions}

In summary, we have studied neutral and charged Higgs boson
production in association with all possible combinations of
 stop and sbottom squarks at the LHC, in the
context  of the SUGRA inspired MSSM. Our interest in such reactions
was driven not only by the fact that they can act as new
sources of Higgs particles but also because they carry 
a strong dependence on the five inputs of the SUSY model, so that
they can possibly be used to constrain the latter. In a sense, 
this note (along with Ref.~\cite{short}) completes previous analyses on 
the subject \cite{djouadi,djouadi2,fawzi}, where 
the emphasis was mainly put on the usefulness of the above kind of
reactions as Higgs production modes and the attention consequently
restricted to the case of light squark and Higgs masses.

We have found that the cross sections of many of these
processes should be  detectable at high
collider luminosity for not too small values of $\tan\beta$. Indeed,
their production rates are strongly sensitive to the ratio of
the VEVs of the Higgs fields, this possibly
allowing one to put potent constraints
on such a crucial parameter of the MSSM Higgs sector. 
Furthermore, also the trilinear
coupling $A_0$ intervenes in these events, in such a 
way that visible rates would mainly be possible if this other fundamental
M-SUGRA input is negative. (Indeed, to know the actual value
of $\tan\beta$ from other sources would further help to assess the
magnitude of $A_0$.) As for the sign of the Higgsino mass term,
$\sign(\mu)$, it affects the phenomenology of such events in one
or two channels only, so that it can easily evades the imposition
of experimental bounds. 
Finally, concerning the remaining two parameters (apart from mixing effects) of
the M-SUGRA scenario, one must say that $M_0$ need not  be small 
(it could be as large as 500 GeV) and that $M_{1/2}$ is enough
to be below 220 GeV in order to guarantee sizable cross sections in many 
cases. 

In a few representative examples, we have further 
investigated the decay phenomenology of these 
reactions, by discussing some possible signatures, their rates 
(of the order of tens to hundreds of events per year at high luminosity) and 
peculiar kinematics, as opposed to the  yield of ordinary, non-SUSY
backgrounds. 

In conclusion, we believe these processes to be potentially very  helpful
in putting drastic limits on several M-SUGRA parameters and 
we thus recommend that their phenomenology
is further investigated in the context of dedicated experimental simulations, 
which were clearly  beyond the scope of this note. In this spirit,
we have derived compact analytical formulae of the relevant production
MEs, that can easily be incorporated in existing MC programs. 

\section*{Acknowledgements}

S.M. acknowledges the financial support from the UK PPARC and useful
conversations with Michael Kr\"amer.
A.D is supported from the Marie Curie Research Training Grant
ERB-FMBI-CT98-3438 and thanks Mike Seymour for useful discussions.
We both thank Herbi Dreiner for precious comments and The Old School
in Oxford for kind hospitality.


\newpage

\nappend{Appendix A}

In this additional Section we present in analytic form
the MEs adopted in calculating all our processes.
We identify the external particles as follows:
\begin{equation}\label{reaction}
g(p_1,\lambda_1) + g(p_2,\lambda_2) \longrightarrow
{\tilde{q}}_{\chi} (p_3) + {\tilde{q}}^{'*}_{\chi'} (p_4) + 
\Phi (p_5),
\end{equation}
where $p_i$, $i=1, \dots 5$ 
are the four-momenta\footnote{For our purposes, we take
the initial state momenta, $p_1$ and $p_2$, as incoming and the final state 
ones, $p_3$, $p_4$ and $p_5$, as outgoing.}, 
and $\lambda_j$, $j=1,2$, are the helicities 
of the QCD vector bosons. 

\begin{figure}
~\epsfig{file=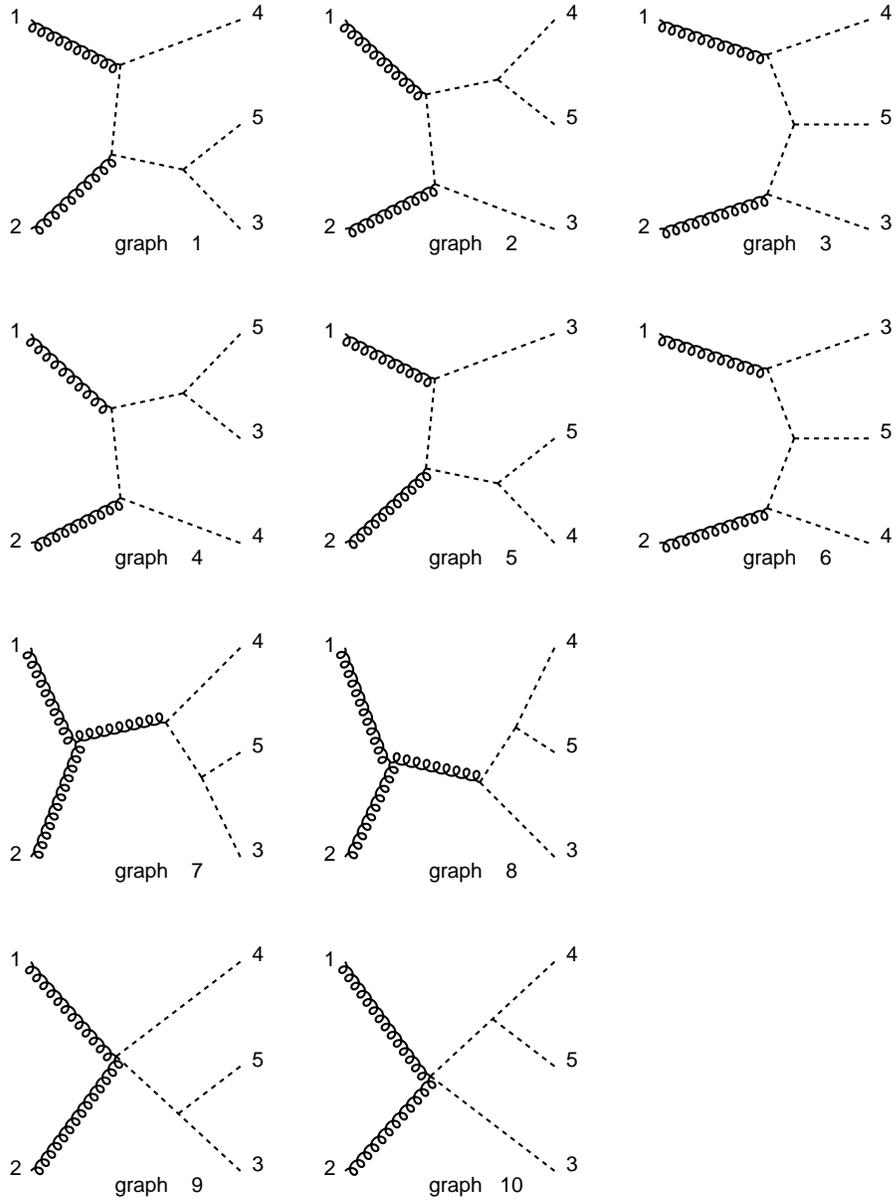,height=22cm,angle=0}
\vskip-4.0cm
\caption{Feynman diagrams contributing at lowest order
to processes (\ref{reaction}). An helical line refers to a gluon whereas
a dashed one symbolises both a squark and a Higgs boson.}
\label{fig:graphs}
\end{figure}

\begin{figure}
\centerline{\epsfig
{figure=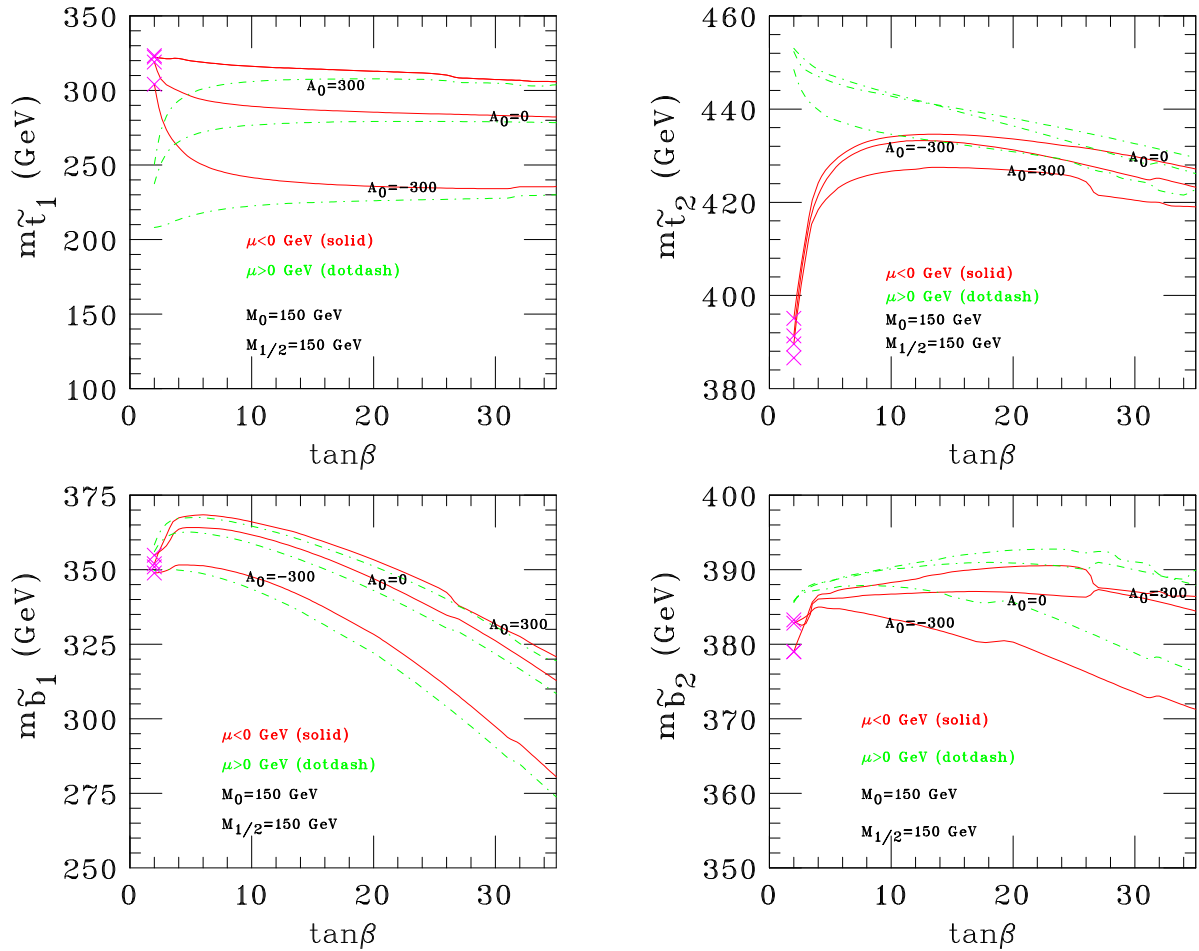,width=5in,angle=90}}
\caption{Resulting masses from \cite{isajet}
for the third generation of squarks 
versus  $\tan\beta$ for three  values  of $A_0$ and
for both positive (dot-dashed) and negative (solid)  $\mu$.}
\label{fig:squarks}
\end{figure}

\begin{figure}
\centerline{\epsfig
{figure=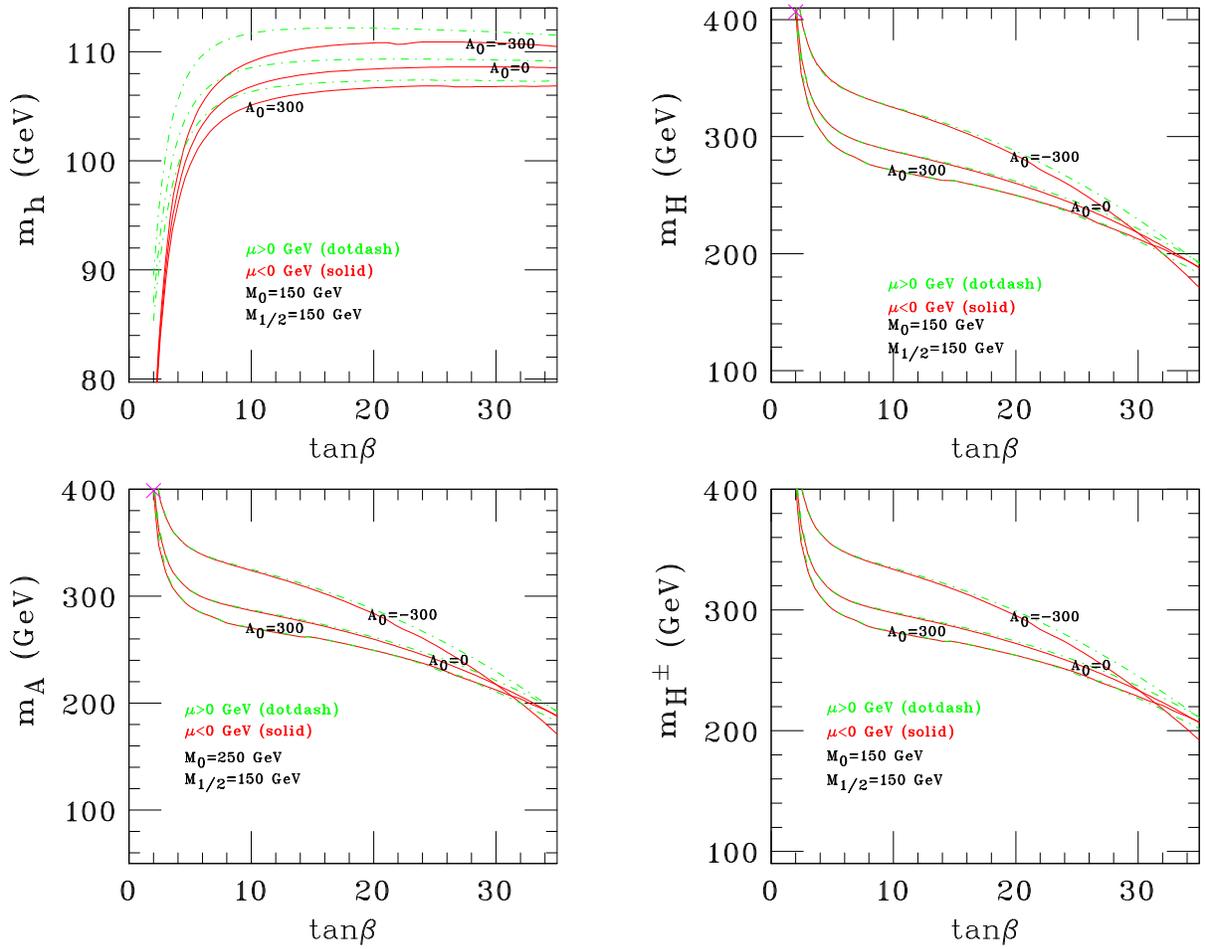,width=5in,angle=90}}
\caption{Resulting masses from \cite{isajet} for the five Higgs bosons 
versus  $\tan\beta$ for three 
 values of $A_0$ and
for both positive (dot-dashed) and negative (solid)  $\mu$.}
\label{fig:higgs}
\end{figure}

It is convenient to rearrange the 
amplitudes corresponding to the graphs in Fig~\ref{fig:graphs}
in terms of their colour structure. For example,
it is trivial to recognise the existence of only two combinations
of colour matrices, namely, $t^A_{ac}t^B_{cb}\equiv(t^At^B)_{ab}$ 
and $t^B_{ac}t^A_{cb}\equiv(t^Bt^A)_{ab}$,
where $A,B=1, \dots 8$ 
are the gluon and $a,b,c=1, \dots 3$ the squark colour indices.
This is immediate for graphs 1 to 6, as one can realise
by explicitly writing down the QCD Feynman rules.
In addition, the triple-gluon diagrams, graphs 7 and 8,
are proportional to the anti-commutator $[t^A,t^B]$, whereas those involving 
quartic couplings, graphs 9 and 10, depend on the commutator $\{t^A,t^B\}$.

Therefore, the total amplitude of processes of the type (\ref{reaction})
can be written as
\begin{equation}\label{colour}
A^{\{\lambda\}}_{{A,B};{{a,b}}} = (t^At^B)_{ab}T_1^{\{\lambda\}} 
                                        +(t^Bt^A)_{ab}T_2^{\{\lambda\}}. 
\end{equation}
The two subamplitudes $T_i^{\{\lambda\}}$, $i=1,2$, are obtained
as follows
\begin{eqnarray}\label{subamp}
T^{\{\lambda\}}_{1}&=&\sum_{i=1}^3    M^{\{\lambda\}}_{i}
                     +\sum_{i=7}^8    M^{\{\lambda\}}_{i}
                     +\sum_{i=9}^{10} M^{\{\lambda\}}_{i},
\\ \nonumber
T^{\{\lambda\}}_{2}&=&\sum_{i=4}^6    M^{\{\lambda\}}_{i}
                     -\sum_{i=7}^8    M^{\{\lambda\}}_{i}
                     +\sum_{i=9}^{10} M^{\{\lambda\}}_{i},
\nonumber
\end{eqnarray}
where the $M^{\{\lambda\}}_{i}$, $i=1, \dots 10$, are the original
Feynman amplitudes associated to the graphs in Fig.~\ref{fig:graphs}, but
deprived of their colour structure (and couplings, see 
eq.~(\ref{M2}) below)\footnote{The notation ${\{\lambda\}}$ refers cumulatively
to the helicities of the two incoming gluons, i.e., $\lambda_i$ with $i=1,2$.}.

This way, the total amplitude squared, summed/averaged over the final/initial
spin and colours, can be expressed in terms of only two colour factors, as
\begin{equation}\label{M2}
{\M}(gg \ar {\tilde{q}}_{\chi} {\tilde{q}}^{'*}_{\chi'} \Phi) =
  |\lambda_{\Phi\tilde{q}_\chi\tilde{q}'_{\chi'}}|^2
  \frac{g_s^4 g_W^2}{256}
\sum_{\{\lambda\}=\pm} \sum_{i=1}^2 \sum_{j=1}^2 {T}^{\{\lambda\} }_i
                                                 {T}^{\{\lambda\}*}_j C_{ij},
\end{equation}
with $g_s^2={4\pi}{\alpha_{\mathrm{s }}}                 $, 
     $g_W^2={4\pi}{\alpha_{\mathrm{em}}}/{s^2_W}$ and
where $C_{ij}$ is a $(2\times2)$ colour matrix with elements
\begin{eqnarray}\label{matrix}
C_{11}&\equiv&C_{22}=\frac{1}{4}(\frac{1}{N_C}-2N_C+N_C^3)=\frac{16}{3}, 
\\ \nonumber
C_{12}&\equiv&C_{21}=\frac{1}{4}(\frac{1}{N_C}-N_C)=-\frac{2}{3},   
\nonumber
\end{eqnarray}
as usual being $N_C\equiv3$ the number of colours in QCD.
In eq.~(\ref{M2}), $\lambda_{\Phi\tilde{q}_\chi\tilde{q}'_{\chi'}}$
represents the strength of the squark-squark-Higgs vertex involved,
as described in Sect.~\ref{sec:parameters}
(apart from an overall phase and the factor $g_W$).

The ten amplitudes $M^{\{\lambda\}}_{i}$ are simply
\begin{eqnarray}\label{M}
{M}^{\{\lambda\}}_{1}&=& 
4 \eps_1(\lambda_1) \cdot p_4 ~\eps_2(\lambda_2) \cdot (p_3+p_5)/P_{14}/P_{35},
\\ \nonumber
{M}^{\{\lambda\}}_{2}&=& 
4 \eps_1(\lambda_1) \cdot (p_4+p_5) ~\eps_2(\lambda_2) \cdot p_3/P_{23}/P_{45},
\\ \nonumber
{M}^{\{\lambda\}}_{3}&=& 
4 \eps_1(\lambda_1) \cdot p_4 ~\eps_2(\lambda_2) \cdot p_3/P_{14}/P_{23},
\\ \nonumber
{M}^{\{\lambda\}}_{4}&=& 
4 \eps_1(\lambda_1) \cdot (p_3+p_5) ~\eps_2(\lambda_2) \cdot p_4/P_{24}/P_{35},
\\ \nonumber
{M}^{\{\lambda\}}_{5}&=& 
4 \eps_1(\lambda_1) \cdot p_3 ~\eps_2(\lambda_2) \cdot (p_4+p_5)/P_{13}/P_{45},
\\ \nonumber
{M}^{\{\lambda\}}_{6}&=& 
4 \eps_1(\lambda_1) \cdot p_3 ~\eps_2(\lambda_2) \cdot p_4/P_{24}/P_{13},
\\ \nonumber
{M}^{\{\lambda\}}_{7}&=&
\eps_{12}(\lambda_1,\lambda_2) \cdot (-2p_4+p_1+p_2)/P_{35},         
 \\ \nonumber
{M}^{\{\lambda\}}_{8}&=& 
\eps_{12}(\lambda_1,\lambda_2) \cdot (+2p_3-p_1-p_2)/P_{45},          
\\ \nonumber
{M}^{\{\lambda\}}_{9}&=& 
\eps_1(\lambda_1) \cdot \eps_2(\lambda_2)/P_{35},
\\ \nonumber
{M}^{\{\lambda\}}_{10}&=& 
\eps_1(\lambda_1) \cdot \eps_2(\lambda_2)/P_{45},
\\ \nonumber
\end{eqnarray}
where we have introduced the propagator functions
\begin{eqnarray}\label{P}
      P_{14}&=&(p_1-p_4)^2-M_{{\tilde{q}'}_{\chi'}}^2,    \\ \nonumber
      P_{24}&=&(p_2-p_4)^2-M_{{\tilde{q}'}_{\chi'}}^2,    \\ \nonumber
      P_{23}&=&(p_2-p_3)^2-M_{{\tilde{q}}_{\chi}}^2,        \\ \nonumber
      P_{13}&=&(p_1-p_3)^2-M_{{\tilde{q}}_{\chi}}^2,        \\ \nonumber
      P_{35}&=&(p_3+p_5)^2-M_{{\tilde{q}'}_{\chi'}}^2
             +{\mathrm{i}} M_{{\tilde{q}'}_{\chi'}} 
                           \Gamma_{{\tilde{q}'}_{\chi'}}, \\ \nonumber
      P_{45}&=&(p_4+p_5)^2-M_{{\tilde{q}}_{\chi}}^2
             +{\mathrm{i}} M_{{\tilde{q}}_{\chi}} 
                           \Gamma_{{\tilde{q}}_{\chi}},     \\ \nonumber
\end{eqnarray}
the gluon polarisation vectors \cite{HZ}, $i=1,2$, 
\begin{eqnarray}\label{vectors}
\eps_i^\mu(\lambda_i=\pm)&=&\frac{1}{\sqrt 2}[\mp\eps_i^\mu(\lambda_i=1)
                                   -{\mathrm{i}}~\eps_i^\mu(\lambda_i=2)],
\\ \nonumber
\eps_i^\mu(\lambda_i=1)&=&(|\vec{p}_i|p_i^T)^{-1}
                          (0, p_i^x p_i^z, p_i^y p_i^z, - {p_i^T}^2),
\\ \nonumber
\eps_i^\mu(\lambda_i=2)&=&(p_i^T)^{-1}
                          (0, - p_i^y , p_i^x , 0),
\\ \nonumber
\end{eqnarray}
with
\begin{eqnarray}
p_i^T       &=& \sqrt{ {p_i^x}^2 + {p_i^y}^2 },             \\ \nonumber
|\vec{p}_i| &=& \sqrt{ {p_i^x}^2 + {p_i^y}^2 + {p_i^z}^2 }, \\ \nonumber
\end{eqnarray}
and their contraction over the triple-gluon vertex times the gluon
propagator
\begin{eqnarray}\label{ggg}
\eps_{12}^\mu(\lambda_1,\lambda_2)&=&\frac{1}{P_{12}}
\{
(p_1-p_2)^\mu \eps_1(\lambda_1) \cdot \eps_2(\lambda_2) \\ \nonumber
&&\phantom{\frac{1}{P_{12}}}
+[(p_2+p_{12}) \cdot \eps_1(\lambda_1)] \eps_2^\mu(\lambda_2)  
-[(p_1+p_{12}) \cdot \eps_2(\lambda_2)] \eps_1^\mu(\lambda_1) \}, \\ \nonumber 
\end{eqnarray}
with $P_{12}=p_{12}^2\equiv(p_1+p_2)^2$.

\vfill\clearpage\thispagestyle{empty}

\vfill\clearpage\thispagestyle{empty}
\vfill\clearpage\thispagestyle{empty}
\vfill\clearpage\thispagestyle{empty}
\begin{figure}
\centerline{
\vbox{
\epsfig{figure=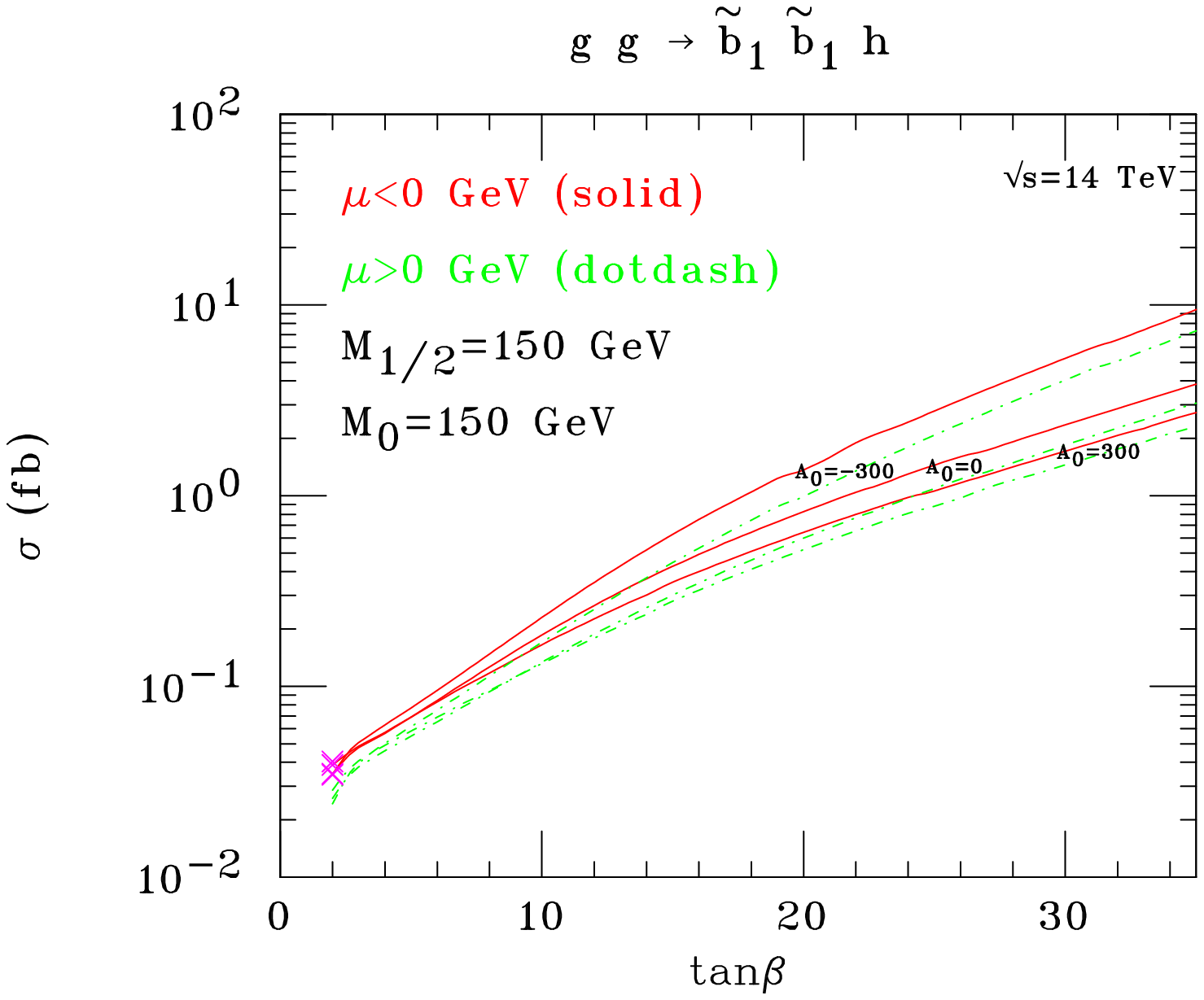,
width=2.7in,angle=90}
\epsfig{figure=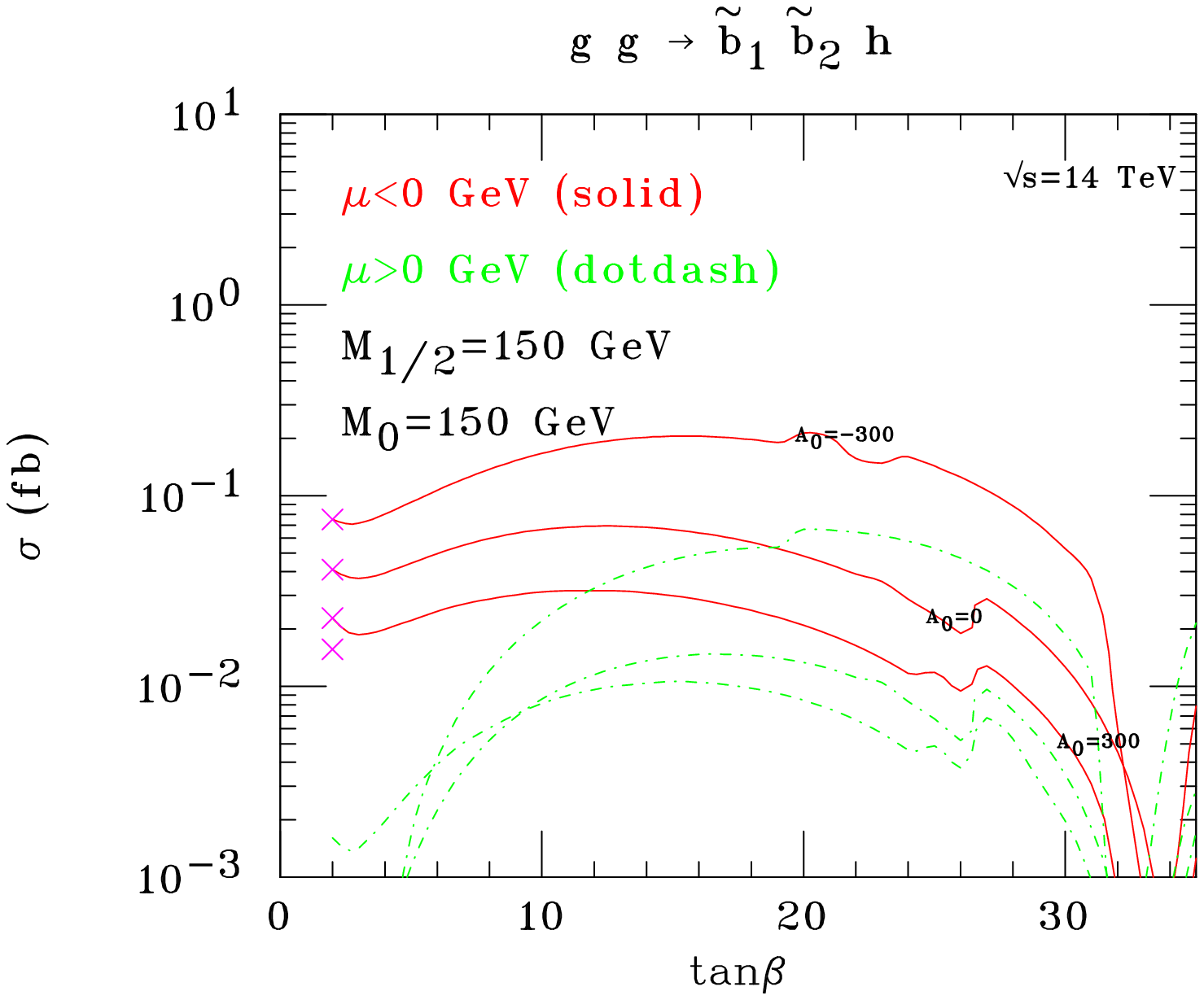,
width=2.7in,angle=90}\\[0.2cm]
\epsfig{figure=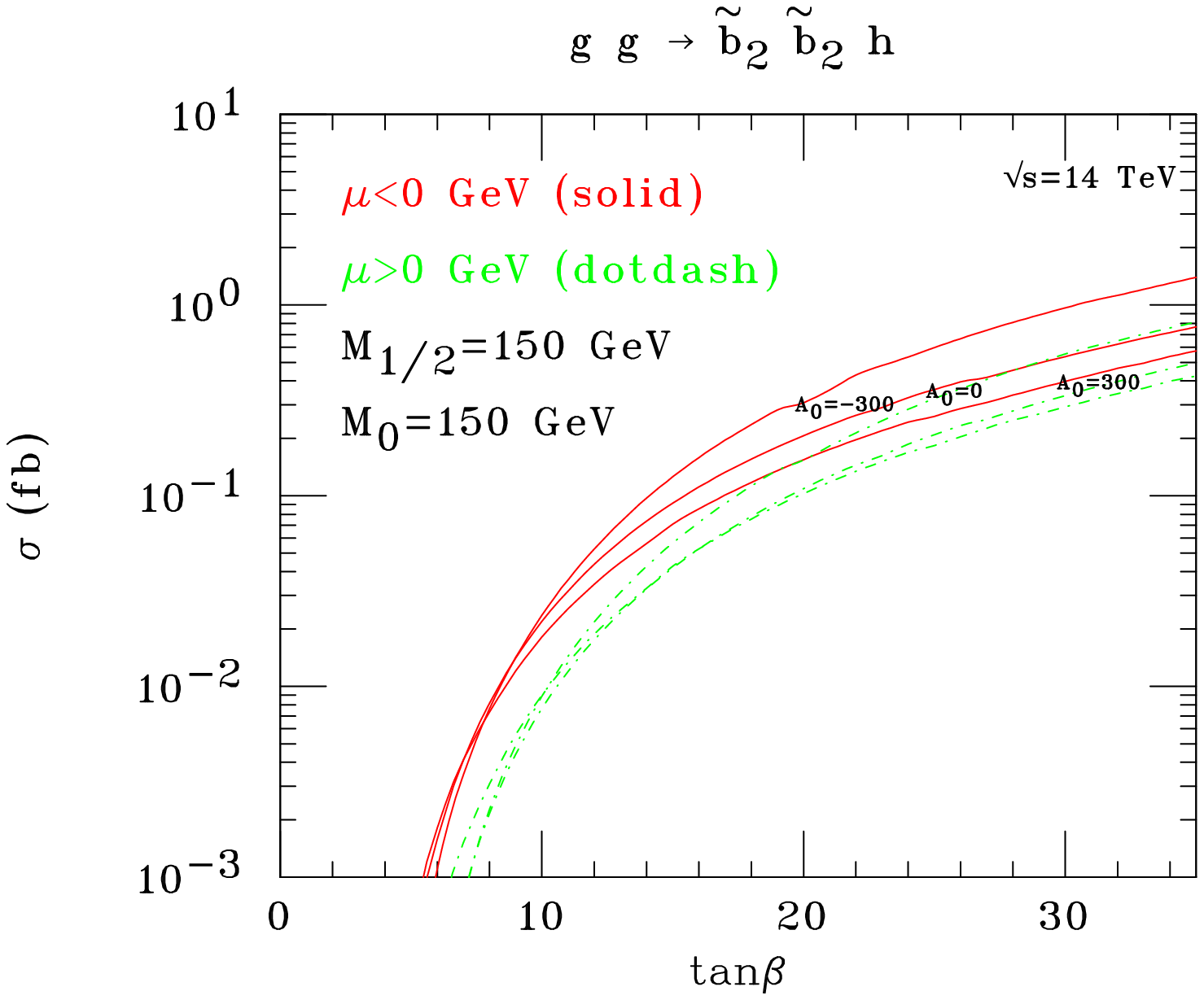,
width=2.7in,angle=90}
\epsfig{figure=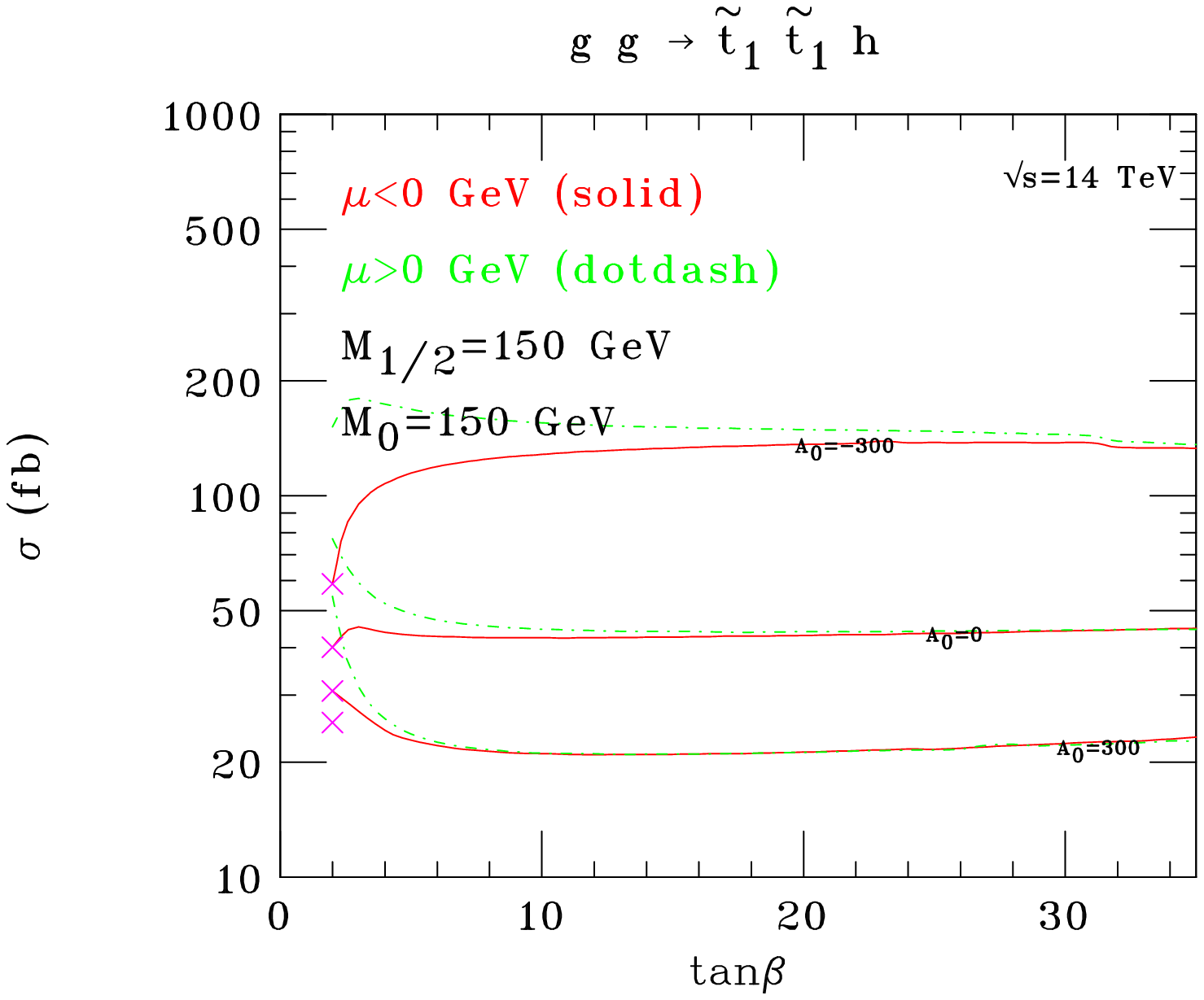,
width=2.7in,angle=90}\\[0.2cm]
\epsfig{figure=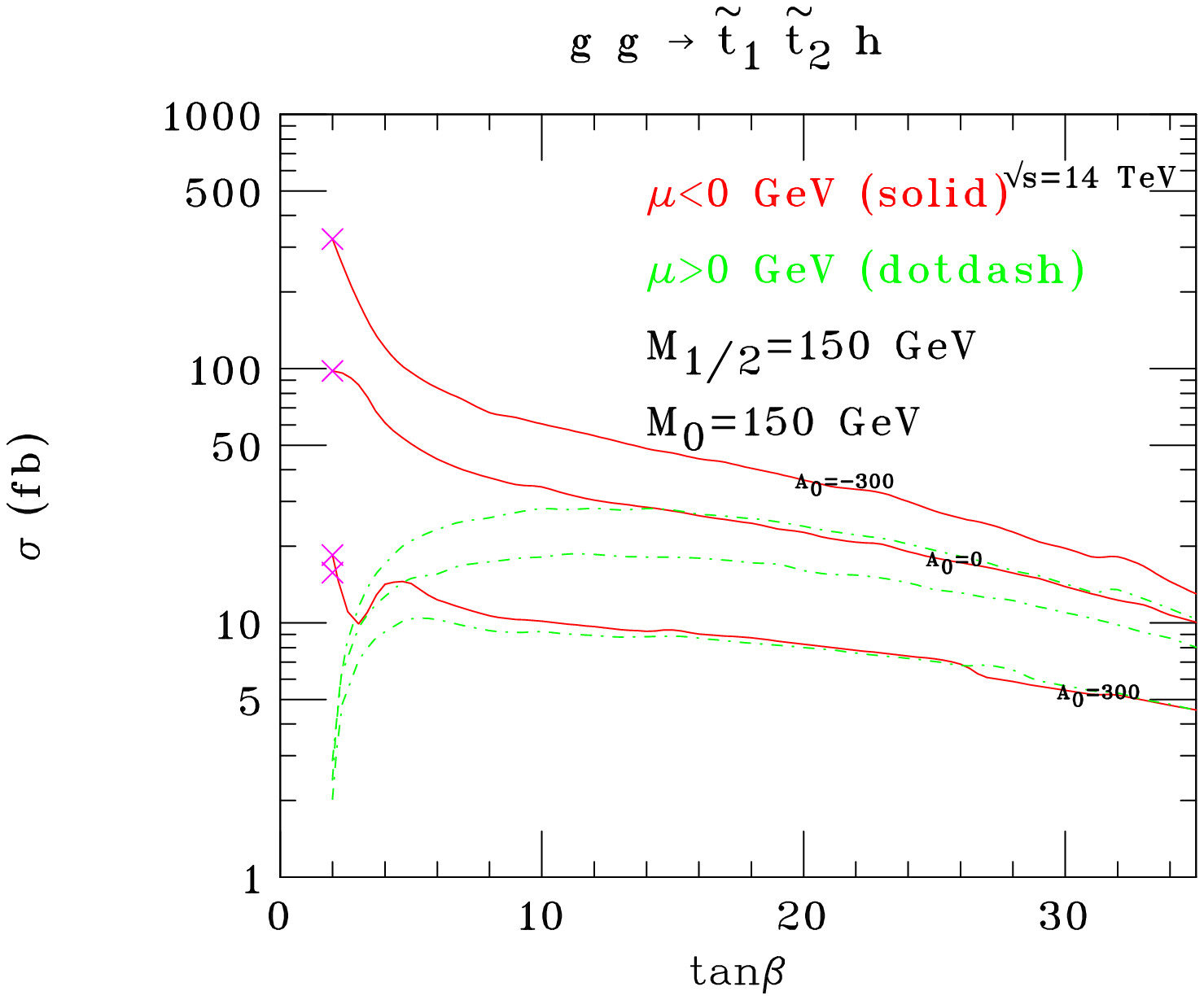,
width=2.7in,angle=90}
\epsfig{figure=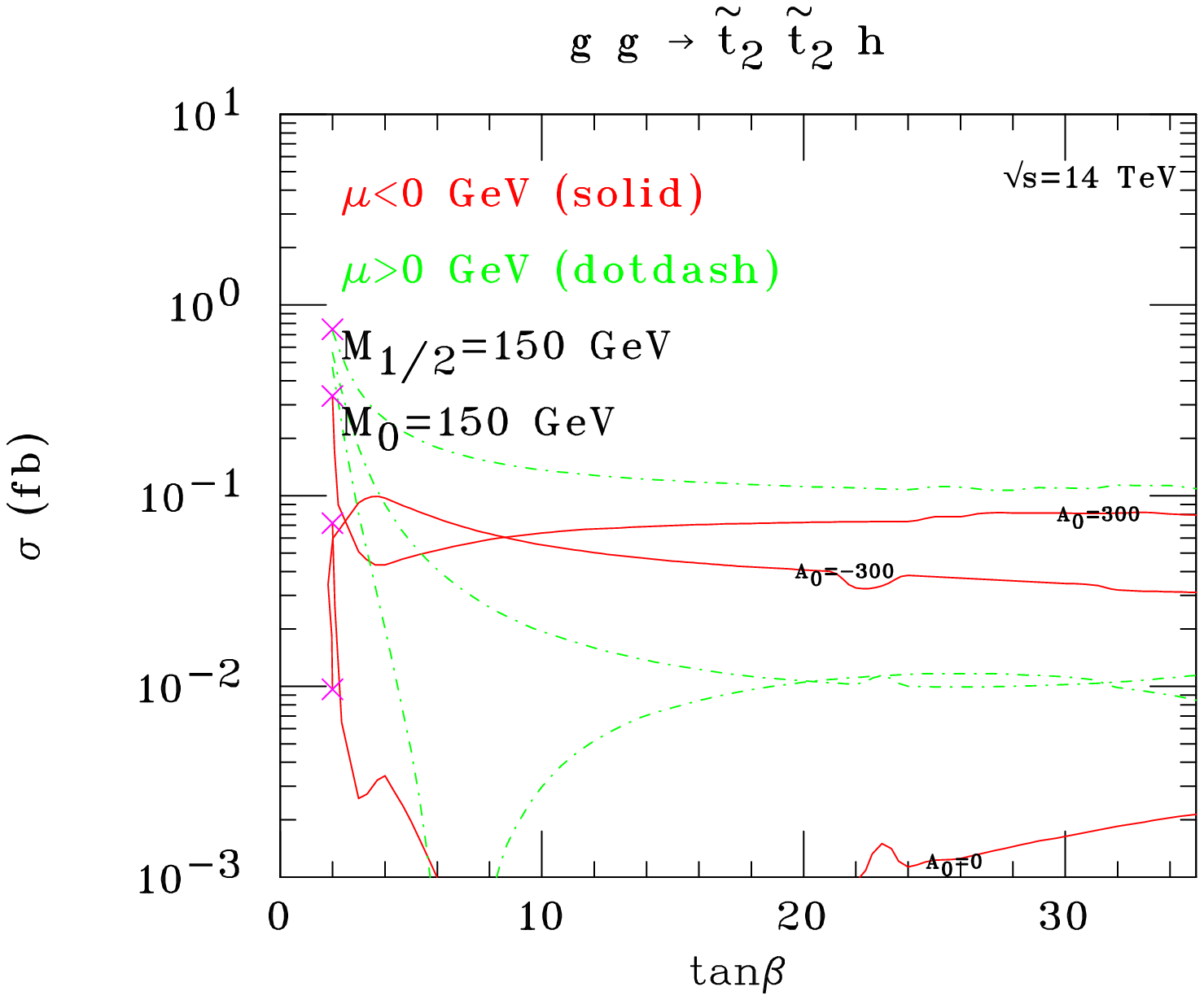,
width=2.7in,angle=90}
}}
\caption{Total cross section of $gg\rightarrow \tilde{q}_{1,2}
\tilde{q}_{1,2} h$ processes (with $q=t,b$) as a function of $\tan\beta$ for
the characteristic input values $M_0=150 $ GeV and $M_{1/2}=150$ GeV.
Both positive (dot-dashed) and negative (solid) $\mu$
as well several $A_0$ contour lines
are shown. The symbol ``$\times$'' is used to indicate parameter
areas forbidden by direct Higgs boson searches.}
\label{fig:lighthiggs}
\end{figure}
\vfill\clearpage\thispagestyle{empty}
\begin{figure}
\centerline{
\vbox{
\epsfig{figure=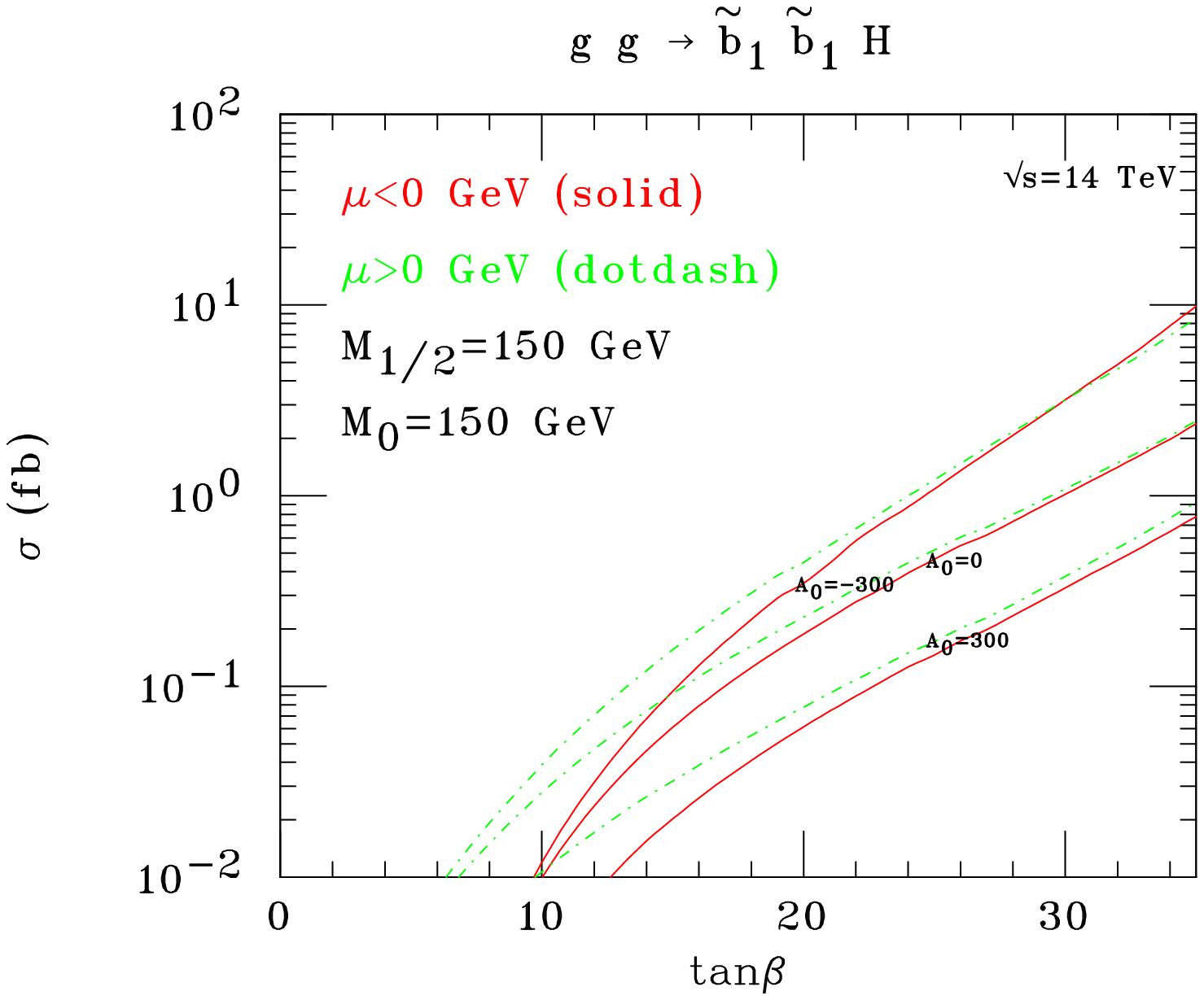,
width=2.7in,angle=90}
\epsfig{figure=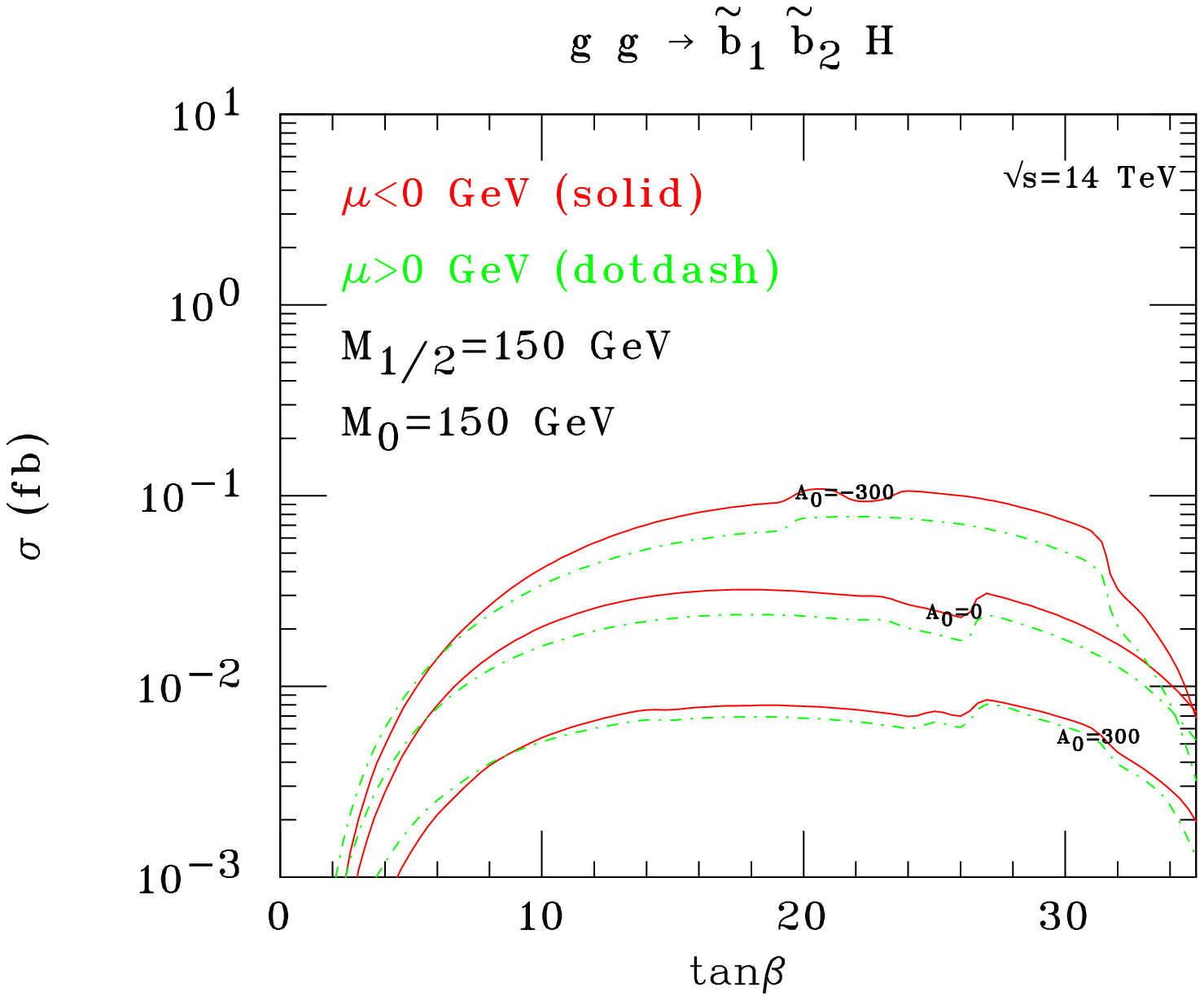,
width=2.7in,angle=90}\\[0.2cm]
\epsfig{figure=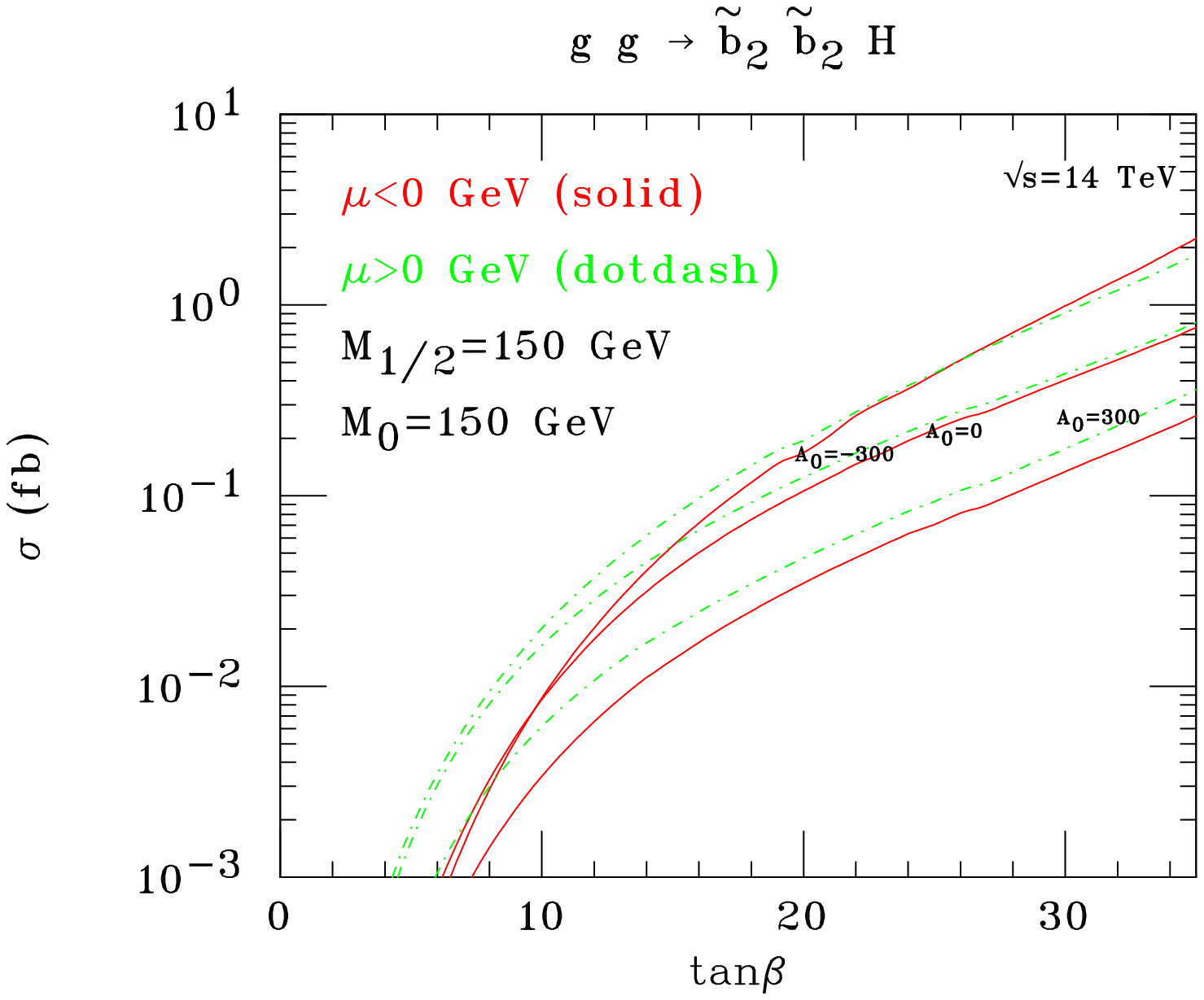,
width=2.7in,angle=90}
\epsfig{figure=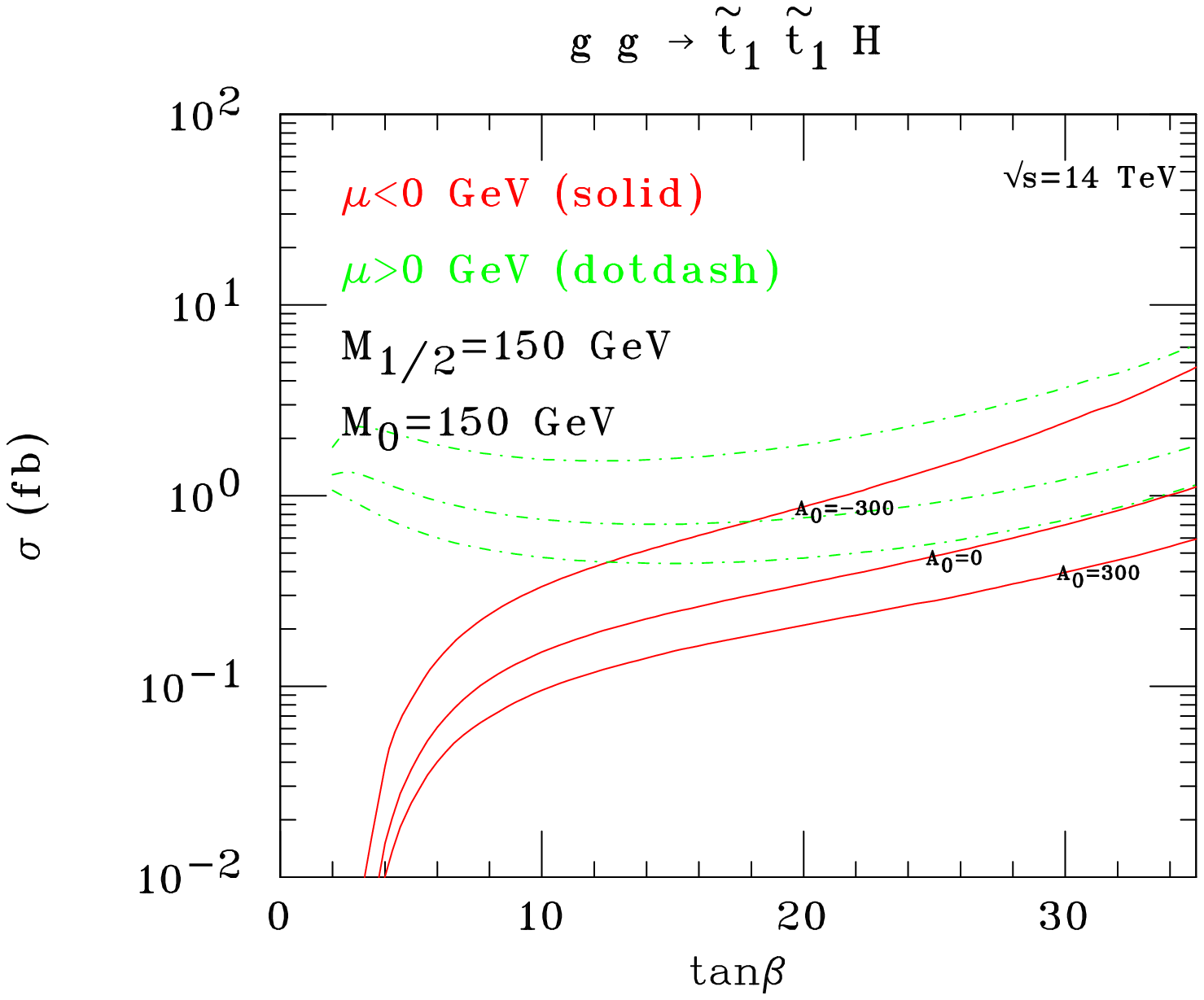,
width=2.7in,angle=90}\\[0.2cm]
\epsfig{figure=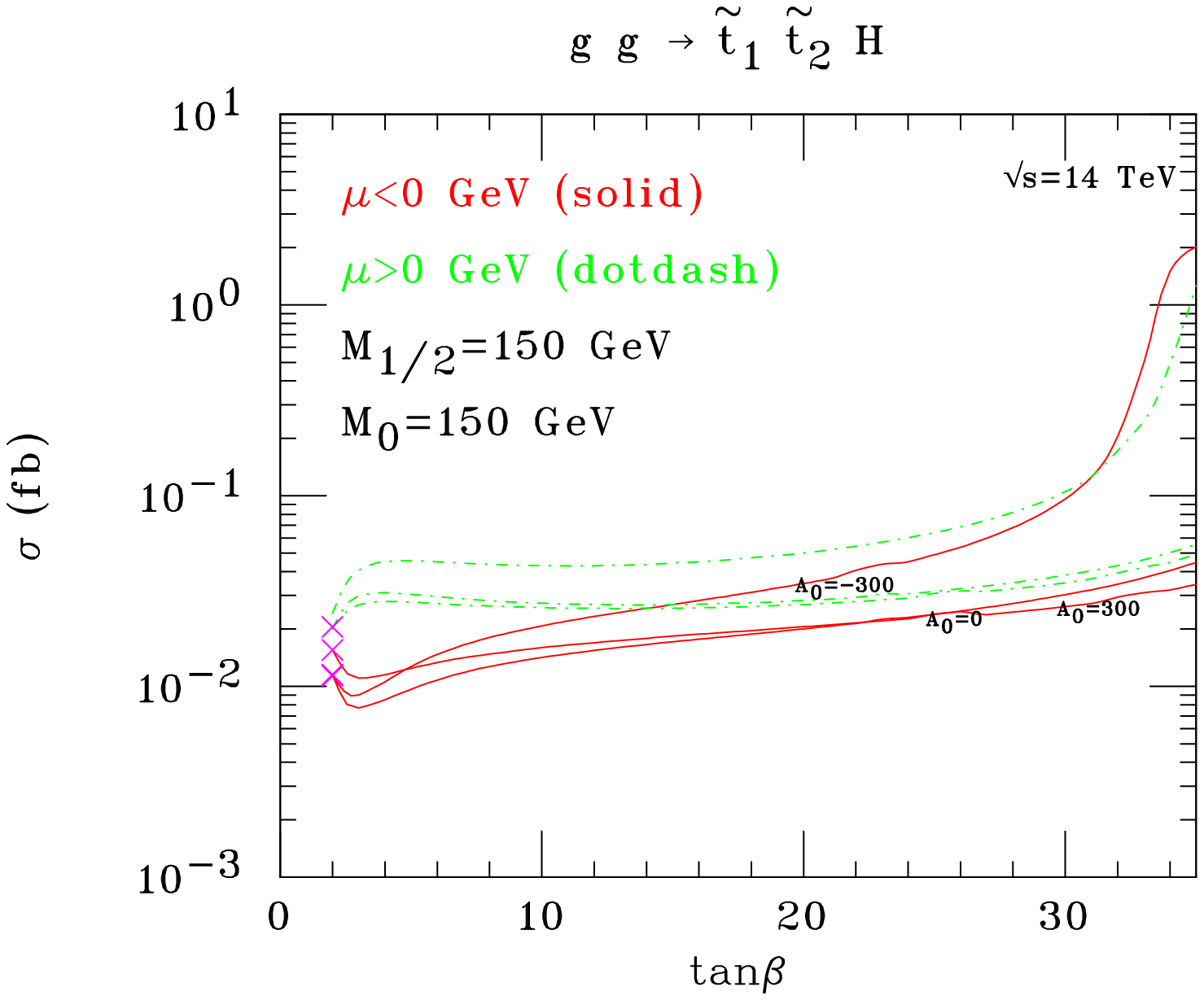,
width=2.7in,angle=90}
\epsfig{figure=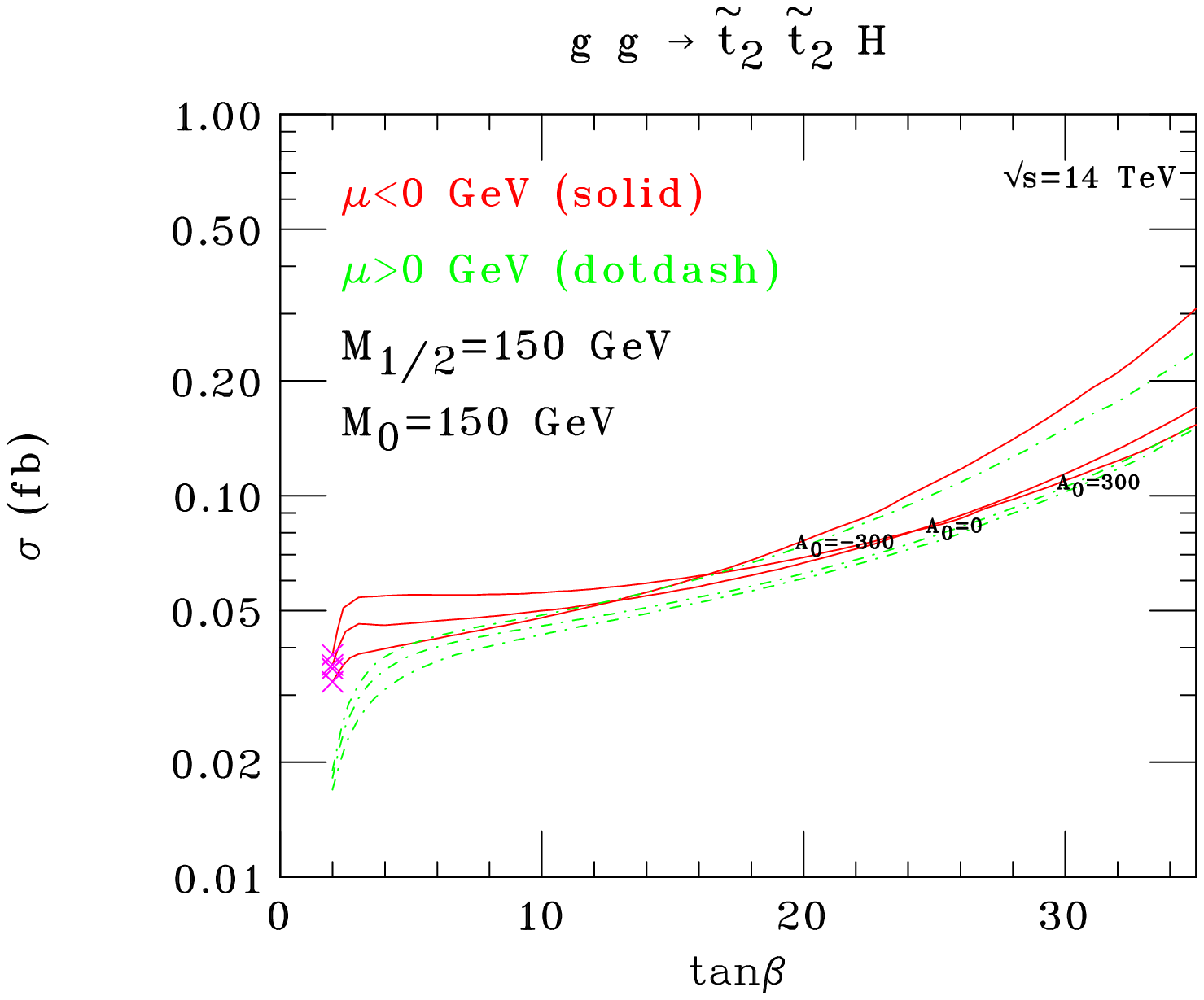,
width=2.7in,angle=90}
}}
\caption{Total cross section of $gg\rightarrow \tilde{q}_{1,2}
\tilde{q}_{1,2} H$ processes (with $q=t,b$) as a function of $\tan\beta$ for
the characteristic  input values $M_0=150 $ GeV and $M_{1/2}=150$ GeV.
Both positive (dot-dashed) and negative (solid) $\mu$
as well several $A_0$ contour lines
are shown. The symbol ``$\times$'' is used to indicate parameter
areas forbidden by direct Higgs boson searches.}
\label{fig:heavyhiggs}
\end{figure}
\vfill\clearpage\thispagestyle{empty}
\begin{figure}
\begin{center}
\centerline{\vbox{\epsfig
{figure=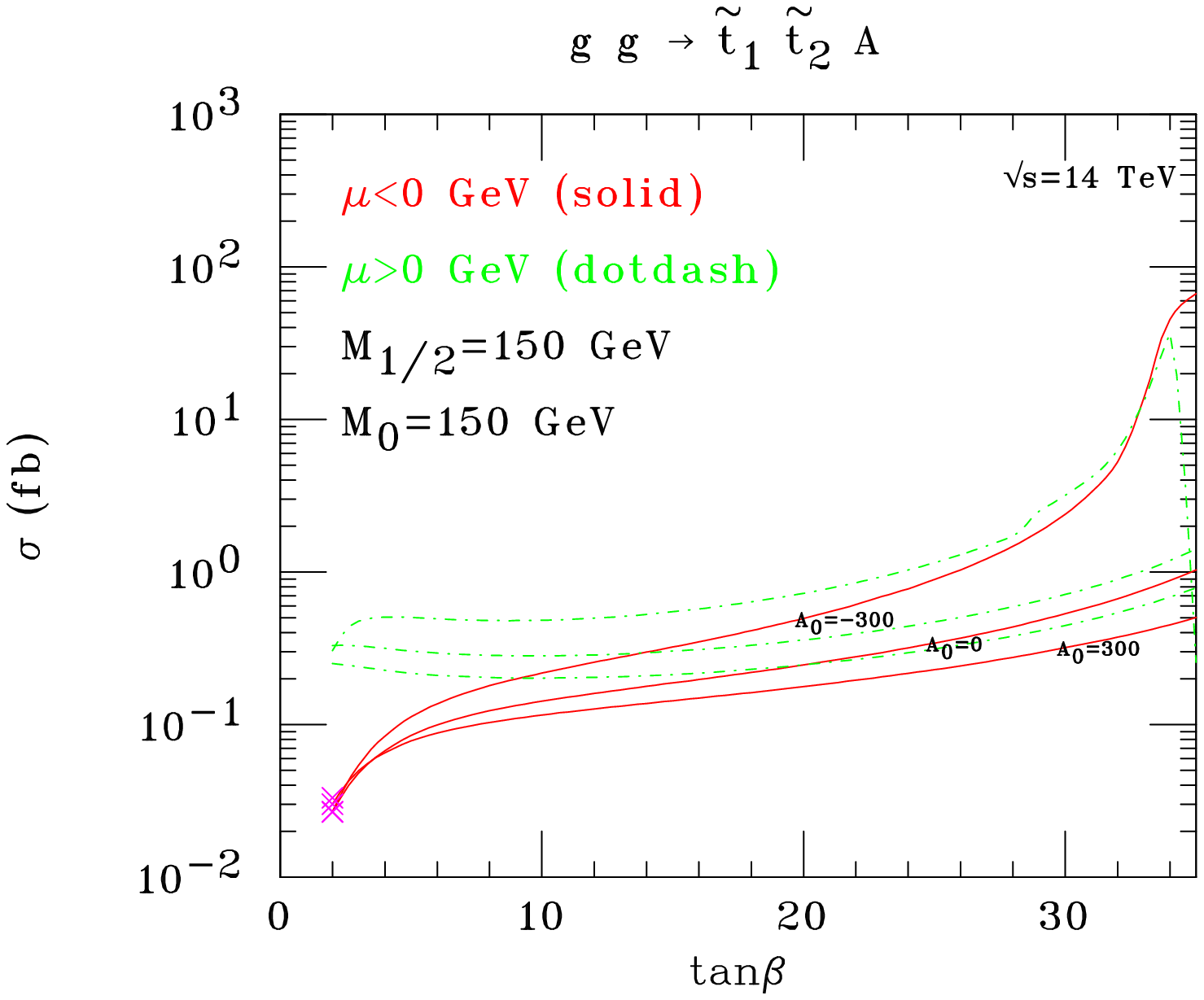,width=4.0in,angle=90}\\[0.3cm]
\epsfig
{figure=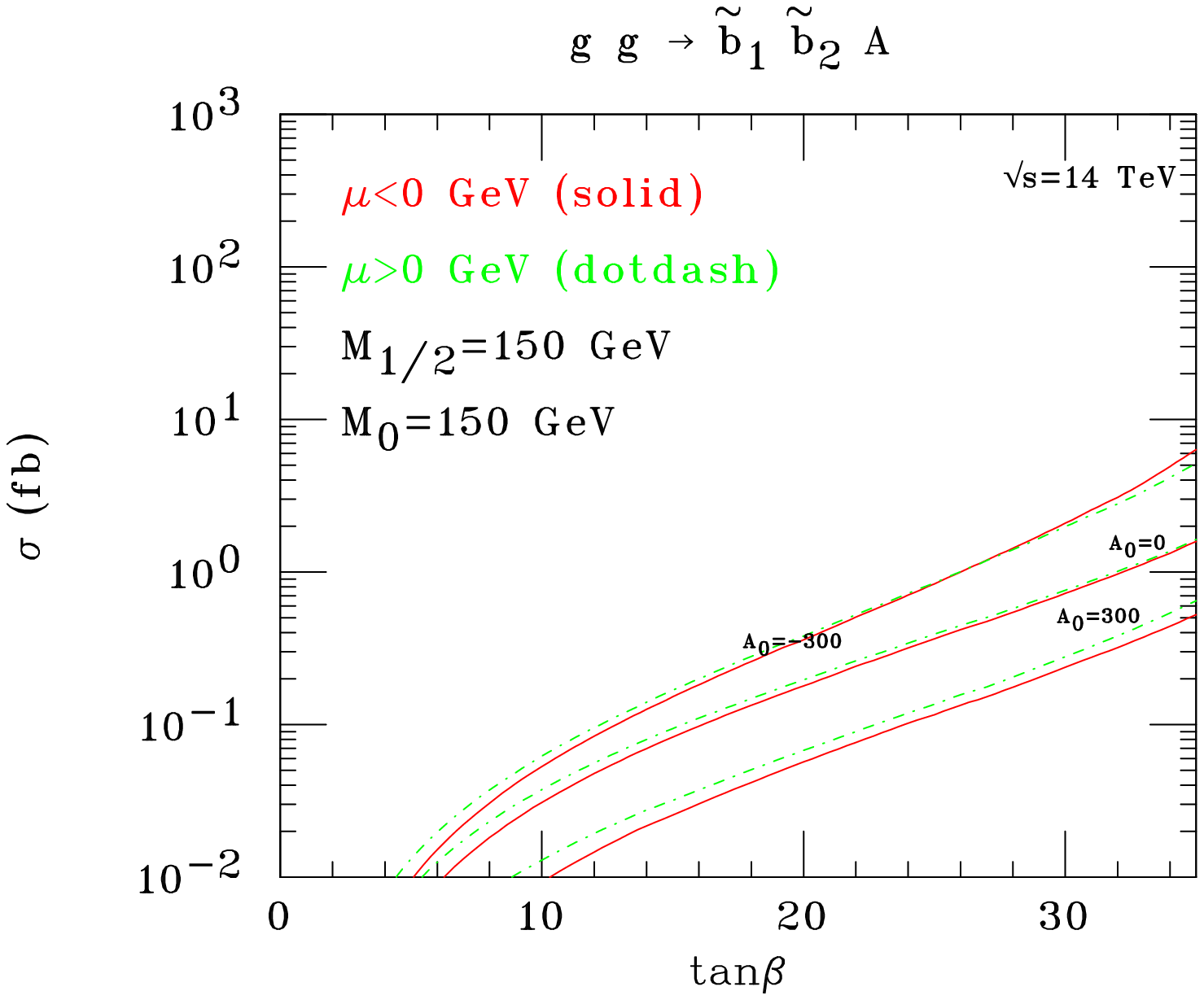,width=4.0in,angle=90}}}
\caption{Total cross section of $gg\rightarrow \tilde{q}_{1}
\tilde{q}_{2} A$ processes (with $q=t,b$) as a function of $\tan\beta$ for
the characteristic  input values $M_0=150 $ GeV and $M_{1/2}=150$ GeV.
Both positive (dot-dashed) and negative (solid) $\mu$
as well several $A_0$ contour lines
are shown. The symbol ``$\times$'' is used to indicate parameter
areas forbidden by direct Higgs boson searches.}
\label{fig:cpoddhiggs}
\end{center}
\end{figure}
\vfill\clearpage\thispagestyle{empty}
\begin{figure}
\centerline{
\vbox{
\epsfig{figure=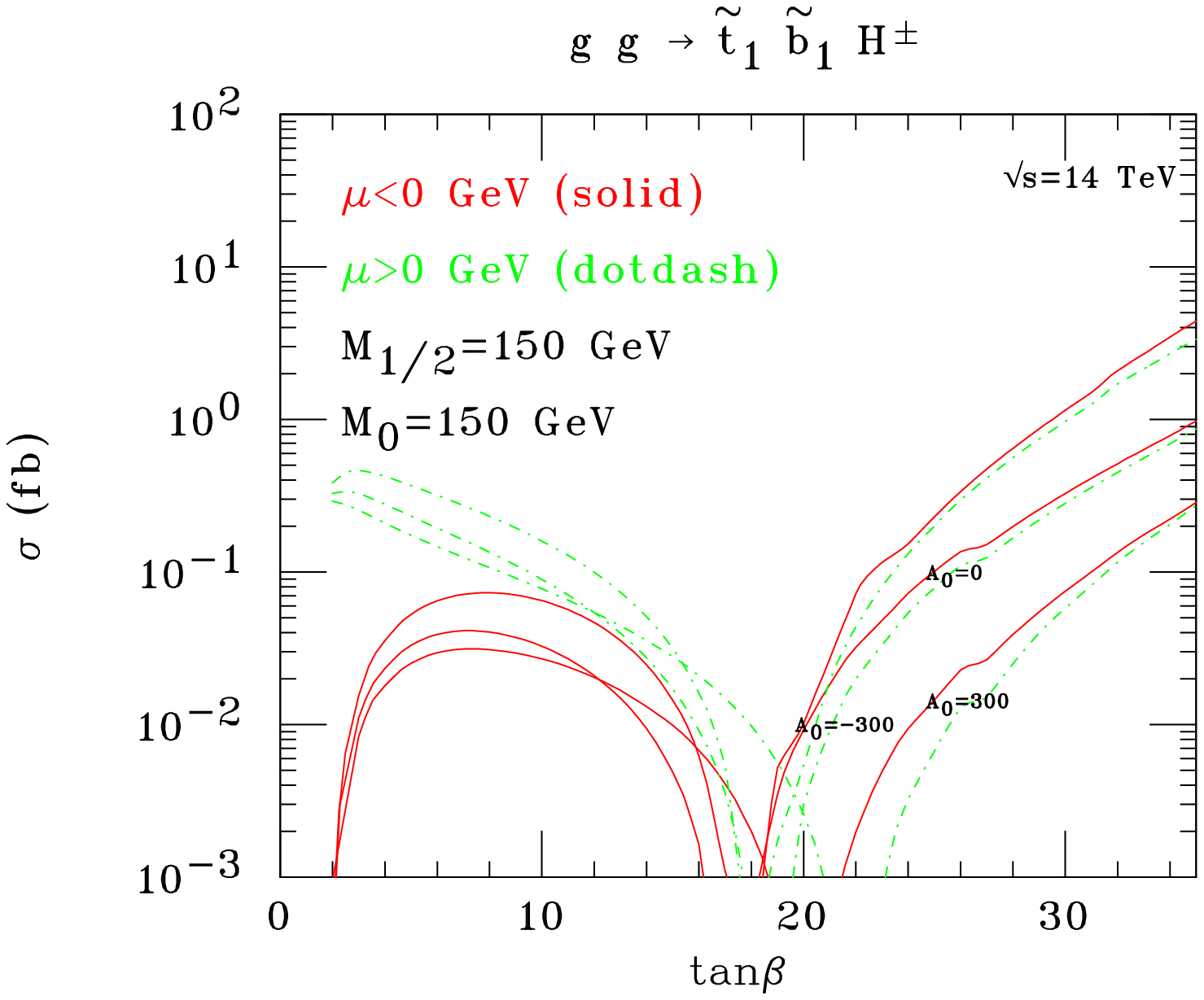,
width=2.7in,angle=90}
\epsfig{figure=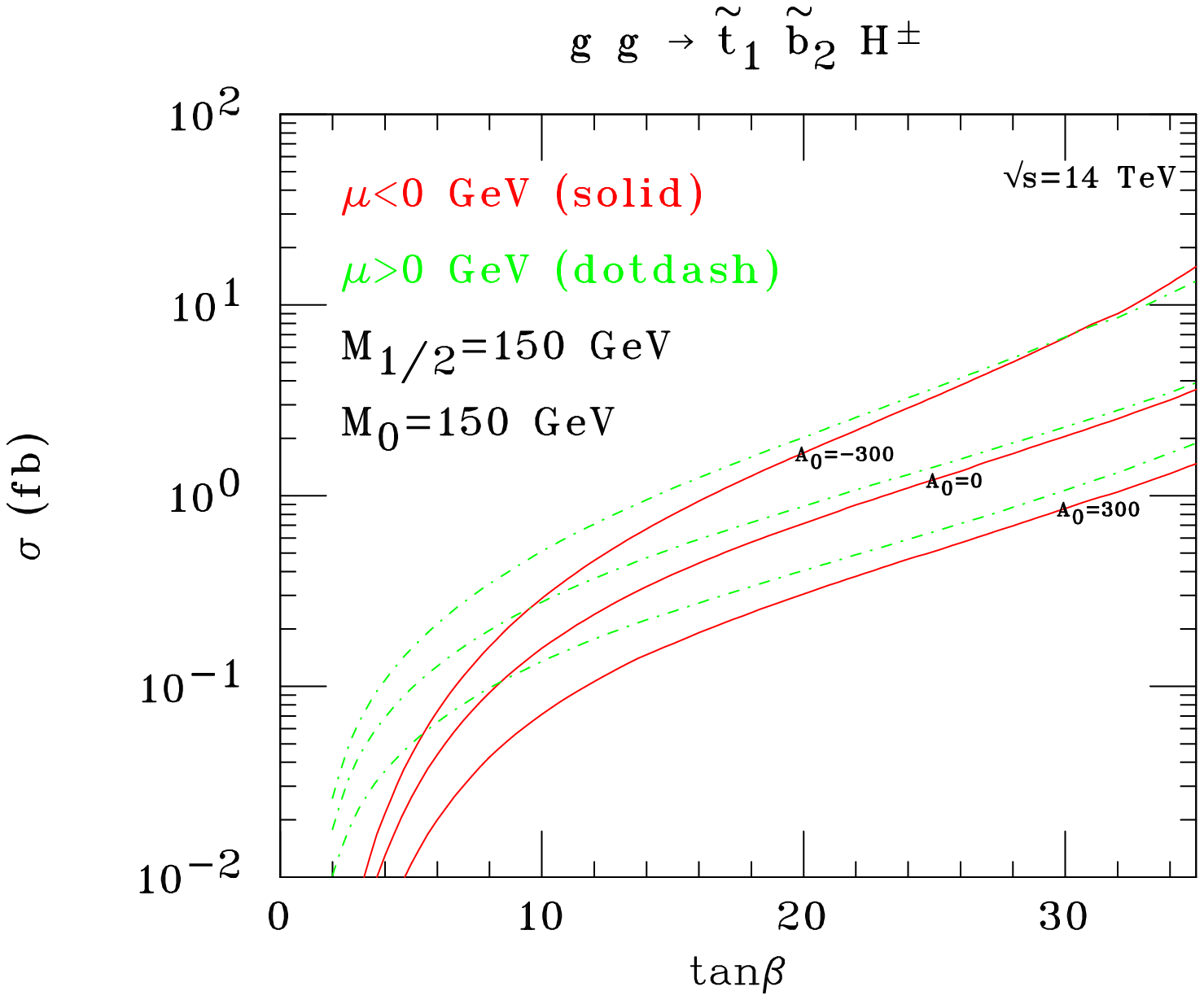,
width=2.7in,angle=90}\\[1cm]
\epsfig{figure=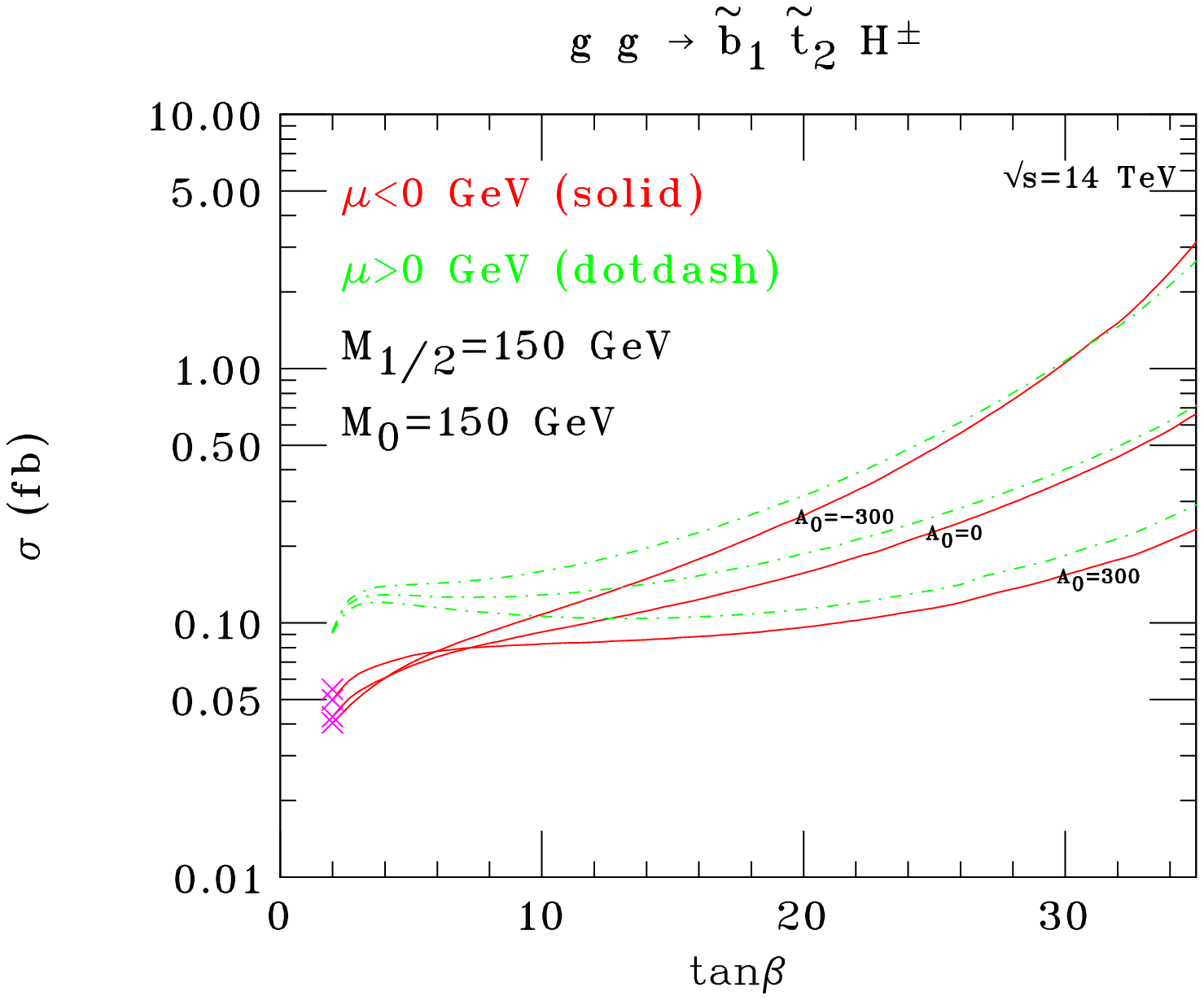,
width=2.7in,angle=90}
\epsfig{figure=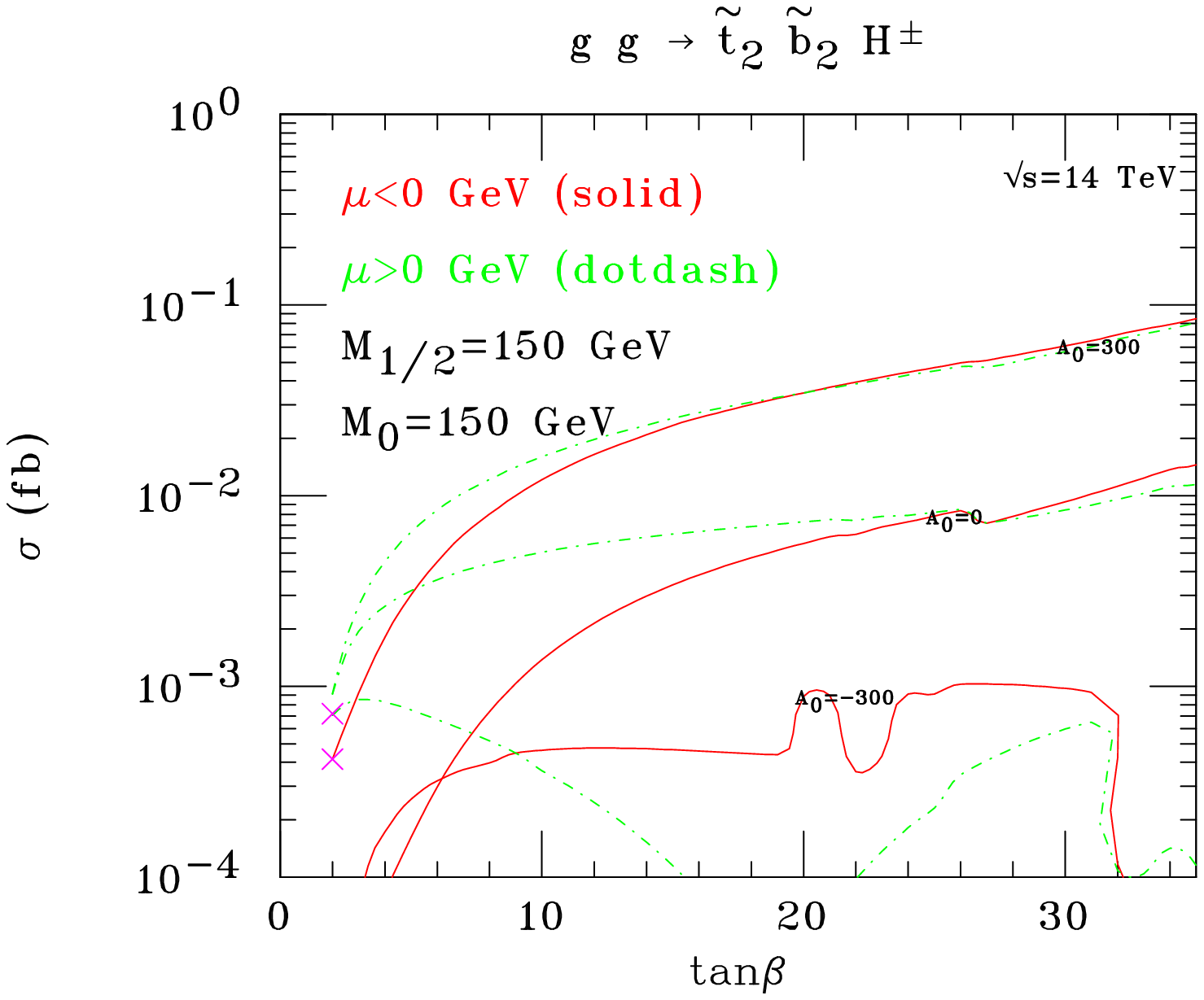,
width=2.7in,angle=90}
}}
\caption{Total cross section of $gg\rightarrow \tilde{q}_{1,2}
\tilde{q}'_{1,2} H^\pm$ processes (with $q^{(')}=t,b$) 
as a function of $\tan\beta$ for
the characteristic  input values $M_0=150 $ GeV and $M_{1/2}=150$ GeV.
Both positive (dot-dashed) and negative (solid) $\mu$
as well several $A_0$ contour lines
are shown. The symbol ``$\times$'' is used to indicate parameter
areas forbidden by direct Higgs boson searches.}
\label{fig:chargedhiggs}
\end{figure}
\vfill\clearpage\thispagestyle{empty}
\begin{figure}
\centerline{\epsfig
{figure=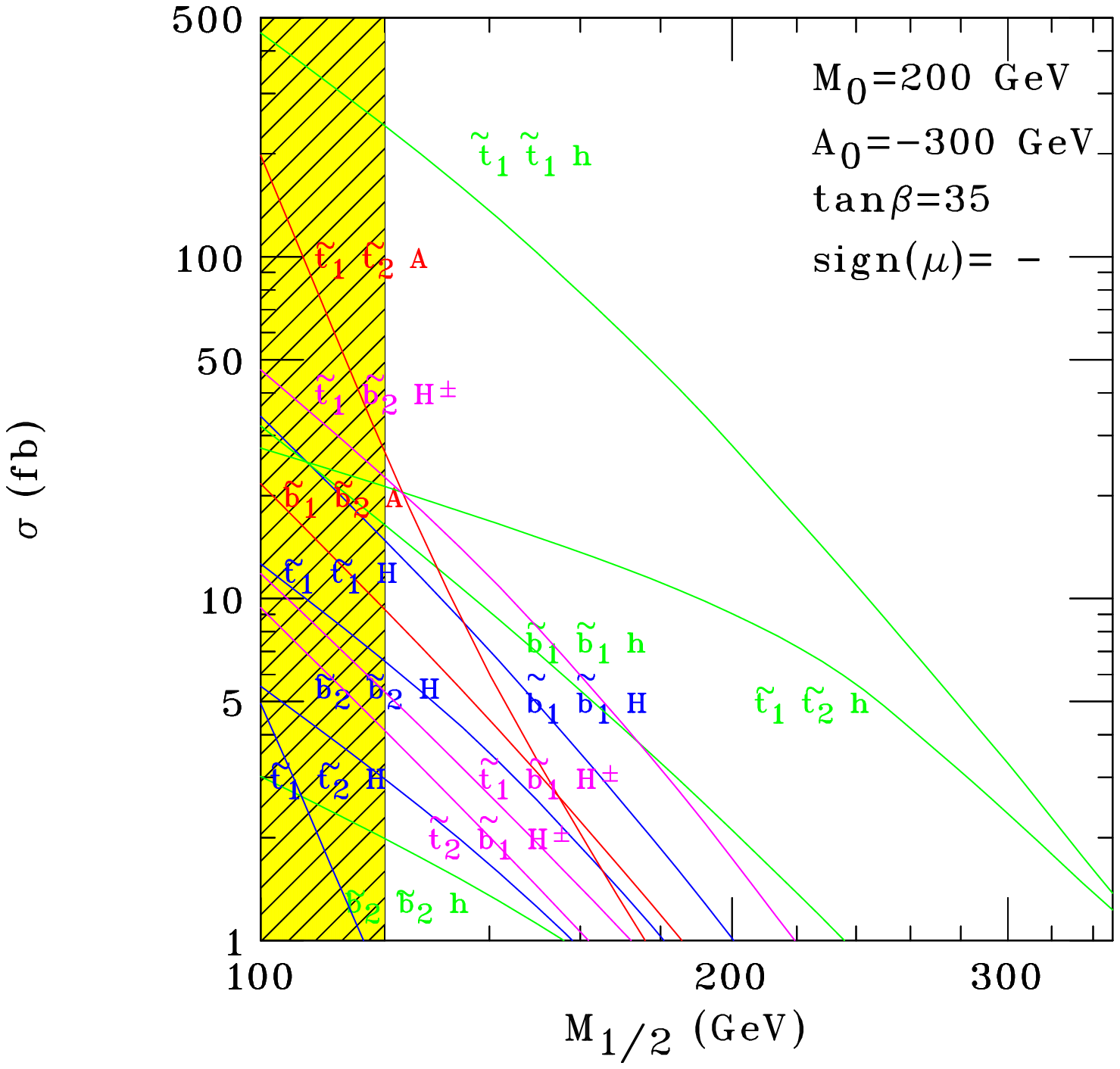,width=3.5in,angle=90}}
\vskip0.05cm\noindent
\centerline{\epsfig
{figure=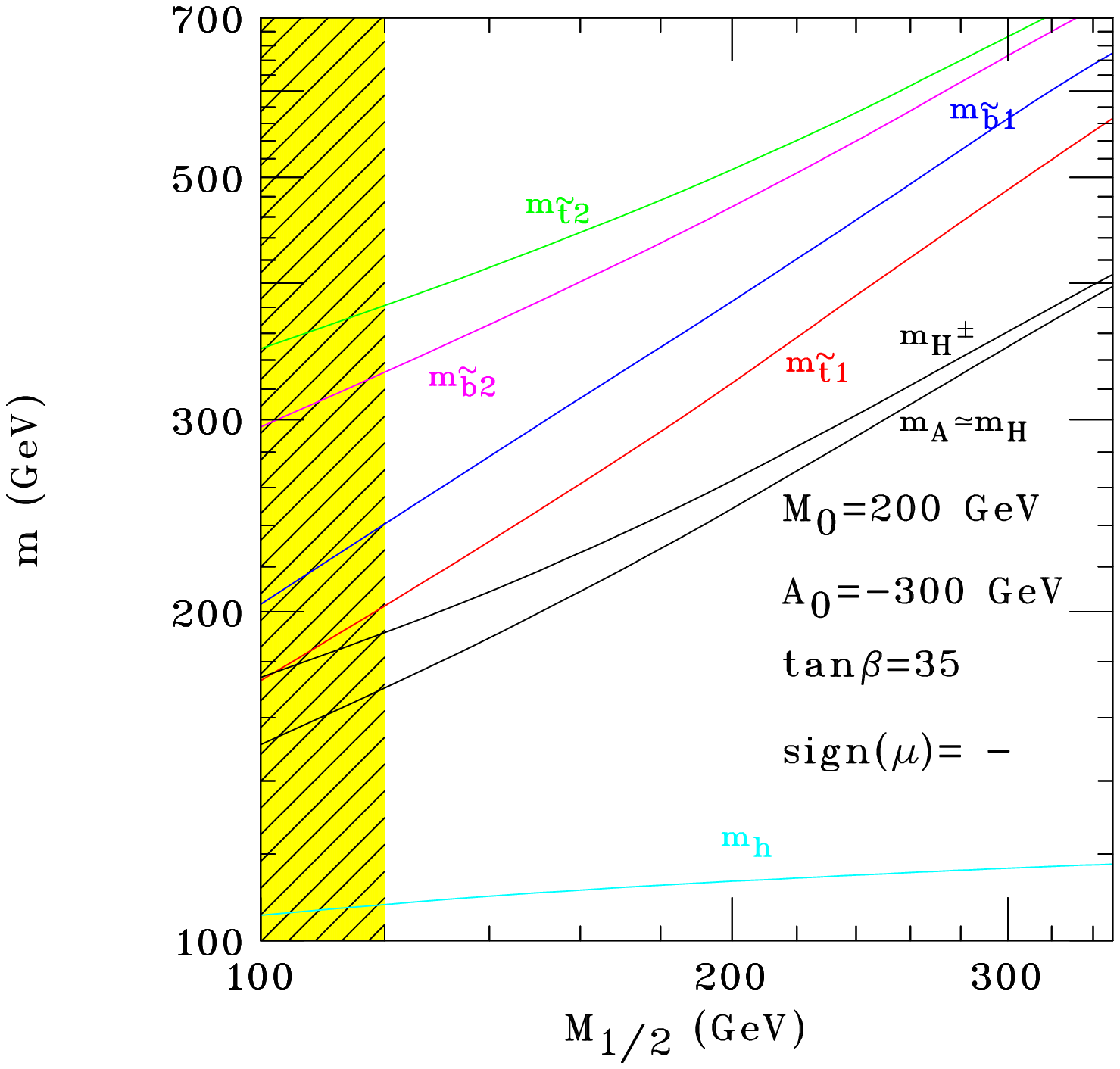,width=3.5in,angle=90}}
\caption{The most significant cross sections which survive
the choice of a light SUSY spectrum (above), alongside the
values for the  masses entering the corresponding production 
processes (below). Shaded regions indicate areas excluded by direct
searches.}
\label{fig:light}
\end{figure}

\vfill\clearpage\thispagestyle{empty}
\begin{figure}
\centerline{\epsfig
{figure=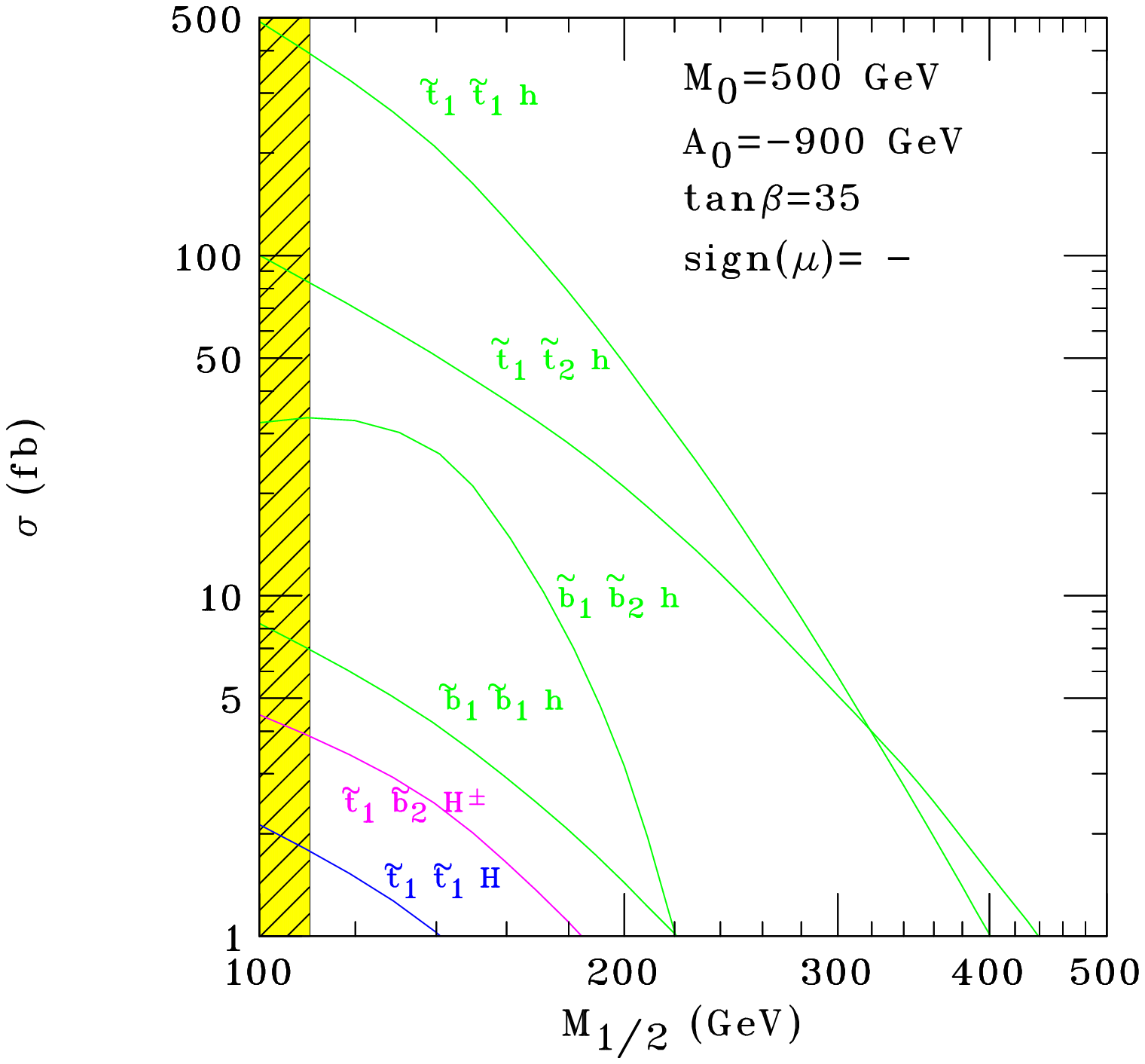,width=3.5in,angle=90}}
\vskip0.05cm\noindent
\centerline{\epsfig
{figure=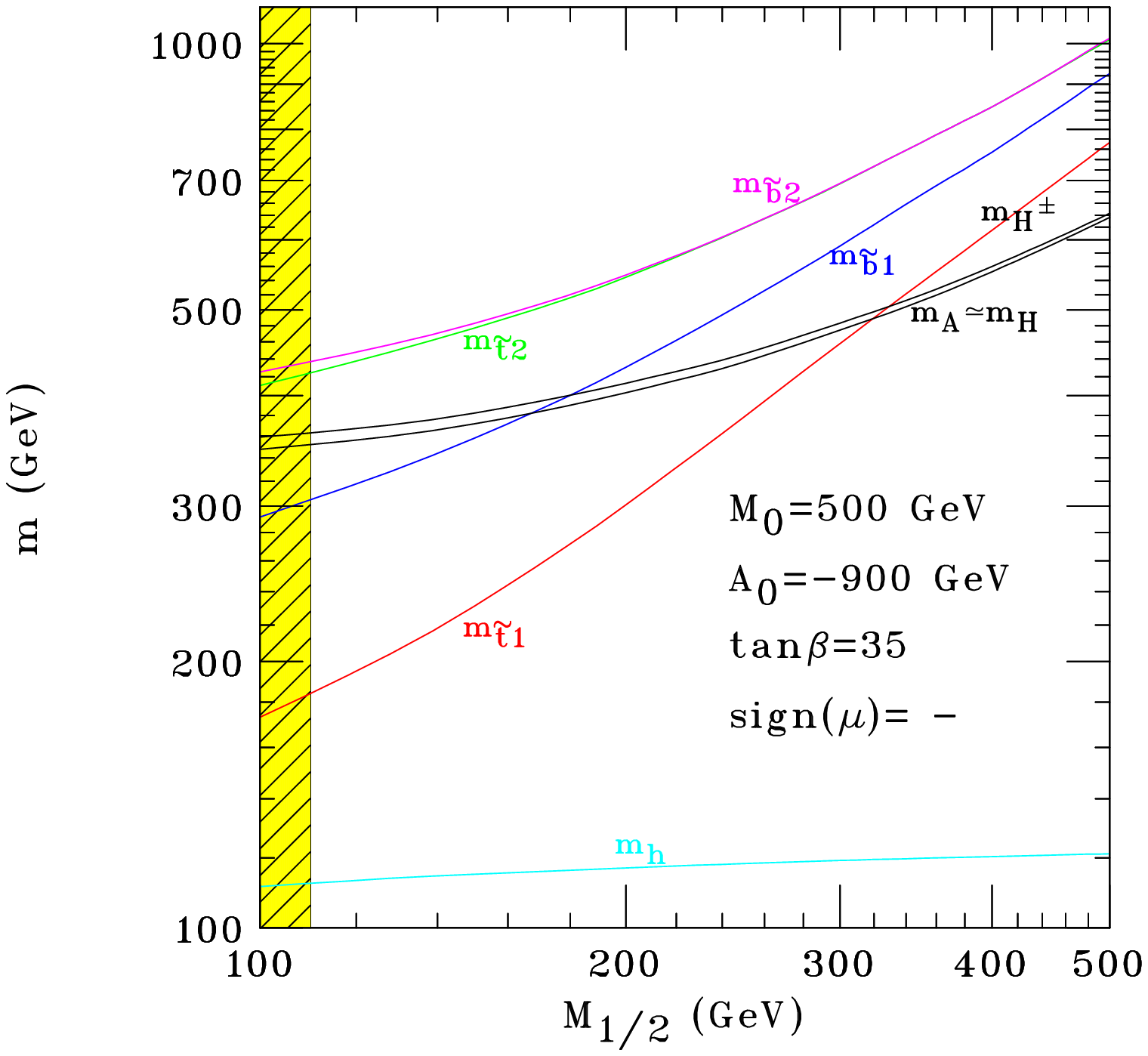,width=3.5in,angle=90}}
\caption{The most significant cross sections which survive
the choice of a heavy SUSY spectrum (above), alongside the 
values for the masses entering the corresponding production 
processes (below). Shaded regions indicate areas excluded by direct
searches.}
\label{fig:heavy}
\end{figure}
\end{document}